\documentclass[usenatbib]{mnras}
\usepackage[T1]{fontenc}
\DeclareRobustCommand{\VAN}[3]{#2}
\let\VANthebibliography\thebibliography
\def\thebibliography{\DeclareRobustCommand{\VAN}[3]{##3}\VANthebibliography}
\usepackage{graphicx}
\usepackage{newtxtext,newtxmath}
\usepackage{amsmath} 
\usepackage{caption}
\usepackage{hyperref}
\usepackage{threeparttable}
\usepackage{float}
\usepackage{placeins} 
\usepackage{titlesec}
\usepackage{enumitem}  
\usepackage{layout}
\usepackage{multirow}
\titleformat*{\subsubsection}{\bfseries}
\DeclareUnicodeCharacter{05BF}{\textsuperscript{}}
\usepackage{tabularx}
\newcommand{\Rvir}[1][]{R_{\mathrm{vir}#1}}
\newcommand{\Mvir}[1][]{M_{\mathrm{vir}#1}}
\newcommand{\mach}{\mathcal{M}}
\newcommand{\msun}{\,{\rm M}_\odot}
\newcommand{\tcool}{t_{\rm cool}^{\rm (s)}}
\newcommand{\tff}{t_{\rm ff}}
\newcommand{\dex}{\,{\rm dex}}
\newcommand{\kms}{\,{\rm km}\,{\rm s}^{-1}}
\newcommand{\kpc}{\,{\rm kpc}}
\newcommand{\Tvir}{T_{\rm vir}}
\newcommand{\trident}{{\sc trident}}
\newcommand*{\ditto}{\texttt{"}}

\newcommand{\EWmgii}{\langle W_{\rm MgII} \rangle}

\newcommand{\plus}{\texttt{+}}

\newcommand{\NH}{N_{\rm H}}

\newcommand{\EWciv}{\langle W_{\ion{C}{iv}}\rangle}
\newcommand{\cloudy}{{\sc cloudy}}
\newcommand{\bt}{b_{\rm t}}

\title[The origin of strong UV absorbers]{Turbulence-dominated CGM: the origin of UV absorbers with equivalent widths of $\sim1$\AA}

\author[A. Kakoly et al.]{\Large \vspace{5pt} \! \! Aharon Kakoly$^{1}$\thanks{E-mail: aharonkakoly@mail.tau.ac.il}, 
Jonathan Stern$^{1}$, 
Claude-André Faucher-Giguère$^{2}$, 
Drummond B. Fielding$^{3}$, \\ 
\Large \rm Roy Goldner$^{1}$, Guochao Sun$^{2}$ and Cameron B.~Hummels$^{4}$
\\
$^{1}$School of Physics and Astronomy, Tel Aviv University, Tel Aviv 69978, Israel\\
$^{2}$Department of Physics, Astronomy and CIERA, Northwestern University, 1800 Sherman Avenue, Evanston, IL 60201, USA\\
$^{3}$Department of Astronomy, Cornell University, Ithaca, NY 14853, USA\\
$^{4}$TAPIR, California Institute of Technology, Mailcode 350-17, Pasadena, CA 91125, USA}

\pubyear{2024} 

\date{Accepted XXX. Received YYY; in original form ZZZ}
\begin{document}
\label{firstpage}
\pagerange{\pageref{firstpage}--\pageref{lastpage}}
\maketitle
\begin{abstract}
Theoretical arguments and observations suggest that in massive halos ($>10^{12}\msun$), the circumgalactic medium (CGM) is dominated by a `hot' phase with gas temperature near the virial temperature ($T\approx T_{\rm vir}$) and a quasi-hydrostatic pressure profile. Lower-mass halos are however unlikely to be filled with a similar quasi-static hot phase, due to rapid radiative cooling. 
Using the FIRE cosmological zoom simulations, we demonstrate that the hot phase is indeed sub-dominant at inner radii ($\lesssim0.3 R_{\rm vir}$) of $\lesssim10^{12}\msun$ halos, and the inner CGM is instead filled with $T\ll T_{\rm vir}$ gas originating in outflows and inflows, with a turbulent velocity comparable to the halo virial velocity. The turbulent velocity thus exceeds the mass-weighted sound speed in the inner CGM, and the turbulence is supersonic. 
UV absorption features from such CGM trace the wide lognormal density distributions of the predominantly cool and turbulent volume-filling phase, in contrast with tracing localized cool `clouds' embedded in a hot medium. We predict equivalent widths of $W_\lambda\sim2\lambda v_{\rm c}/c\sim 1$\AA\ for a broad range of strong UV and EUV transitions (Mg\,II, C\,II, C\,IV, Si\,II--IV, O\,III--V) in sightlines through inner CGM dominated by turbulent pressure of $\lesssim L^\star$ galaxies at redshifts $0\leq z\lesssim2$, where $\lambda$ is the transition wavelength, $v_{\rm c}$ is the circular velocity and $c$ is the speed of light. 
Comparison of our predictions with observational constraints suggests that star forming $\lesssim$\,$L^\star$ and dwarf galaxies are generally dominated by turbulent pressure in their inner CGM, rather than by thermal pressure. The inner CGM surrounding these galaxies is thus qualitatively distinct from that around quenched galaxies and massive disks such as the Milky-Way and M31, in which thermal pressure likely dominates.
\end{abstract}


\section{INTRODUCTION}
At which halo mass and redshift does a quasi-static ‘hot’ phase, with temperature comparable to the halo virial temperature $T_{\rm vir}$, form in the circumgalactic medium (CGM)?
The development of the hot CGM phase and its implications for galaxy evolution  have been the subject of classic galaxy evolution papers since the late 1970s \citep{silk77, rees77, white78, white91, Birnboim03,keres05, dekel06}. These studies argued that a hot CGM that contracts slowly onto the galaxy can form only if its cooling time exceeds the halo free-fall time, corresponding to the mass of the dark matter halo exceeding a threshold in the range $\sim10^{11}-10^{12}\msun$. 
Since the formation of the hot CGM alters the physics of galaxy accretion and feedback, its effects on galaxy evolution are plausibly profound. Indeed, 
the interplay between hot CGM formation and black hole (BH) feedback has been a leading explanation for quenching of star-formation, due to the increased susceptibility of the hot phase to energy deposition by BH feedback \citep{dekel06,Croton06,Bower06,Somerville08}, or due to the confinement of stellar-driven outflows by the hot CGM, which allows the central BH to grow thus enabling BH feedback \citep{Bower17,Byrne23}. More recent studies have argued that hot CGM formation may also facilitate the formation of thin galactic disks observed mainly at redshift $z\lesssim1$  \citep{Sales12, Stern21A, Yu21, Yu23, Hafen22, Gurvich23, Byrne23}, due to the ability of hot accreting gas to develop an aligned angular momentum distribution prior to joining the galaxy \citep[see especially][]{Hafen22}. These scenarios demonstrate that the hot CGM likely has a significant role in galaxy evolution, and it would thus be useful to identify its formation in observations. 

Recent theoretical advances on the nature of hot CGM formation provide insight into its observational signatures.
First, CGM simulations that account for stellar feedback have demonstrated that hot $T\sim\Tvir$ gas also exists at halo masses below the threshold, but is limited to transient and localized radiative shocks, in contrast with a long-lived and volume-filling hot medium at high masses \citep{vandeVoort16, Fielding17, Stern20, Stern21A, Lochhaas20, Gurvich23, Pandya23}. This is in contrast with earlier studies which did not include feedback and associated hot CGM formation with whether a virial shock occurs at all \citep[e.g.,][]{Birnboim03}.  In the presence of feedback, `hot CGM formation' is thus  more accurately defined as a transition from a kinetic energy-dominated CGM (i.e., turbulence and coherent inflow/outflow) to a thermal energy-dominated CGM \citep{Fielding17}. 
Second, recent studies have shown that the mass threshold for hot CGM formation depends on CGM radius, with a larger mass threshold of $\sim10^{12}\msun$ at inner CGM radii where cooling times are relatively short and a lower threshold of $\sim10^{11}\msun$ at outer CGM radii where cooling times are long
\citep{Stern20, Stern21A, Stern21B, Gurvich23}. Thus, in a given halo the quasi-static hot CGM forms first at large CGM radii and later at inner radii. This `outside-in' hot CGM formation process is in contrast with the `inside-out' picture suggested by the simulations of \cite{Birnboim03} which did not include feedback. The transition in the inner CGM which has a higher mass threshold, dubbed `inner CGM virialization' or ICV, appears especially important for the evolution of the central galaxy since it changes the immediate environment in which the galaxy evolves \citep{Stern21A}. 

How can we identify if a quasi-static and long-lived hot phase exists in the inner CGM of star forming galaxies? Addressing this question via hot phase observations can be challenging, since one needs to determine whether this phase constitutes a quasi-static volume-filling medium as expected after ICV, or rather the transient bursts originating in feedback expected prior to ICV. In the Milky-Way, observations of \ion{O}{vii} and \ion{O}{viii} absorption lines in the CGM appear to require that a quasi-static hot phase indeed extends down to the galaxy \citep{Pezzulli17, Bregman18, Sormani18, Stern19, Stern24, Faerman20, Singh24, Sultan24}, i.e., the Milky-Way is post-ICV. A similar conclusion can be drawn for nearby massive spirals, based both on resolved X-ray emission  \citep[e.g.,][]{Anderson16} and on measurements of the thermal Sunyaev Zeldovich (tSZ) effect \citep{Bregman22,Oren2024}. For more distant or lower mass galaxies where these X-ray and tSZ observations are less constraining, and especially around the low-mass and high-redshift galaxies expected to be pre-ICV, the nature of the hot phase in the inner CGM is still unclear. 
 
An alternative method to identifying a quasi-static hot CGM directly is to infer its existence / absence from the properties of the cool CGM ($\sim10^4$\,K), a gas phase considerably more accessible to observations. Indeed, the first  observational indications for a hot CGM were the inferred high thermal pressures of cool \ion{Ca}{ii}-absorbing gas around the Milky-Way \citep{Spitzer56}. This classic result raises the question: \textit{what are the properties of the cool CGM when the hot phase is sub-dominant}?  \cite{Theuns21} and \cite{Stern21B} addressed this question for high-redshift CGM ($z\gtrsim 3$), focusing on implications for observations of Damped Ly$\alpha$ absorbers (DLAs).
In the present study we focus on UV absorption signatures of pre-ICV CGM at lower redshifts of $z\lesssim2$, at which large samples of circumgalactic absorbers and their associated galaxies are available (see below). 
Based on the properties of pre-ICV CGM mentioned above, we expect UV absorption features in such systems to indicate the presence of large cool gas columns and high velocity dispersions. In the present study we quantify these predictions using the FIRE-2 cosmological zoom simulations \citep{Hopkins18}.

The inner CGM of $\lesssim L^*$ galaxies, which is predicted to be pre-ICV, has a relatively small cross-section for sightlines to background sources in which UV absorption can be detected. Large samples of UV absorber-galaxy pairs are thus required for robust statistics, and more so given the large halo-to-halo and sightline-to-sightline variability. On the other hand, the large cool gas columns and velocity dispersions expected in pre-ICV halos imply large absorption equivalent widths, which are detectable even in relatively low resolution ($\sim100\kms$) spectra. Such large samples of UV absorber-galaxy pairs are available from the Sloan Digital Sky Survey (SDSS, \citealt{Lan14,Lan20,Anand21}) and more recently in larger numbers from the Dark Energy Survey Instrument (DESI, \citealt{Wu24}), mainly in the \ion{Mg}{ii}\,$\lambda\lambda2796,2803$ transition. These samples provide the main observational constraint to which we compare our predictions. 

This paper is structured as follows. In section~\ref{s:methods} we describe the FIRE simulations and our analysis technique. Section~\ref{s:results} presents the predicted UV absorption properties before and after a volume-filling hot phase forms in the inner CGM. These predictions are compared to observational constraints in section~\ref{s:obs}. We discuss our results in section~\ref{s:discussion} and provide a summary of our main conclusions in section~\ref{s:summary}.
Throughout the paper we use cosmological parameters from the Planck 2018 results \citep{Plank2018} with Hubble parameter $H_0=70\,\text{km}\,\text{s}^{-1}\,\text{Mpc}^{-1}$, $\Omega_{\rm m}=0.3$, and $\Omega_{\rm b}=0.05$.

\section{METHODS}\label{s:methods}
We analyze FIRE-2 simulations \citep{Hopkins18} in order to deduce the properties and observational signatures of circumgalactic UV absorbers prior to the formation of a quasi-static hot phase. In these simulations a hot phase forms in the inner CGM ($\sim0.1\,\Rvir$, where $\Rvir$ is the halo virial radius) when the halo mass reaches $\approx10^{12}\msun$ \citep{Stern21A}. We utilize the \trident-v1.3\ code \citep{Hummels2017} to generate synthetic UV absorption spectra.

\subsection{FIRE simulations}\label{s:FIRE Simulations}

The Feedback In Realistic Environments project\footnote{See the FIRE project website: \hyperlink{http://fire.northwestern.edu}{http://fire.northwestern.edu}.} (FIRE, \citealt{hopkins14, Hopkins18, hopkins23}) investigates how feedback mechanisms influence galaxy formation in cosmological simulations. We utilize the second iteration of the FIRE cosmological `zoom' simulations (FIRE-2), which allow investigating physical processes within the CGM including the distribution and dynamics of cold and hot gas. FIRE-2 use the GIZMO code for gravity and hydrodynamics computations \citep{hopkins15} in its meshless finite-mass mode (MFM). Gravity calculations are conducted using a modified version of the Tree-PM solver, similar to the one used in GADGET-3 \citep{Springel2005}, with adaptive softening for gas resolution elements. Gas heating and cooling processes include metal line cooling, free-free emission, photoionization, recombination, Compton scattering with the cosmic microwave background, collisional and photoelectric heating by dust grains, and cooling processes at low temperatures ($<10^4 \, \rm{K}$), including molecular and fine-structure cooling. The relevant ionization states are derived from precomputed \cloudy\ tables \citep{Ferland1998}, accounting for the effects of the cosmic UV background from \cite{FaucherGiguere2009} and galactic radiation sources (see \citealt{Hopkins18}). 
FIRE-2 also includes a sub-grid turbulent diffusion model that captures unresolved mixing of metals and thermal energy between gas resolution elements, which is important for modeling the distribution and observability of metal absorbers \citep{Hopkins17,Escala18}.
Star formation and stellar feedback are treated in a sub-grid manner. Star formation occurs within self-gravitating, self-shielded molecular gas with $n_{\rm H}>1000 \, \rm{cm^{-3}}$ \citep{hopkins13}. Feedback from stars is implemented through radiation pressure, heating via photoionization and photoelectric processes, and the deposition of energy, momentum, mass, and metals from supernovae and stellar winds. Feedback parameters and their time dependence are based on the stellar evolution models of \cite{Leitherer1999}, assuming a \cite{Kroupa2001} initial mass function. The simulations analyzed in this work do not include feedback from active galactic nuclei (AGN), a caveat we address in the discussion. A full description of the FIRE-2 simulations is provided in \cite{Hopkins18}.

\subsection{Simulation selection}

We analyze nine representative FIRE-2 simulations, as listed in table \ref{table:FIRE-2Simulations}. Five simulations (m12i, m12b, m12w, m12c, m12f) have masses similar to the Milky-Way, with $M_{\rm halo} \sim 10^{12} \,\rm M_\odot$ and an $\sim L^*$ galaxy at $z=0$. We focus on this mass range in this paper. To explore the dependence on halo mass history we analyze also a simulation of a dwarf galaxy halo with $M_{\rm halo}\sim 10^{11}\,\rm M_\odot$ at $z=0$ (m11d) and a massive halo with $M_{\rm halo} \sim 0.4 \cdot 10^{13}\,\rm M_\odot$ at $z=1$ (m13A1). This allows us to demonstrate that the formation of a quasi-static hot phase occurs earlier in more massive halos. The typical baryonic resolution element is $m_{\rm b}=7100 - 33\,000\, \mathrm{M}_\odot$, ensuring that sub-grid prescriptions are applied at the level of giant molecular clouds or better, while $\sim L^*$ galaxy discs are resolved with $\sim 10^6$ particles. We analyze also m12i simulations with different resolutions in order to test the dependence of our main results on resolution. The virial radius $\Rvir$ and virial mass ($\Mvir$) of the central halo in each snapshot are determined using the Amiga Halo Finder (AHF, \citealt{Knollmann2009}), based on the \cite{Bryan98} criterion. 

\subsection{Ion fractions}\label{s:ion fraction}

Ion fractions $f_{\rm ion}$ in the CGM were calculated as a function of gas density, temperature, metallicity, and redshift in the simulation, using the \trident\ ionization tables which assume collisional ionization equilibrium and photoionization equilibrium from the meta-galactic ultraviolet background in \citet{Madau2012}. 
Ion volume densities were then calculated using the formula:
\begin{equation}
\label{eq:IonVolumeDensity}
n_{\text{ion}} = n_{\rm H} \cdot \frac{N_{\rm X}}{N_{\rm H}} \cdot f_{\text{ion}},
\end{equation}
where \( n_{\rm H} \) is the hydrogen number density and $N_{\rm X}/N_{\rm H}$ represents the abundance of element X relative to hydrogen as tracked in FIRE.
We focus on ions with UV transitions commonly observed in the CGM. The transitions are listed in Table \ref{table:IonsTable}, which includes the ion name, transition wavelength ($\lambda$), oscillator strength ($f_{\rm lu}$), and $N_{\rm X}/N_{\rm H}$ assuming $Z=0.3\,{\rm Z}_\odot$. 

Our calculation of $f_{\rm ion}$ does not include the contribution of ionizing photons from young stars in the galaxy. The effect of these local sources on the ionization state of the CGM in FIRE was explored by \cite{Holguin24} using a Monte Carlo radiation transfer code. At the radius of $0.2\,\Rvir$ which we focus on below, they deduced a small contribution of local sources at $z<1$ and a contirubion comparable to the UV background at $1<z<2$. Such a factor of $\lesssim2$ uncertainty in the ionizing background does not affect are conclusions, as discussed below.  

\newcommand{\machavg}{\tilde{\mach}_{\rm turb}}

\begin{table}
\centering
\begin{tabular}{|c|c|c|c|c|c|c|}
\hline
Name & $m_{\mathrm{b}}$ & $z_{\text{min}}$ & $M_{\mathrm{halo}}$($z_{\text{min}}$) & $M_*$($z_{\text{min}}$) & $z_{\text{ICV}}$ &Ref. \\
&  [$\mathrm{M}_\odot$] & & [$10^{12}\,\mathrm{M}_\odot$] & [$10^{10}\,\mathrm{M}_\odot$] & &\\
(1) & (2) & (3) & (4) & (5) & (6) & (7) \\
\hline
m12i & 7100 & 0 & $1.1$ & $7.3$ & 0.31 & D \\
m12b & 7100 & 0 & $1.3$ & $10$ & 0.64 & C \\
m12w & 7100  & 0 & 0.95 & 6.5  & 0.27 & G \\
m12c & 7100  & 0 & 1.3 & 6.8  & 0.6 & C \\
m12f & 7100  & 0 & 1.5 & 9.7  & 0.63 & B \\
m11d & 7100 & 0 & $0.3$ & $0.51$ & -- & A \\
m13A1 & 33000 & 1 & $4$ & $28$ & 3.1 & E \\
m12i & 880 & 0 & $0.9$ & $ 5.7$ & 0.62 & F \\
m12i & 57000 & 0 & $1.2$ & $9.1$ & 0.98 & D \\
\hline
\end{tabular}
\caption{FIRE-2 cosmological zoom simulations used in this work. 
(1) galaxy name; (2) mass of baryonic resolution element; (3) final redshift of the simulation; (4) central halo mass at the final redshift; (5) stellar mass of central galaxy at the final redshift; (6) redshift of ICV at which inner CGM turbulence becomes subsonic; (7) Reference papers for simulations: A: \protect\cite{El-Badry2018A}, B: \protect\cite{Garrison-Kimmel2017}, C: \protect\cite{Garrison-Kimmel2019}, D: \protect\cite{Wetzel2016}, E: \protect\cite{Angles17b}, F: \protect\cite{Bhattarai22}, G: \protect\cite{Samuel2020}.}
\label{table:FIRE-2Simulations}
\end{table}

\subsection{Characteristic velocities}

We demonstrate below that the formation of a quasi-static hot CGM phase implies a transition in the nature of CGM turbulence, from supersonic to subsonic. To this end we calculate several radius and time-dependent characteristic velocities in the simulation. 

The circular velocity \( v_{\rm c} \) at a given time and radius is calculated using:
\begin{equation}
\label{eq:CircularVelocity}
v_{\rm c}(r,t) = \sqrt{\frac{GM\left(<r\right)}{r}},
\end{equation}
where \( G \) is the gravitational constant, and \( M\left(<r\right) \) is the total enclosed mass within a radius \( r \) in the snapshot corresponding to time $t$ including contributions from gas, dark matter, and stars. 
The mass-weighted sound speed \( \langle c_{\rm s} \rangle_{\rho} \) at a given time $t$ and radius $r$ is calculated via
\begin{equation}
\label{eq:SoundSpeedVelocity}
\langle c_{\rm s} \rangle_{\rho} =  \cdot \frac{\sum_i m_i \sqrt{k_{\rm B} T_i/(\mu m_{\rm p})}}{\sum_i m_i},
\end{equation}
where \( k_{\rm B} \) is the Boltzmann constant, \( \mu \) is the mean molecular weight (taken to be 0.62), \( m_{\rm p} \) is the proton mass, \( m_i \) and \( T_i \) are the mass and temperature of the \( i \)-th gas resolution element. The summations are done on all resolution elements within a radial shell centered on radius $r$ with thickness  \( \delta r = 0.01R_{\rm vir} \), in a simulation snapshot corresponding to time $t$. 
Finally, the 3D turbulent velocity $\sigma_{\rm turb}$ at radius $r$ and time $t$ is calculated via
\begin{equation}
\label{eq:SigmaTurb}
\sigma_{\text{turb}} = \sqrt{\sigma_{r}^2 + \sigma_{\theta}^2 + \sigma_{\phi}^2},
\end{equation}
where \(\sigma_{r}\), \(\sigma_{\theta}\), and \(\sigma_{\phi}\) are the velocity dispersions in the radial, polar, and azimuthal directions, respectively. The angle $\theta$ is defined with respect to the total angular momentum vector of stars within $0.2 R_{\rm vir}$, and $\phi$ is the corresponding azimuthal angle.  
To obtain the velocity dispersion in each direction, we calculate the variance of the velocity components as follows:
\begin{equation}
\label{eq:SigmaTurbComponents}
\sigma_{j}^2=\overline{(u_j-\langle u_j \rangle_{\rho})^2}
\end{equation}
where $u_j$ represents the velocity component in the \(j\)-th direction for an individual resolution element ($j=r,\theta,\phi$), and as above $\langle\rangle_\rho$ indicates a mass-weighted average. Summation is done over all gas resolution elements in a shell centered on radius $r$ with thickness $0.01\,R_{\rm vir}$. Our choice to calculate the dispersion in spherical coordinates avoids contributions to $\sigma_{\rm turb}$ from coherent motions in the radial or rotational directions.  

\begin{table}
\centering
\small
\begin{tabular}{|c|c|c|c|c|}
\hline
Ion & $\lambda$ [Å] & $f_{\rm lu}$ & $N_{\rm X}/N_{\rm H}$ $(Z=0.3\,{\rm Z}_\odot)$ & $\chi$ [eV] \\
(1) & (2) & (3) & (4) & (5) \\
\hline
Si II & 1260 & 1.22 & $9.7 \times 10^{-6}$ & 8.2 \\
Si III & 1206 & 1.67 & \ditto & 16.3 \\
Si IV & 1403 & 0.255 & \ditto & 33.5 \\
Mg II & 2796 & 0.608 & $1.2 \times 10^{-5}$ & 7.6 \\
Mg II & 2803 & 0.303 & \ditto & \ditto \\
S III & 1190 & 0.61 & $4.0 \times 10^{-6}$ & 23.3 \\
O III & 833 & 0.106 & $1.5 \times 10^{-4}$ & 35.1 \\
O IV & 788 & 0.111 & \ditto & 54.9 \\
O V  & 630 & 0.512  & \ditto & 77.4 \\
O VI & 1031 & 0.133 & \ditto & 113.9 \\
C II & 1335 & 0.127 & $8.1 \times 10^{-5}$ & 11.3 \\
C IV & 1548 & 0.19 & \ditto & 47.9 \\
\hline
\end{tabular}
\caption{Absorption features discussed in this work. (1) Ion; (2) wavelength; (3) oscillator strength; (4) element abundance relative to hydrogen assuming a CGM metallicity of $0.3\,{\rm Z}_\odot$; (5) ionization potential.}
\label{table:IonsTable}
\end{table}

\subsection{Cooling time and free-fall time}\label{s:tcool and tff}

The formation of a long-lived and quasi-static hot phase occurs when  $\tcool$, the cooling time of shocked gas at the halo virial temperature (`s' for shocked) exceeds the free-fall time $\tff$ \citep{white78,Birnboim03,Mo24}. \cite{Stern20} and \cite{Stern21A} demonstrated that $\tcool/\tff$ increases with CGM radius in a given halo, and thus hot phase formation occurs above a higher halo mass threshold of $\approx10^{12}\msun$ at inner CGM radii than at outer CGM radii (see Introduction). In this work we focus on this transition in the inner CGM which is expected to occur at $z<1$ for Milky-Way mass galaxies.
We measure $\tcool$ and $\tff$ in the FIRE simulations at inner CGM radii as described in \cite{Stern21A} and summarized here. We calculate $\tcool$ at a given radius $r$ via:
\begin{equation}
\label{eq:t_cool}
t_{\rm cool}^{\rm (s)} \equiv t_{\rm cool}\left(T^{\rm (s)}\right) = \frac{(3/2) \cdot k_{\rm B} T^{\rm (s)}}{ \mu X\langle n_{\rm H}\rangle \Lambda},
\end{equation}
where $\langle n_{\rm H}\rangle$ is the volume-weighted average density in a shell with radius $r$, $X$ is the hydrogen mass fraction, and $T^{({\rm s})}$ is the temperature of a cooling flow in the subsonic limit
\begin{equation}\label{eq:T^s}
    T^{\rm (s)} = \frac{3}{5}\frac{\mu m_{\rm p}v_{\rm c}^2}{k_{\rm B}} \approx T_{\rm vir}
\end{equation}
(see equation 24 in \citealt{Stern19}). The cooling function \( \Lambda \) is defined such that $n_{\rm H}^2\Lambda$ is the radiative energy loss rate per unit volume and is calculated using the tables in  \cite{Wiersma09}, using $T^{({\rm s})}$, $Z$, $\langle n_{\rm H}\rangle$ and $z$ as imput parameters.  
The free-fall time is calculated as:
\begin{equation}
\label{eq:t_ff}
t_{\rm ff} = \frac{\sqrt{2}r}{v_{\rm c}}.
\end{equation}
Note that since $\Lambda(T)$ decreases roughly as $T^{-0.5}$ at $T\approx10^6\,{\rm K}$, the ratio $t_{\rm cool}^{\rm (s)}/\tff$ scales strongly with $v_{\rm c}$, as $\sim v_{\rm c}^4$.

\begin{figure*}
    \includegraphics[width=\textwidth]{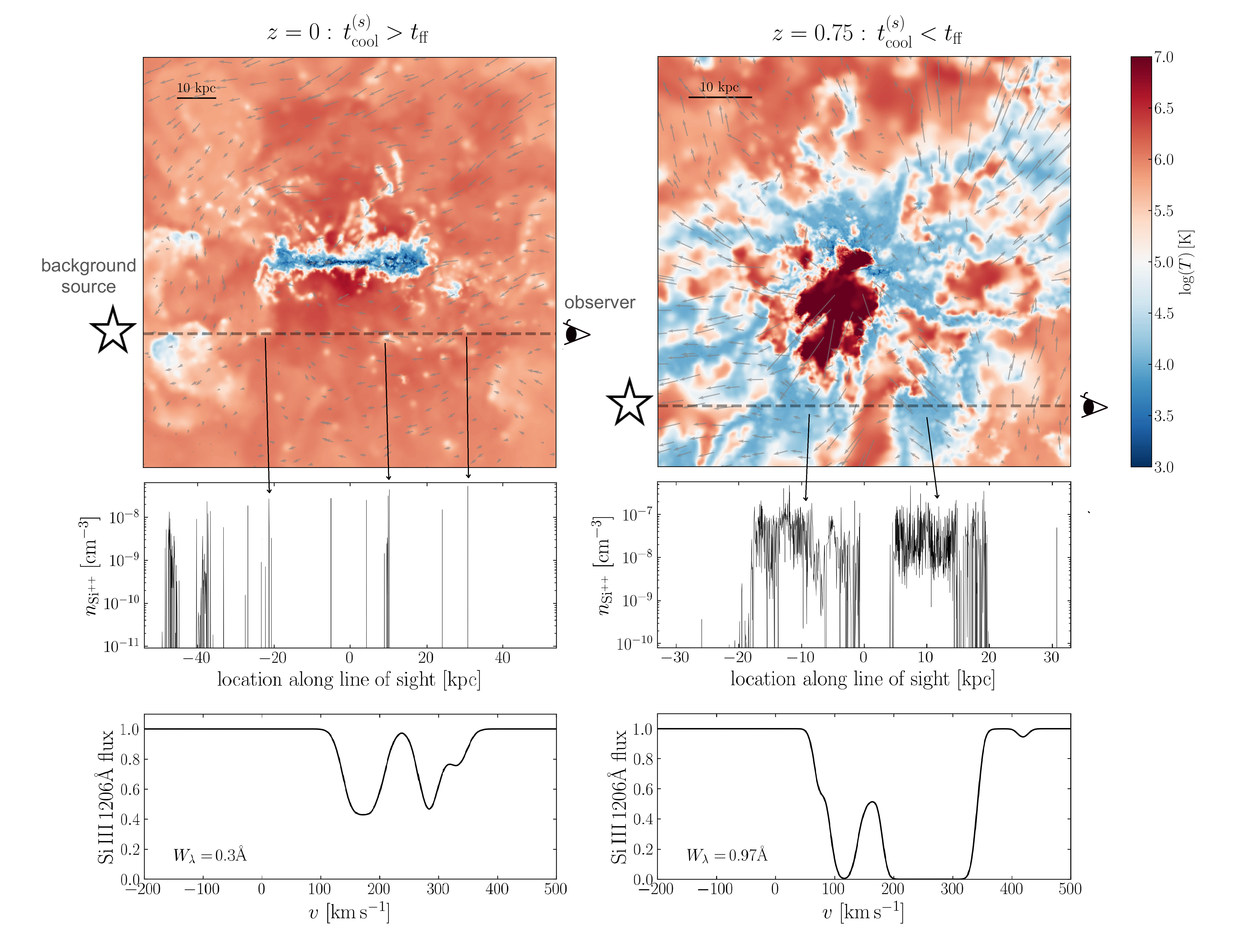}
    \caption{A qualitative difference between UV absorbers before and after a quasi-static hot phase forms in the CGM.
    Columns show two snapshots of a Milky Way-mass galaxy simulated in FIRE (`m12i'). Left panels show the $z=0$ snapshot where hot gas cools slowly so the inner halo is filled with a quasi-static hot phase, while right panels show the $z=0.75$ snapshot where hot gas cools rapidly so the hot phase is limited to localized and transient clouds. Top panels show gas temperature maps extending to $0.2\Rvir$ and the trajectories of mock sightlines through them (dashed lines). Middle panels show Si$^{++}$ volume density along the mock sightlines, with $0$\,kpc defined as the location nearest to the halo center, and arrows connecting cool gas in the images with peaks in $n_{\rm Si^{++}}$. Bottom panels show  predicted \ion{Si}{iii} absorption spectra for spectral resolution $\lambda/\Delta \lambda=20\,000$. At $z=0$, \ion{Si}{iii} traces localized cool clouds embedded in a hot $T\approx T_{\rm vir}$ medium, consistent with the common paradigm for UV absorbers. In contrast at $z=0.75$ \ion{Si}{iii} traces the cool volume filling phase of the inner CGM.}
    \label{fig: TemperatureProjectionsWithSi_IIISpectrum_m12i_z_0_z_0.75}
\end{figure*}

\section{RESULTS: UV ABSORPTION SIGNATURES OF TURBULENCE-DOMINATED INNER CGM}\label{s:results}

ֿ\subsection{UV absorbers trace the volume-filling medium}

Figure~\ref{fig: TemperatureProjectionsWithSi_IIISpectrum_m12i_z_0_z_0.75} plots the temperature distribution and absorption characteristics in the inner CGM of the m12i FIRE simulation, for an example absorption line Si\,III~$\lambda1206$. The left column shows the $z=0$ snapshot after the quasi-static hot phase has formed in the inner CGM ($\tcool=9\,\tff$), while the right snapshots shows $z=0.75$ (lookback time $t_{\rm lookback}\approx6.6 \,{\rm Gyr}$), before the quasi-static hot phase has formed ($\tcool=0.3\,\tff$). The top row shows temperature maps of the ISM and inner CGM in these two snapshots, produced using a mass-weighted projection of $\log T$ in a slice with depth 1\,kpc centered on the halo center. Panel size is $0.4\Rvir$ in each axis. Gas velocities relative to the halo center in the projection plane are overplotted as gray arrows. 

In the \( z=0 \) snapshot shown on the left of Fig.~\ref{fig: TemperatureProjectionsWithSi_IIISpectrum_m12i_z_0_z_0.75} the temperature map shows a prominent hot, diffuse medium with localized cooler regions, or `clouds'. 
In contrast, the \( z=0.75 \) snapshot shown on the right reveals a more disordered temperature distribution with widespread cooler regions, as found by previous studies of FIRE simulations \citep{Stern21A, Gurvich23}. 

The dashed lines in the top panels represent trajectories of mock sightlines used for analyzing UV absorption. The middle panels plot the volume density of Si$^{++}$ ions $n_{\rm Si^{++}}$ along these sightlines, where $0$ is defined as the location along the sightline which is nearest to the halo center. At $z=0$ after the hot quasi-static phase has formed, $n_{\rm Si^{++}}$ is characterized by isolated peaks, corresponding to localized cool clouds embedded in the hot CGM as in the common paradigm for UV absorbers \citep[e.g.,][]{Tumlinson17}. In contrast, at $z=0.75$, the Si$^{++}$ density profile is more continuous and widespread along the sightlines, indicating that Si$^{++}$ ions are part of the volume-filling phase rather than limited to localized clouds. We show below that this alternative paradigm, where circumgalactic UV absorbers trace the volume filling phase, is applicable to all UV ions originating in inner CGM without a quasi-static hot phase.  

The bottom panels of Fig.~\ref{fig: TemperatureProjectionsWithSi_IIISpectrum_m12i_z_0_z_0.75} show the absorption spectrum of Si\,III~1206\AA\ in the mock sightline, generated with \trident\ assuming an instrumental spectral resolution of $R=20\,000$. The absorption line equivalent widths $W_\lambda$ are noted in each panel. 
At $z=0$ the absorption feature is optically thin and relatively weak with $W_\lambda=0.3\text{\AA}$, due to the less extensive presence of cool gas. Conversely, at $z=0.75$, the absorption feature is saturated over a range of $\gtrsim100\kms$ with $ W_\lambda=0.97\text{\AA}$, due to the large amount of cool gas in the inner CGM. 

\begin{figure}
    \includegraphics[width=\columnwidth]{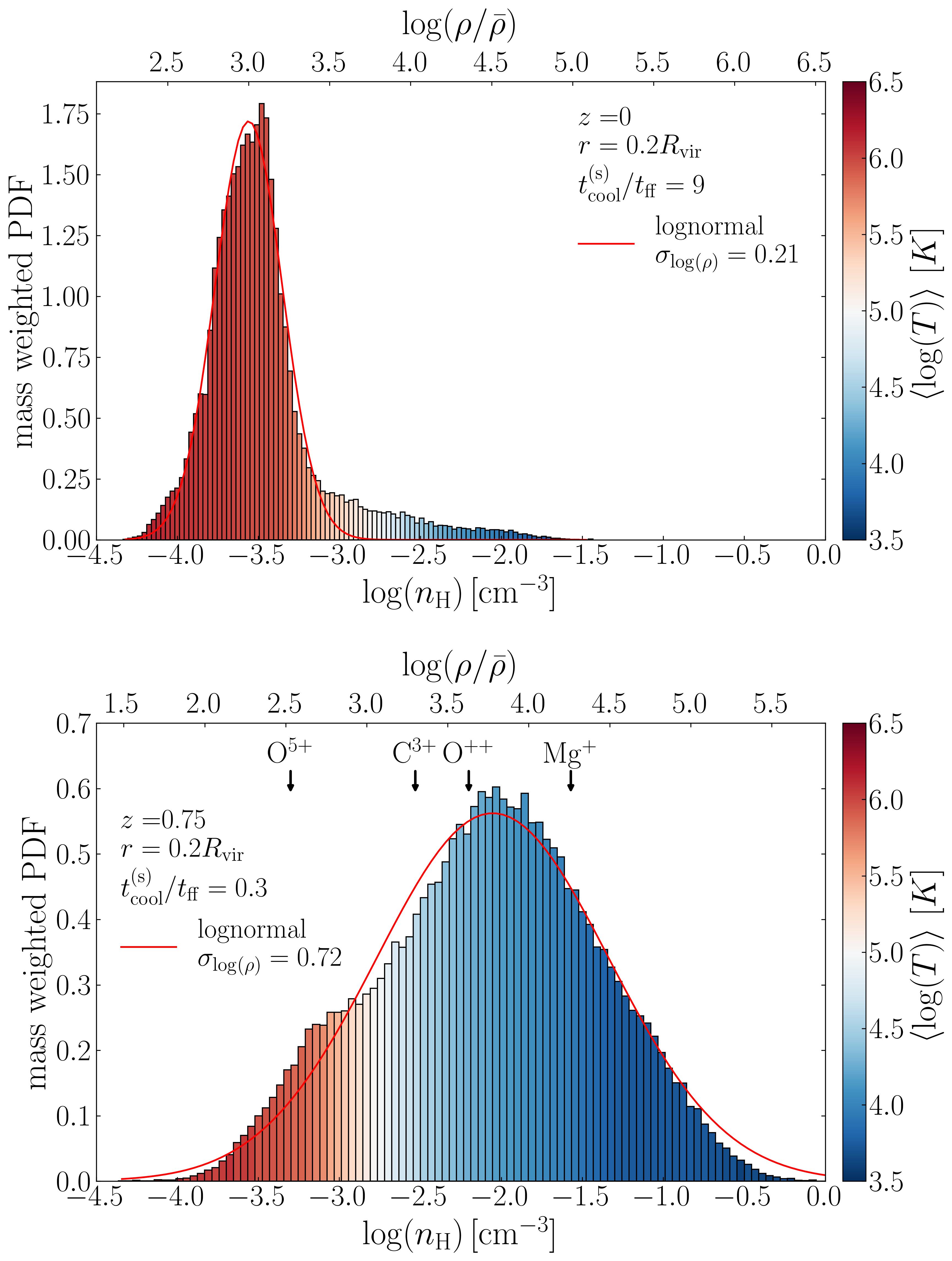}
    \caption{Density distribution of gas in a thin shell with $r=0.2\Rvir$, for the two snapshots shown in Fig.~\ref{fig: TemperatureProjectionsWithSi_IIISpectrum_m12i_z_0_z_0.75}. Color indicates the mean temperature at each density, and the top axes denote the density in units of the cosmic mean. The ratio of hot gas cooling time to free-fall time is noted in the panels. At $z=0$ (top) the distribution is narrow with $\sigma_{\log \rho}=0.21$ and a tail towards high densities, as expected in the common CGM paradigm with a low density hot phase and denser cool clouds. Conversely, at $z=0.75$ (bottom) the gas is predominantly cool and the distribution is a broad lognormal with $\sigma_{\log \rho}=0.72$, as expected in isothermal turbulent gas with Mach $\sim10$. Labels denote the median density of gas which produces each ion.}
    \label{fig: AllGasMassWeightedlog(rho)Distribution_m12i_z_0_z_0.75}
\end{figure}

\begin{figure}
    \centering
    \includegraphics[width=0.98\columnwidth]{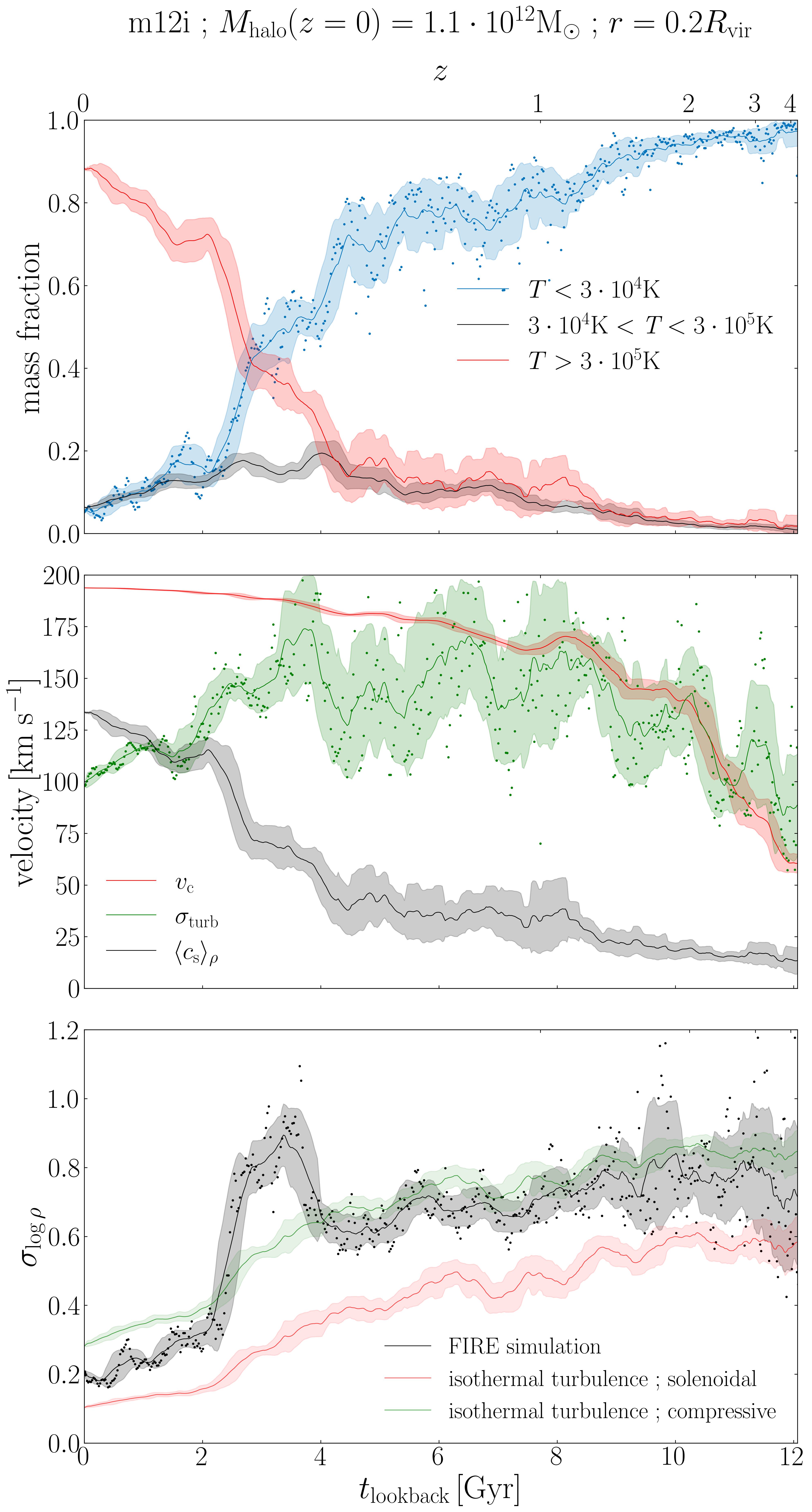}
    \caption{(Top:) Evolution of mass fractions of gas with different temperatures at $r=0.2\,\Rvir$ in the m12i simulation. Dots denote individual snapshots (shown only for cool gas for clarity), while lines and bands denote means and dispersions within a running $600\,{\rm Myr}$ window. The mass fraction of hot gas (red) sharply increases after $t_{\rm lookback} \approx 4 \rm \, Gyr$. (Middle:) Evolution of 3D turbulent velocity (green), circular velocity (red), and mass-weighted sound speed (black) at $0.2\,\Rvir$. CGM turbulence is highly supersonic when the cool phase dominates. 
    (Bottom:) Black dots and line denote the width of the density distribution at $0.2\,R_{\rm vir}$, derived using a lognormal fit to the distribution in each snapshot (see Fig.~\ref{fig: AllGasMassWeightedlog(rho)Distribution_m12i_z_0_z_0.75}). Red lines indicate the expected $\sigma_{\log\rho}$ based on isothermal  turbulence simulations with two types of driving mechanisms and the same Mach number as in FIRE (eqns.~\ref{eq:SigmaLnRho} -- \ref{eq:MachTurb}). A wide density distribution is apparent when turbulence is supersonic ($> 4 \rm \, Gyr$) consistent with isothermal turbulence simulations with compressive driving.   }
    \label{fig: MassFractionandSigmaRhoandMuVAgainstLookbackTime_m12i}
\end{figure}

\subsection{Wide lognormal gas density distributions}

Figure~\ref{fig: AllGasMassWeightedlog(rho)Distribution_m12i_z_0_z_0.75} presents mass-weighted gas density distributions in the two snapshots shown in Fig.~\ref{fig: TemperatureProjectionsWithSi_IIISpectrum_m12i_z_0_z_0.75}. 
We include in the distribution all gas resolution elements within a thin shell with a radius of $0.2\,R_{\rm vir}$, without any selection by gas temperature. This radius is chosen as a representative radius of the inner CGM that is beyond regions where angular momentum support is dominant (typically $\lesssim 0.05\,\Rvir$, see \citealt{Stern21A}). Shell thickness is chosen to be $\delta r = 0.01R_{\rm vir}$, in order to avoid the range in density induced by radial gradients. 
The color indicates the average $\log T$ of gas at each density, while the thin curves plot lognormal fits to the density distributions, defined via:
\begin{equation}
\label{eq:LogNormalDistribution}
\text{PDF}(\log \rho) = \frac{1}{\sqrt{2 \pi} \sigma_{\log \rho}} \exp \left( -\frac{\left(\log \rho - \mu_{\log\rho}\right)^2}{2 \sigma_{\log \rho}^2} \right),
\end{equation}
where \(\rho\) is the density, $\mu_{\log\rho}$ is the mean of \(\log \rho\), and $\sigma_{\log \rho}$ is the standard deviation of \(\log \rho\). Here and henceforth we use $\log$ as shorthand for $\log_{10}$.

At $z=0$, the density distribution shown in the top panel of Fig.~\ref{fig: AllGasMassWeightedlog(rho)Distribution_m12i_z_0_z_0.75} has a Gaussian-like shape with $\sigma_{\log \rho} \approx 0.21\dex$ and an asymmetric tail extending towards higher densities. The Gaussian is dominated by temperatures of $\gtrsim10^6\,{\rm K}$, while the tail has lower temperatures of $10^4-10^5\,{\rm K}$. This distribution demonstrates that the gas is predominantly in a hot diffuse phase with some cooler regions of higher density, as indicated also by Fig.~\ref{fig: TemperatureProjectionsWithSi_IIISpectrum_m12i_z_0_z_0.75} and consistent with the common `multi-phase' paradigm of the CGM. 

At $z=0.75$, the density distribution shown in the bottom panel of Fig.~\ref{fig: AllGasMassWeightedlog(rho)Distribution_m12i_z_0_z_0.75} is substantially broader and more symmetric than at $z=0$. A lognormal distribution with $\sigma_{\log \rho} \approx 0.72\dex$ provides a reasonable fit to the entire distribution, regardless of gas temperature. Most of the gas is cool, with $\approx 74\%$ of the mass having $T<3\cdot10^4\,{\rm K}$. The hot $T>10^{5.5}\,{\rm K}$ gas accounts only for $\approx 14\%$ of the total gas mass and $58\%$ of the volume. The origin of this wide lognormal distribution, which is distinct from the hot peak \plus\ cool tail distribution shown in the top panel, is further discussed below.

The hot phase being sub-dominant at $r=0.2\,\Rvir$ in the $z=0.75$ snapshot is a result of the short cooling time ($\tcool<\tff$) at this redshift and radius, which implies that any hot gas formed via accretion or feedback shocks rapidly cools \citep{Stern21A}. 
This is in contrast with cases where $\tcool>\tff$ and hence the hot shocked gas is long-lived, such as the $z=0$ snapshot at $r=0.2\,\Rvir$ (top panel of Fig.~\ref{fig: AllGasMassWeightedlog(rho)Distribution_m12i_z_0_z_0.75}) and the $z=0.75$ snapshot at a larger CGM radius of $r=0.5\,\Rvir$, shown in Appendix~\ref{sec:halfRvir}.

The evolution of hot gas mass fraction in the inner CGM is further explored in the top panel of Figure~\ref{fig: MassFractionandSigmaRhoandMuVAgainstLookbackTime_m12i}, which plots the mass fraction in each CGM phase versus $t_{\rm lookback}$. Blue shows the cool phase with temperatures $T < 3\cdot10^4$ K, red shows the hot phase with temperatures $T > 3\cdot10^5$ K, and intermediate temperatures are shown in black. As in Fig.~\ref{fig: AllGasMassWeightedlog(rho)Distribution_m12i_z_0_z_0.75}, we sum only gas resolution elements within a thin shell centered at $0.2\,R_{\rm vir}$, to avoid the effects of radial gradients. Each data point represents an individual snapshot from the simulation, while the lines and shaded regions indicate the average values and their dispersions within a window spanning 25 snapshots ($\approx600\,{\rm Myr}$). The top axis plots the corresponding redshift. 
There is a marked increase in the mass fraction of the hot gas phase at $2\lesssim t_{\rm lookback}\lesssim5\,{\rm Gyr}$. This is the formation of the quasi-static hot phase in the inner CGM as noted by previous FIRE studies, and is coincident with $\tcool$ exceeding $t_{\text{ff}}$ \citep{Stern21A,Gurvich23}.

Note that the mode of the density distributions in Fig.~\ref{fig: AllGasMassWeightedlog(rho)Distribution_m12i_z_0_z_0.75} shifts from $n_{\rm H}\approx 10^{-3.5}\,{\rm cm}^{-3}$ at $z=0$ to $n_{\rm H} \approx 10^{-2}\,{\rm cm}^{-3}$ at $z=0.75$. This shift to higher density at higher redshift is partially due to the increase in mean cosmic density. In units of the cosmic mean $\bar\rho=3\Omega_b H^2/8\pi G$ the mode is $\rho/\bar{\rho}=5\cdot10^3$ at $z=0.75$, compared to $\rho/\bar{\rho}=10^3$ at $z=0$  (see top-axes of Fig.~\ref{fig: AllGasMassWeightedlog(rho)Distribution_m12i_z_0_z_0.75}). 
The remaining difference arises from our choice to weigh the density distribution by mass. 
The volume-weighted mean density, i.e., the total shell mass divided by its volume, are similar in the two snapshots in units of $\bar{\rho}$.

\subsection{Supersonic turbulence}

The middle panel of Fig.~\ref{fig: MassFractionandSigmaRhoandMuVAgainstLookbackTime_m12i} shows the evolution of different velocity components at $0.2\,R_{\rm vir}$, calculated according to eqns.~(\ref{eq:CircularVelocity}) -- (\ref{eq:SigmaTurb}): turbulent velocity (green), circular velocity (red), and mass-weighted sound speed (black). 
At early times the mass-weighted sound speed is less than $50\kms$ due to the dominance of cool $\sim10^4\,{\rm K}$ gas as seen in the top panel. The turbulent velocity is $\approx150\kms$ and comparable to $v_{\rm c}$ especially at early times, indicating the CGM is dominated by turbulence. This result demonstrates that $\sigma_{\rm turb}$ significantly exceeds the mass-weighted sound speed, i.e., CGM turbulence is supersonic. When the hot phase becomes dominant at $t_{\rm lookback}\lesssim 2\,{\rm Gyr}$ the mass-weighted sound speed rises above $100\kms$, while the turbulent velocity declines from $\approx150\kms$ to $100\kms$. The formation of the hot quasi-static phase thus corresponds to turbulent velocities dropping below the mean sound speed, i.e., the turbulences become subsonic. 

Idealized simulations of isothermal turbulence demonstrate that the width of the density distribution increases with turbulent Mach number, roughly as (e.g.\ \citealp{Krumholz2014}):
\begin{equation}
\label{eq:SigmaLnRho}
    \sigma_{s}^2 = \ln\left(1 \, \text{+} \, \bt^2\mach_{\rm turb}^2\right)~,
\end{equation}
where $\sigma_{s}=0.43\,\sigma_{\log\rho}$ is the width of the lognormal density distribution for a natural logarithm, $\mach_{\rm turb}$ is the turbulent Mach number and $\bt$ is a unit-less parameter which equals $\approx 1$ for pure compressive turbulence driving and $\approx 1/3$ for pure solenoidal driving\footnote{The subscript `$\rm t$' differentiates $\bt$ from absorption line width $b$ used below.}. We check here whether equation~(\ref{eq:SigmaLnRho}) can explain the density distribution in the CGM of the non-isothermal FIRE simulations. 

The bottom panel of Fig.~\ref{fig: MassFractionandSigmaRhoandMuVAgainstLookbackTime_m12i} plots the evolution of $\sigma_{\log \rho}$ versus $t_{\rm lookback}$, derived by fitting the density distribution at $0.2\,R_{\rm vir}$ in each snapshot with a lognormal as done in Fig.~\ref{fig: AllGasMassWeightedlog(rho)Distribution_m12i_z_0_z_0.75}. 
Before the formation of the hot phase when turbulent velocities are larger than the mean sound speed we find $\sigma_{\log \rho}\approx 0.6-0.8$, as shown for the $z=0.75$ snapshot in the bottom panel of Fig.~\ref{fig: AllGasMassWeightedlog(rho)Distribution_m12i_z_0_z_0.75}. This dispersion is significantly larger than after the hot phase forms where the value of $\sigma_{\log \rho}\approx 0.2-0.3$, as shown for the $z=0$ snapshot in the top panel of Fig.~\ref{fig: AllGasMassWeightedlog(rho)Distribution_m12i_z_0_z_0.75}. At the transition itself $\sigma_{\log \rho}$ is highest with values of $0.8-1.0$, since both the cold and hot gas mass fractions are $\approx50\%$ and hence the density dispersion spans both phases, i.e.~a unimodal fit is no longer justified. 

To compare the density distribution in FIRE with that predicted by isothermal turbulence simulations we measure the following average turbulent Mach number in the simulations
\begin{equation}
\label{eq:MachTurb}
\machavg \equiv \frac{\sigma_{\rm turb}}{\langle c_{\rm s} \rangle_{\rho}}~.
\end{equation}
The bottom panel of Fig.~\ref{fig: MassFractionandSigmaRhoandMuVAgainstLookbackTime_m12i} plots the expected $\sigma_{\log \rho}(\machavg)$ based on eqn.~(\ref{eq:SigmaLnRho}) for $\bt=1/3$ and $\bt=1$.
The panel shows that eqn.~(\ref{eq:SigmaLnRho}) reasonably captures the relation between turbulence Mach number and the density distribution, both at early times when the cold phase dominates and at late times when the hot phase dominates. The similarity is best for compressive driving ($\bt=1$) mainly at the early supersonic stage, though note that this conclusion may be affected by our choice of how to average $c_{\rm s}$ for the calculation of $\machavg$. Only during the transition from cool phase dominance to hot phase dominance at $t_{\rm lookback}\approx3\,{\rm Gyr}$ the value of $\sigma_{\log \rho}$ is significantly larger than expected based on isothermal simulations, since the width spans both peaks of a bi-modal density distribution.  

The similarity between FIRE and isothermal turbulence simulations shown in the bottom panel of Fig.~\ref{fig: MassFractionandSigmaRhoandMuVAgainstLookbackTime_m12i} suggests that (1) at a given radius, the density dispersion in the dominant CGM phase is set by turbulence physics; and (2) the formation of a quasi-static hot phase causes a transition in the nature of CGM turbulence, from supersonic turbulence with wide density distributions pre-ICV, to subsonic CGM turbulence with narrow density distributions post-ICV. 

\begin{figure}  
    \centering    
    \includegraphics[width=\columnwidth]{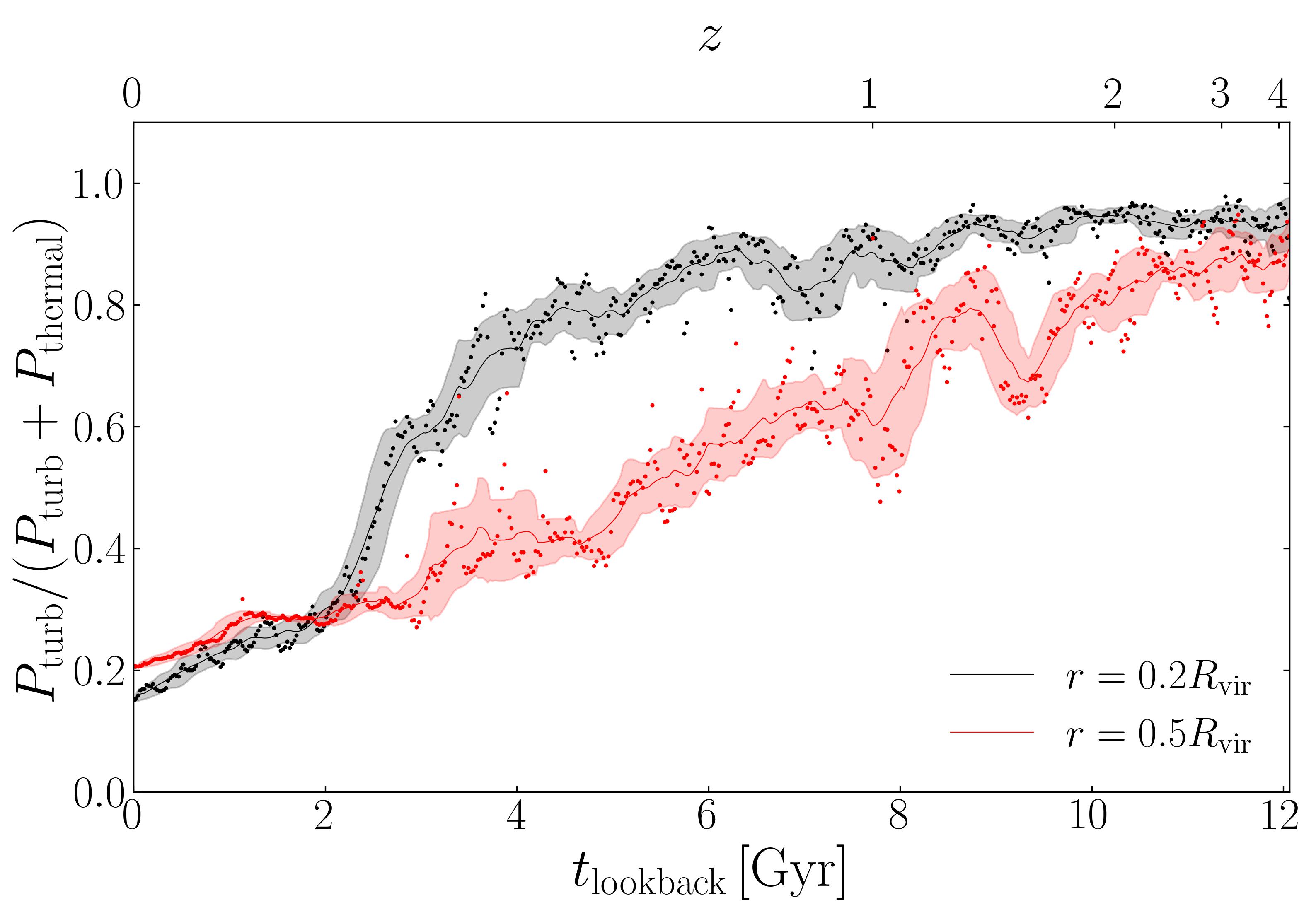}
    \caption{Evolution of the ratio of turbulent pressure ($P_{\rm turb}$) to total pressure ($P_{\rm turb} + P_{\rm thermal}$) in the CGM of the FIRE m12i simulation. Different colors denote gas at $r = 0.2R_{\rm vir}$ (black) and at $r = 0.5R_{\rm vir}$ (red). Dots represent individual data points, while lines and shaded bands represent running averages and dispersions within a $600\,{\rm Myr}$ window. The inner CGM exhibits a relatively sharp transition (within $\approx2\,{\rm Gyr}$) between turbulent pressure dominance to thermal pressure dominance, corresponding to the formation of a quasi-static hot phase in the inner CGM as seen in Fig.~\ref{fig: MassFractionandSigmaRhoandMuVAgainstLookbackTime_m12i}. At larger radii the importance of turbulence decreases more gradually with time.}
    \label{fig: (P_turb):(P_turb+P_thermal)AgainstLookbackTime _m12i.png}
\end{figure} 

\begin{figure*}
    \includegraphics[width=\textwidth]{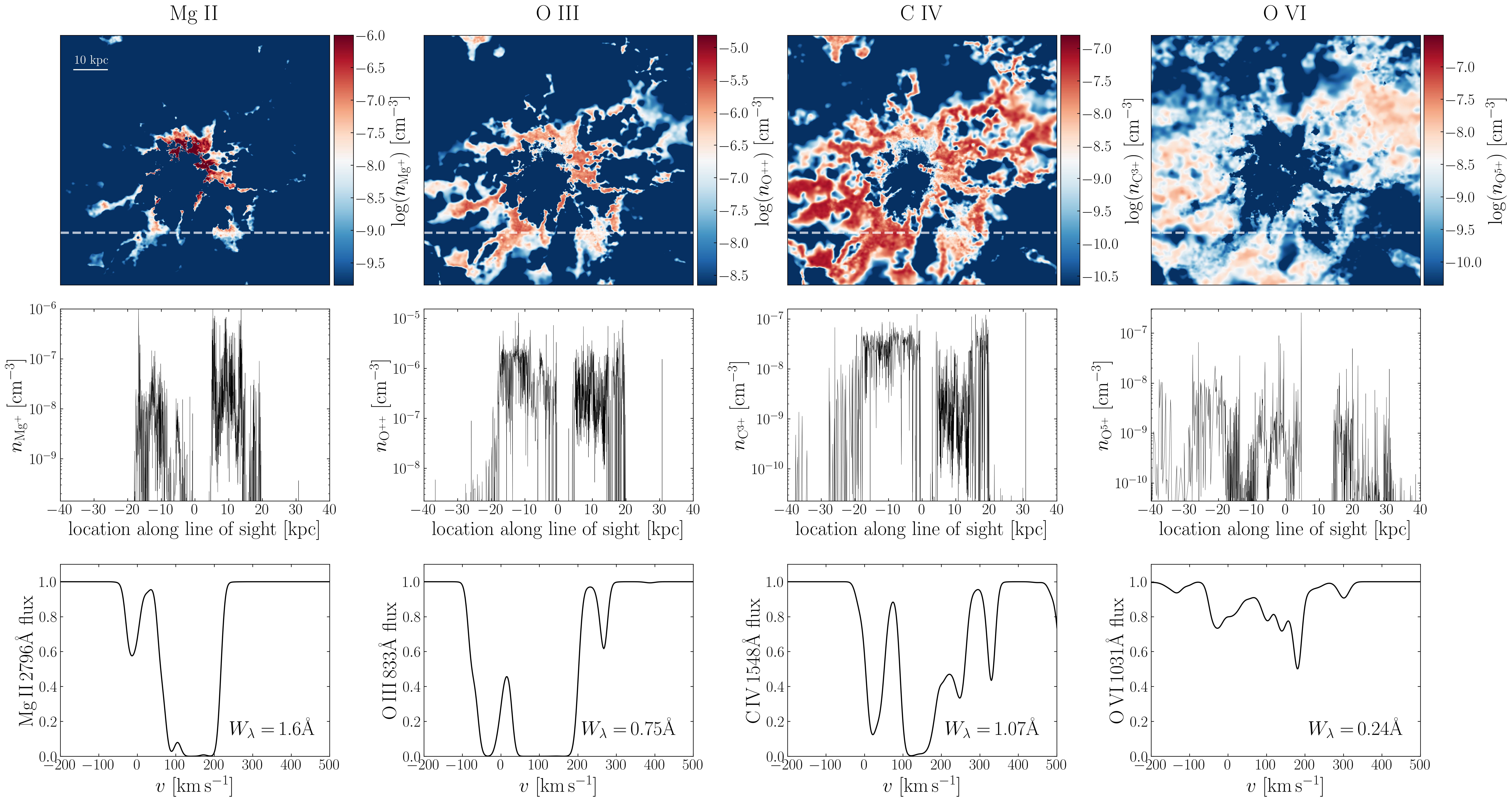}
    \caption{(Top row:) Maps of ion volume density in the turbulence-dominated $z=0.75$ snapshot shown in Figs.~\ref{fig: TemperatureProjectionsWithSi_IIISpectrum_m12i_z_0_z_0.75}--\ref{fig: AllGasMassWeightedlog(rho)Distribution_m12i_z_0_z_0.75}, for Mg$^{+}$ (left), O$^{++}$ (middle-left), C$^{3+}$ (middle-right), and O$^{5+}$ (right).  
    (Middle row:) Ion density as a function of location along the mock sighline marked as a dashed line in the top panels. (Bottom row:) Absorption spectra for the three ions assuming $R=20\,000$. Equivalent widths are noted in the panels. All ions shown are spread over a significant fraction of the line of sight, indicating that collectively they trace the volume-filling phase of the inner CGM, rather than tracing localized clouds and interfaces with a hot phase as usually assumed. \ion{Mg}{II} traces the high density part of the lognormal density distribution shown in Fig.~\ref{fig: AllGasMassWeightedlog(rho)Distribution_m12i_z_0_z_0.75}, and thus appears at the highest density peaks. \ion{O}{III} and \ion{C}{IV} originate from the center of the distribution and are thus widespread, while \ion{O}{VI} traces the low density part of the density distribution.}
    \label{fig: EUVIons(Mg_II,O_III,C_IV,O_VI)_m12i_z_0.75}
\end{figure*}

Figure~\ref{fig: (P_turb):(P_turb+P_thermal)AgainstLookbackTime _m12i.png} shows the evolution of the ratio of turbulent pressure ($P_{\rm turb}$) to the total pressure ($P_{\rm turb} + P_{\rm thermal}$) in the same simulation (m12i), calculated as:
\begin{equation}
\frac{P_{\rm turb}}{P_{\rm turb}+P_{\rm thermal}} = \frac{\frac{1}{3}\sigma_{\rm turb}^2}{\frac{1}{3}\sigma_{\rm turb}^2 + \frac{3}{5}\langle c_{\rm s} \rangle_{\rho}^2} ~.
\end{equation}
where the factor of $1/3$ is because we defined $\sigma_{\rm turb}$ as the 3D velocity dispersion (eqn.~\ref{eq:SigmaTurb}). The ratio is calculated at two different radial distances: $r = 0.2R_{\rm vir}$ (black) and $r = 0.5R_{\rm vir}$ (red). 
For $r = 0.2R_{\rm vir}$, the ratio remains relatively constant and high until $t_{\rm lookback} \approx 4$ Gyr, after which there is a rapid decline, as suggested by Fig.~\ref{fig: MassFractionandSigmaRhoandMuVAgainstLookbackTime_m12i}. 
At $r = 0.5R_{\rm vir}$, the ratio starts at a lower value and decreases more gradually over time, without a sharp transition. The ratio remains lower than at $r = 0.2R_{\rm vir}$ until the hot phase forms at $t_{\rm lookback}\approx2\,{\rm Gyr}$, reflecting the higher importance of thermal pressure and longer $\tcool$ at larger radii, consistent with previous results \citep{Stern21A,Gurvich23}.

\subsection{\texorpdfstring{Equivalent widths of $\sim1$\AA\ in strong UV transitions}
                            {Equivalent widths of ~1 Å in strong UV transitions}}

In Figs.~\ref{fig: TemperatureProjectionsWithSi_IIISpectrum_m12i_z_0_z_0.75} -- \ref{fig: MassFractionandSigmaRhoandMuVAgainstLookbackTime_m12i} we showed that most of the inner CGM is cool prior to the formation of a quasi-static hot phase, with a turbulent velocity of order $150\kms$ and a wide density distribution of $\sigma_{\log \rho}\approx 0.7\dex$ at a given radius. In this section we demonstrate that these properties imply high equivalent widths of $\sim1$\AA\ for commonly observed UV absorption features.

\subsubsection{Analytic estimate of equivalent widths}\label{sec:analytic}

We first estimate the absorption equivalent width $W_\lambda$ in inner CGM dominated by turbulence analytically, by assuming saturated absorption in strong transitions due to the near-unity mass fraction of cool gas.  
In saturated absorbers the equivalent width is  determined by the velocity range of the absorption, which in turn is set by the turbulent velocity. We thus get a rough estimate of the equivalent width of
\begin{equation}
    \label{eq:ExpectedEW}
   {W_\lambda} \sim \frac{2b}{c}\lambda \approx \sqrt{\frac{8}{3}}\frac{\sigma_{\rm turb}}{c}\lambda= 1.0\text{\AA}\cdot\left(\frac{\sigma_{\rm turb}}{150\kms}\right) \left(\frac{\lambda}{1206\text{\AA}}\right)~,
\end{equation}
where $b$ is the absorption line width, equal to $\sqrt{2}$ times the line-of-sight velocity dispersion so $b\approx\sqrt{2/3}\sigma_{\rm turb}$, and $\lambda$ is the wavelength of the transition normalized to that of \ion{Si}{iii}.
We thus expect $W_\lambda\sim1$\AA\ in UV absorption features which originate in inner CGM dominated by turbulence. 

To demonstrate that strong UV absorption lines are indeed saturated in turbulence-dominated CGM, we use the equation for optical depth at line center \citep[e.g.,][]{Draine2011}:
\begin{eqnarray}
    \label{eq:OpticalDepth}
    \tau &=& 0.015\frac{\rm cm^2}{\rm s}\cdot\frac{N_{\rm ion}\lambda f_{\rm lu}}{b} =\nonumber \\ 
    &\approx&5.8 \left(\frac{N_{\rm H}}{10^{20.5}\,{\rm cm}^{-2}}\right)\left( \frac{f_{\rm lu}\frac{N_{\rm X}}{N_{\rm H}}}{10^{-5}}\right)
    \left(\frac{f_{\rm ion}}{0.1}\right) 
    \left(\frac{\lambda}{1206\text{\AA}}\right)
    \left(\frac{150\frac{\rm km}{\rm s}}{\sigma_{\rm turb}}\right)\nonumber\\
\end{eqnarray}
where $N_{\rm ion}=N_{\rm H}(N_{\rm X}/N_{\rm ion})f_{\rm ion}$ is the ion column (see eqn.~\ref{eq:IonVolumeDensity}), and the total hydrogen column density $N_{\rm H}$ is normalized by typical CGM values \citep[e.g., eqn.~B6 in][]{Stern21B}:
\begin{equation}
   N_{\rm H} = 7\cdot10^{20} f_{\rm CGM} M_{12}^{1/3}\left(\frac{1+z}{2}\right)^2\left(\frac{R_\perp}{0.2\,\Rvir}\right)^{-1}\,{\rm cm}^{-2}
\label{eq:NH}
\end{equation}
where $f_{\rm CGM}$ is the CGM mass in units of the halo baryon budget, assumed to be $0.5$ for the estimate in eqn.~(\ref{eq:OpticalDepth}). 
The product $f_{\rm lu}\cdot N_{\rm X}/N_{\rm ion}$ in eqn.~(\ref{eq:OpticalDepth}) is normalized by $10^{-5}$, characteristic of strong UV absorbers assuming $Z=0.3\,{\rm Z}_\odot$ (Table \ref{table:IonsTable}). The estimate of $f_{\rm ion}\sim0.1$ is based on the result that the CGM is predominantly cool and has a broad range of gas densities at each radius during the turbulence-dominated stage, spanning a factor of $\approx 50$ (full-width-half-max) in the example shown in the bottom panel of Fig.~\ref{fig: AllGasMassWeightedlog(rho)Distribution_m12i_z_0_z_0.75}. Since the ionization state in photoionized cool gas depends on the ratio of ionizing photon density to gas density, this broad range in density implies that many UV-absorbing ions have optimal formation conditions in a significant fraction of the gas along the sightline (say, $\gtrsim10\%$), and we thus expect $f_{\rm ion}\gtrsim 0.1$ for these ions. The broad range in density implies also that the predicted $f_{\rm ion}$ are robust to the factor of $\approx2$ uncertainty in the ionizing photon density (see section~\ref{s:ion fraction}). 
Taken together, these considerations indicate that strong UV transitions have $\tau\gg1$, i.e.~highly saturated absorption features as assumed in the estimate of $W_\lambda$ in eqn.~(\ref{eq:ExpectedEW}). 

How does the predicted $W_\lambda$ depend on $M_{\rm halo}$ and $z$ for fixed $R_\perp/R_{\rm vir}$? We show below that $\sigma_{\rm turb}\approx\,v_{\rm c}$ in CGM dominated by turbulence. We thus expect $W_\lambda$ to scale roughly as $v_{\rm c}\propto M_{\rm halo}^{1/3}$ for fixed $z$ and as $v_{\rm c}\propto\left(1+z\right)^{0.5}$ for fixed halo mass. Different halo masses and redshifts can also change the predicted $\tau$, though this will significantly change $W_\lambda$ only if the transition becomes optically thin. 

\subsubsection{Equivalent widths in FIRE}

Figure~\ref{fig: EUVIons(Mg_II,O_III,C_IV,O_VI)_m12i_z_0.75} shows the spatial distributions and absorption characteristics of Mg$^+$, O$^{++}$, C$^{3+}$, and O$^{5+}$, in the $z=0.75$ snapshot of m12i shown above. 
The top row maps the ion volume densities in a $1\,{\rm kpc}$-thick slice, while the middle row shows the ion volume densities along the mock sightline (plotted as dashed lines in the top panels). 
The maps show that the filling fractions of the ions increase from Mg$^+$ to O$^{++}$ to C$^{3+}$, where Mg$^+$ has the most intermittent structure while C$^{3+}$ has a more continuous and widespread distribution. O$^{5+}$ has a similar filling fraction and pattern as C$^{3+}$ except at inner radii where it is absent. 

The different spatial distributions of the different ions shown in Fig.~\ref{fig: EUVIons(Mg_II,O_III,C_IV,O_VI)_m12i_z_0.75} are a result of higher ions generally tracing lower densities (which imply higher photoionization) and higher temperatures (which imply higher collisional ionization). The different densities correspond to different parts of the wide gas density distribution in supersonically turbulent CGM. Specifically, the mass-weighted density and temperature of Mg$^+$ along the mock sightline are \mbox{$0.03\, {\rm cm}^{-3}$} and \mbox{$0.9 \cdot 10^{4}\, {\rm K}$} with a $16-84$ density percentile range of $0.02 - 0.1 {\rm cm}^{-3}$, corresponding to the high-density part of the lognormal density distribution shown in the bottom panel of Fig.~\ref{fig: AllGasMassWeightedlog(rho)Distribution_m12i_z_0_z_0.75}. For O$^{++}$, the corresponding density and temperature are \mbox{$6\cdot 10^{-3}\, {\rm cm}^{-3}$} and \mbox{$ 1.4 \cdot 10^{4} \,{\rm K}$}, while for C$^{3+}$ we find \mbox{$3\cdot10^{-3} \,{\rm cm}^{-3}$} and \mbox{$ 2.6 \cdot 10^{4}\, {\rm K}$}. These two ions thus trace gas densities near the peak of the density distribution. For O$^{5+}$ the mass-weighted density and temperature along the sightline are \mbox{$5\cdot 10^{-4}\, {\rm cm}^{-3}$} and \mbox{$2 \cdot 10^{5}\, {\rm K}$}, i.e.\ this ion traces the low-density part of the density distribution shown in Fig.~\ref{fig: AllGasMassWeightedlog(rho)Distribution_m12i_z_0_z_0.75}. 

The spatial distributions of different ions shown in Fig.~\ref{fig: EUVIons(Mg_II,O_III,C_IV,O_VI)_m12i_z_0.75} demonstrate the volume-filling nature of UV-absorbing gas during the turbulence-dominated phase, where different ions trace different parts of a broad density distribution which is predominantly cool. This is a qualitatively distinct picture from the common paradigm where low-ion UV absorbers trace `localized clouds' and mid-ions trace interfaces between the clouds and the hot background. 

The bottom panels in Fig.~\ref{fig: EUVIons(Mg_II,O_III,C_IV,O_VI)_m12i_z_0.75} plot the absorption spectra for \ion{Mg}{ii}~$2796\text{\AA}$, \ion{O}{iii}~$833\text{\AA}$, \ion{C}{iv}~$1548\text{\AA}$, and \ion{O}{vi}~$1031\text{\AA}$. Except for \ion{O}{vi}, these absorption features and that of \ion{Si}{iii}~$1206\text{\AA}$ (bottom-right panel of Fig.~\ref{fig: TemperatureProjectionsWithSi_IIISpectrum_m12i_z_0_z_0.75}) indeed exhibit a velocity spread of $\approx 100-200\kms$ and $W_\lambda \sim1$\AA, as expected based on the analytic estimate in eqn.~(\ref{eq:ExpectedEW}). The corresponding values of $f_{\rm ion}$ along the sightline are $0.29$, $0.17$, and $0.24$ for Mg$^+$, O$^{++}$, and Si$^{++}$, respectively, consistent with the order of magnitude estimate of $f_{\rm ion}\sim 0.1$ in the previous section. For C$^{3+}$ we find a lower $f_{\rm ion}=0.008$, and the optical depth is close to unity. For \ion{O}{vi} we find $f_{\rm ion}=10^{-3}$ and thus the transition is optically thin with $W_\lambda=0.24$\AA. The relatively low optical depth and $W_\lambda$ of \ion{O}{vi} are a result of this ion tracing densities which are two standard deviations below the peak of the lognormal distribution, and thus such gas densities are relatively scarce.   

The absorption profiles shown in the plot are limited to contributions from $\pm 60\,{\rm kpc}$ along the sightline. We verified that the contribution from larger scales to the absorption is small. For example $W_\lambda (\ion{O}{vi})$ increases from $0.24$\AA\ to $0.3$\AA\ when accounting for scales up to $\pm 500\,{\rm kpc}$, compareable to the size of the region fully simulated by the zoom simulation. 

\begin{figure}
    \includegraphics[width=\columnwidth]{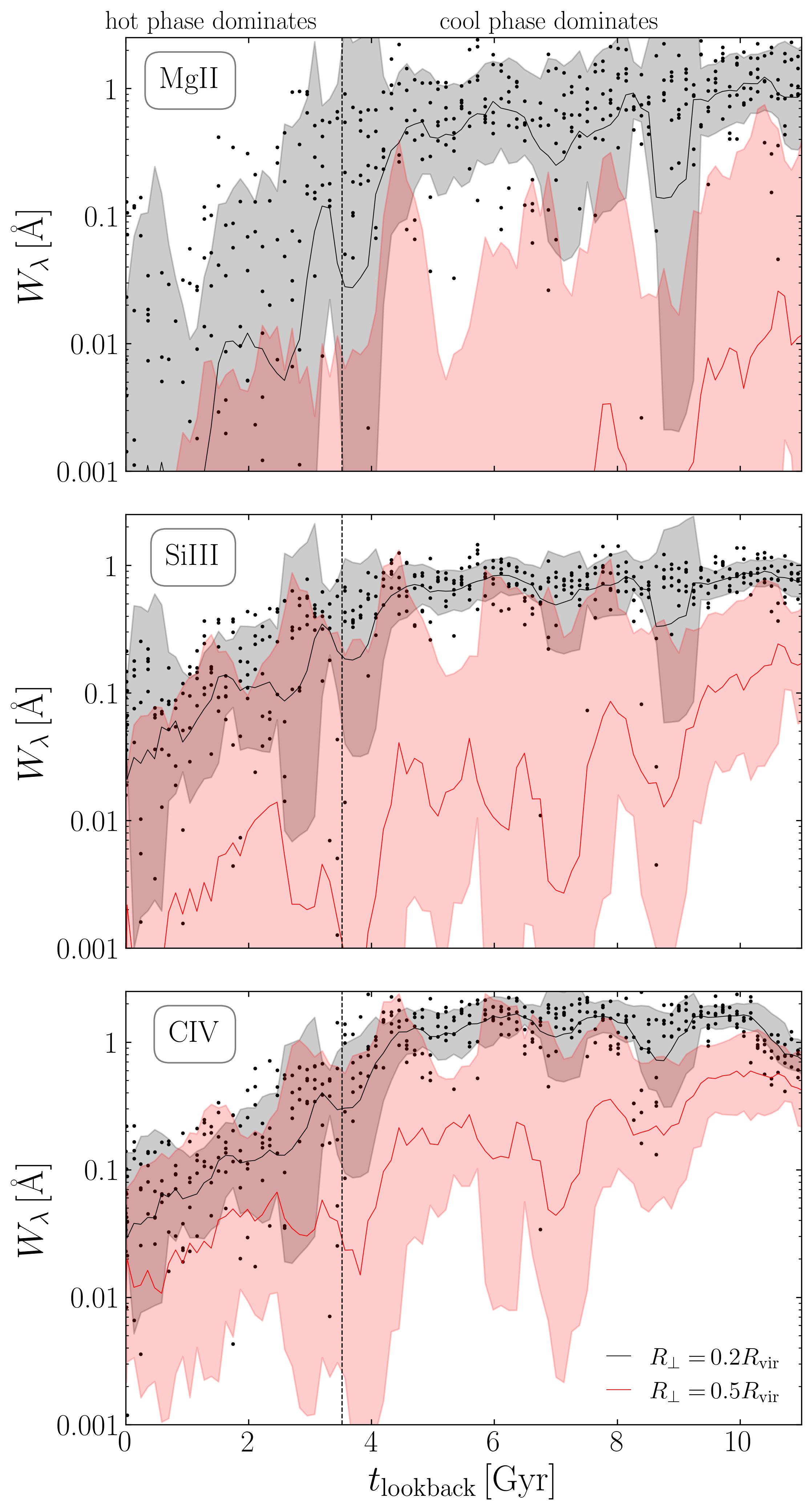}
    \caption{Predicted evolution of CGM absorption equivalent widths in m12i. 
    Different panels show different ions, while impact parameters are denoted by color. Dots represent individual mock sightlines (shown  for $R_\perp=0.2\,R_{\rm vir}$), while lines and bands represent averages and dispersions within a running $600\,{\rm Myr}$ window. Vertical lines denote the time after which the hot phase dominates (see Fig.~\ref{fig: MassFractionandSigmaRhoandMuVAgainstLookbackTime_m12i}). At $t_{\rm lookback}>4\,{\rm Gyr}$ when the cool phase dominates we predict $W_\lambda(0.2\,\Rvir)\approx1$\AA\ in all ions, consistent with the analytic approximation (eqn.~\ref{eq:ExpectedEW}). After the hot phase forms $W_\lambda(0.2\,\Rvir)\ll1$\AA.
    }
    \label{fig: DistributionofTotalWAgainstLookbackTimeforDifferentIons_m12i_Part1}
\end{figure}

\begin{figure*}
    \includegraphics[width=\textwidth]{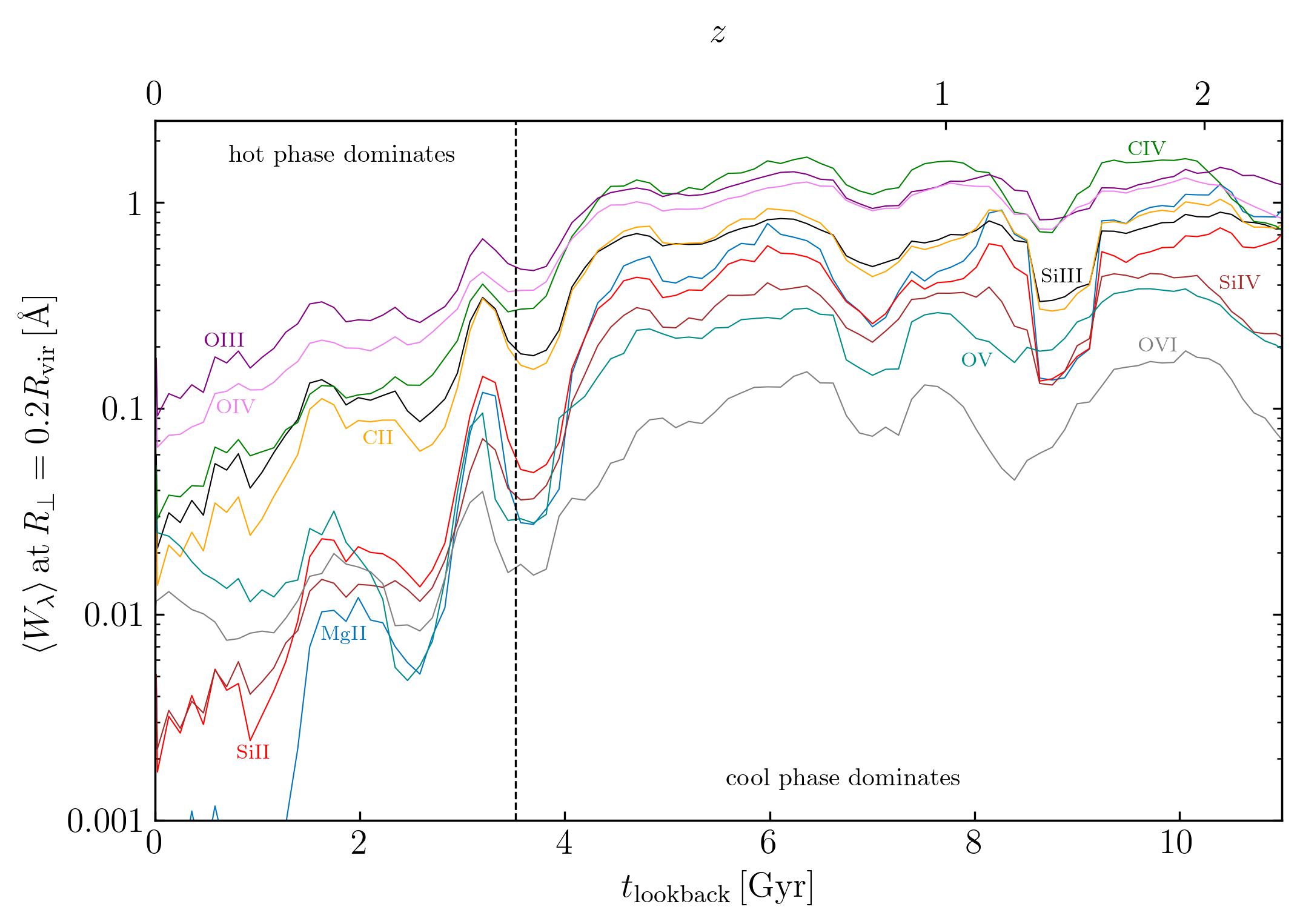}
    \caption{Predicted evolution of mean absorption equivalent widths of UV absorption features at an impact parameter of \(0.2 R_{\mathrm{vir}} \), based on the m12i simulation. Absorption features are listed in Table~\ref{table:IonsTable}, while plotted lines are labeled by the respective ions. The vertical dashed line indicates when the hot phase becomes dominant by mass. 
    Before this transition cool gas and turbulence dominate and $W_\lambda\sim1$\AA\ is predicted for most plotted features, while afterwards $W_\lambda$ drops.
    }
    \label{fig: DistributionofTotalWAgainstLookbackTimeforDifferentIons_m12i_Part2}
\end{figure*}

\begin{figure*}
    \includegraphics[width=\textwidth]{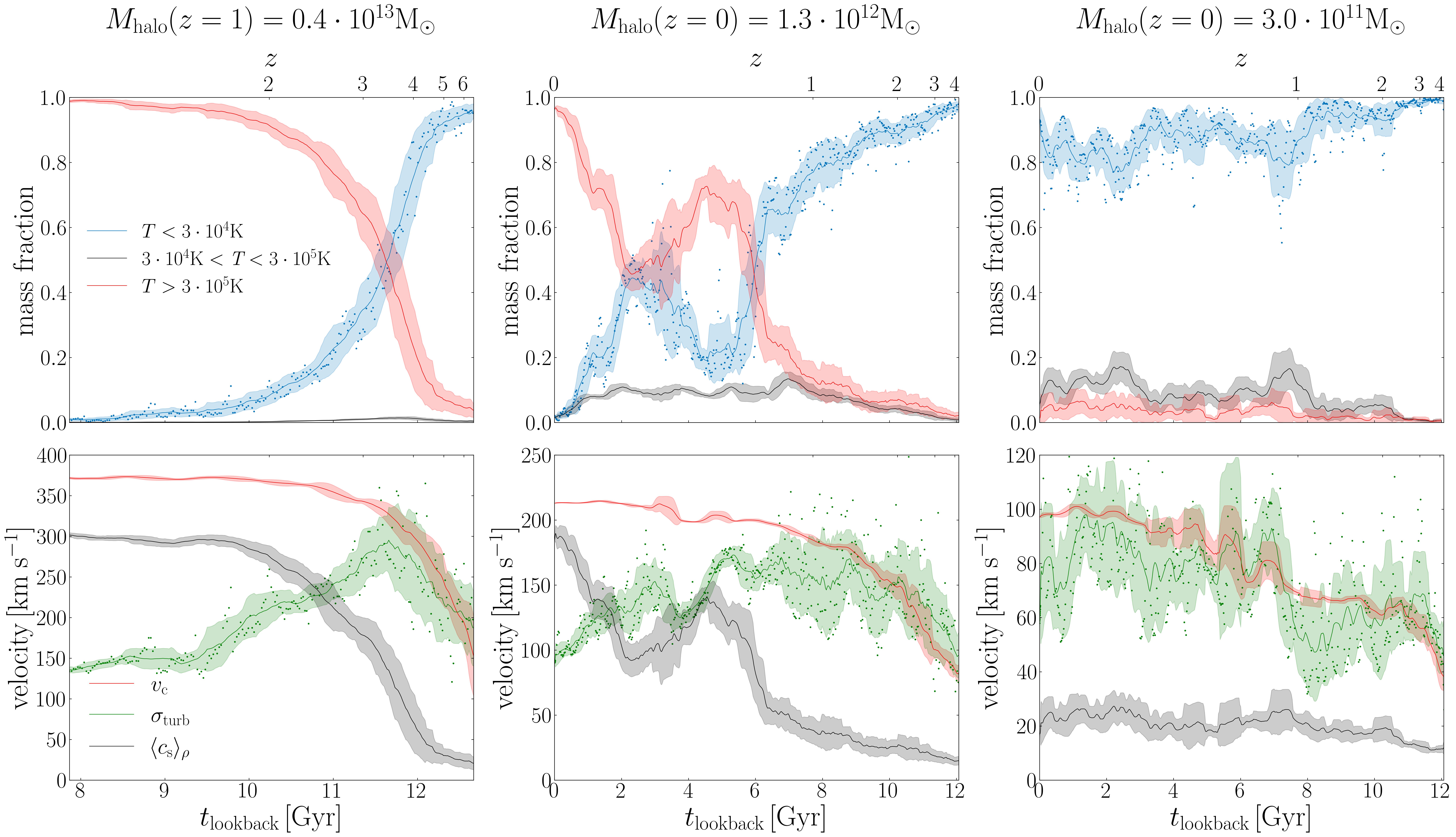}
    \caption{(Top:) Evolution of the mass fractions of different CGM phases at $r=0.2R_{\rm vir}$, in three FIRE simulations of halos of different masses. 
    The panels from left to right correspond to a group-size halo (`m13A1'), a Milky Way-mass galaxy (`m12b'), and a dwarf galaxy (`m11d'). Dots represent individual snapshots (shown only for the cool phase for clarity), while lines and bands represent averages and dispersions within a running $600\,{\rm Myr}$ window. (Bottom:) Evolution of the turbulent velocity (green), circular velocity (red), and mass-weighted sound speed (black) at $r=0.2R_{\rm vir}$. Note the difference in plotted velocity ranges between the panels. A significant increase in hot gas mass fraction is apparent around $t_{\rm lookback} \approx 7 \, \text{Gyr}$ in m12b and $t_{\rm lookback} \approx 12 \, \text{Gyr}$ in m13A1, indicating the formation of a quasi-static hot phase with subsonic turbulence. The short-term decrease in hot gas mass fraction in m12b at $t_{\rm lookback}\approx2\,{\rm Gyr}$ is associated with a major merger. The m11d galaxy on the right remains dominated by the supersonically turbulent cool phase at all times.}
    \label{fig: MassFractionandSigmaRhoandMuVAgainstLookbackTime_m11d_m12b_A1}
\end{figure*}

Figure~\ref{fig: DistributionofTotalWAgainstLookbackTimeforDifferentIons_m12i_Part1} plots the evolution of $W_\lambda$ versus lookback time. 
Each panel presents a different ion at two impact parameters (\(0.2R_{\text{vir}}\) in black and \(0.5R_{\text{vir}}\) in red), where dots indicate individual mock sightlines (five per snapshot), while lines and bands denote averages and dispersions in running windows spanning $600\,{\rm Myr}$. 
Before the transition at $t_{\text{lookback}} > 5$ Gyr we find $W_\lambda\approx1$\AA\ at $0.2\,R_{\rm vir}$ for all three ions shown, as expected from the analytic estimate in eqn.~(\ref{eq:ExpectedEW}). 
At $t_{\text{lookback}} \lesssim 2$ Gyr after the hot phase becomes dominant $W_{\lambda}$ drops in all ions since cool gas is less abundant (see Fig.~\ref{fig: MassFractionandSigmaRhoandMuVAgainstLookbackTime_m12i}). After the transition $W_\lambda$ also differs between different ions and evolves with time, in contrast with the approximately uniform $W_\lambda$ prior to the transition. 

At $R_\perp=0.5\,R_{\text{vir}}$ the values of $W_\lambda$ shown in Fig.~\ref{fig: DistributionofTotalWAgainstLookbackTimeforDifferentIons_m12i_Part1} are consistently lower than at $R_\perp=0.2\,R_{\text{vir}}$, especially for \ion{Mg}{ii}, and exhibit a more gradual decline with time. This reflects both the dominance of the hot phase in the outer CGM, and the overall lower gas columns and lower gas densities at larger radii (see Fig.~\ref{fig: AllGasMassWeightedlog_rho_Distribution_m12i_z_0_z_0.75_For_0.5Rvir} and \citealt{Stern21A}). 

Figure~\ref{fig: DistributionofTotalWAgainstLookbackTimeforDifferentIons_m12i_Part2} plots the evolution of the mean $W_\lambda$ for all strong UV and EUV transitions listed in Table~\ref{table:IonsTable}.
At $t_{\rm lookback}>4\,{\rm Gyr}$ when the inner CGM is turbulence dominated we find $\langle W_\lambda\rangle\sim 0.2-2$\AA\ in all shown absorption features except for \ion{O}{vi} where $\langle W_\lambda\rangle\approx0.02-0.1$\AA. In contrast after $t_{\rm lookback}<2\,{\rm Gyr}$ when the hot phase has formed the values of $\langle W_\lambda\rangle$ are substantially lower, in the range $10^{-3}-0.2$\AA. 

\subsection{Dependence on halo mass history}

In this subsection we explore how the the mass history of the halo affects the transition from turbulence-dominated CGM to thermal-pressure dominated CGM and the implied absorption signatures.

Figure~\ref{fig: MassFractionandSigmaRhoandMuVAgainstLookbackTime_m11d_m12b_A1} repeats the analysis in Fig.~\ref{fig: MassFractionandSigmaRhoandMuVAgainstLookbackTime_m12i} for three simulations: a group-size halo (m13A1, left panels), another Milky Way-mass galaxy halo (m12b, middle panels), and a dwarf galaxy halo (m11d, left panels).
In the massive m13A1 simulation the hot phase becomes dominant, and the turbulence becomes subsonic at an early time of $t_{\rm lookback}\approx11.5\,{\rm Gyr}$ ($z\approx3.2$). This follows since $\tcool$ exceeds $\tff$ earlier in more massive halos \citep{Stern21A}. 
In the m12b simulation shown in the middle column, the hot phase becomes dominant by mass and the turbulence becomes subsonic at $t_{\rm lookback}\approx6\,{\rm Gyr}$ ($z\approx 0.65$), a few Gyr before the transition occurs in m12i (see Fig.~\ref{fig: MassFractionandSigmaRhoandMuVAgainstLookbackTime_m12i}). This demonstrates the range in formation times of the inner hot CGM phase in Milky-Way mass FIRE galaxies, which typically span $0<z<1$ \citep[see][and below]{Stern21A}. The m12b simulation also exhibits a bump in the cool gas mass fraction at $t_{\rm lookback}\approx 2\,{\rm Gyr}$ after the hot gas formed, coincident with a major merger occuring at this time as noted by \cite{Yu21}.   
In the dwarf galaxy halo shown in the left column the cool gas phase dominates the mass at all times. Correspondingly, the turbulent velocity remains higher than the mean sound speed, i.e.~turbulent velocities remain supersonic. This result reflects the \cite{Stern21A} result that a quasi-static hot CGM phase does not form in the inner CGM of FIRE dwarfs, since $\tcool$ remains lower than $t_{\text{ff}}$ down to $z=0$.

\begin{figure}
    \includegraphics[width=\columnwidth]{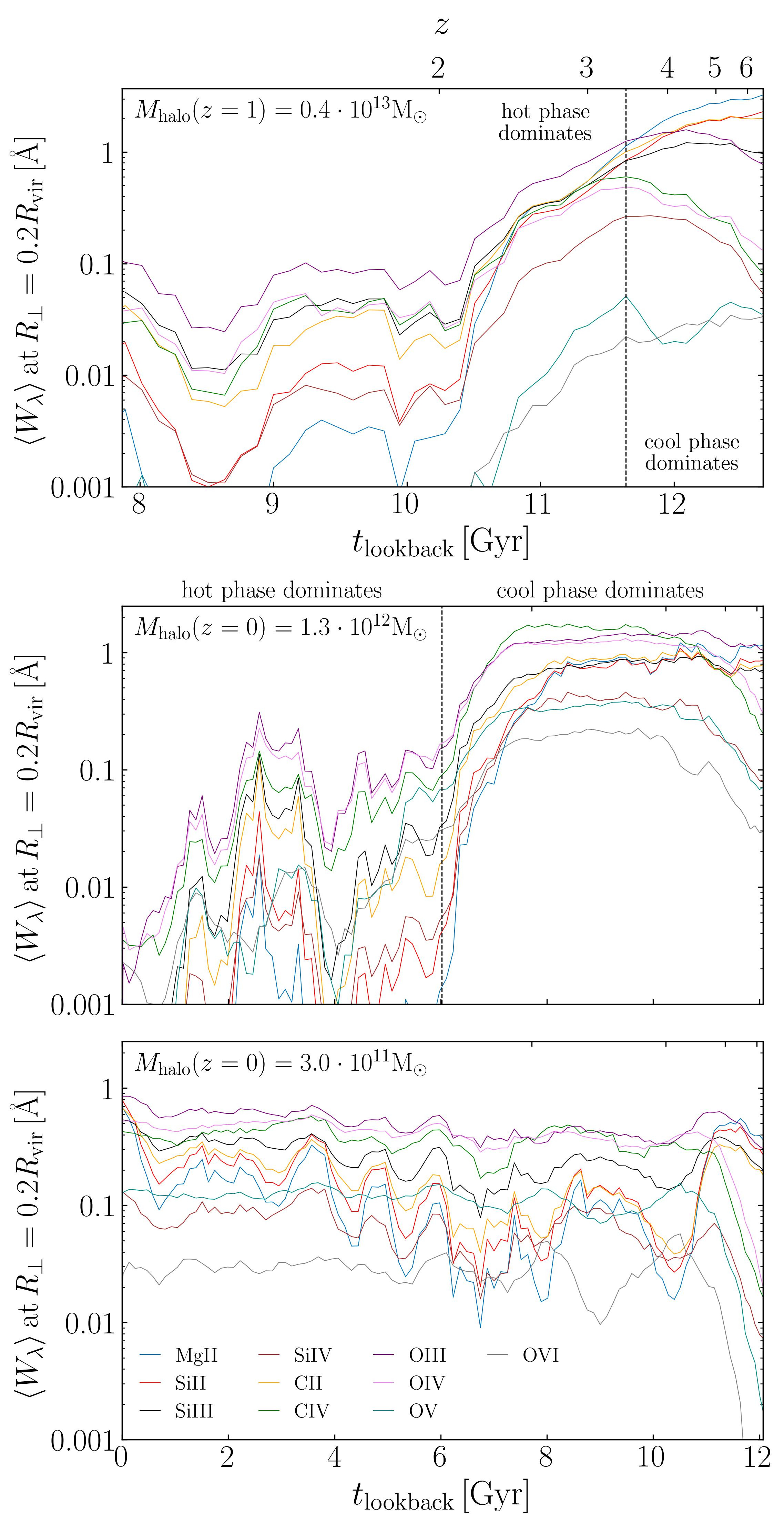}
    \caption{Evolution of mean $W_\lambda$ at \( R_\perp = 0.2 R_{\mathrm{vir}} \) for the three  simulations shown in Fig.~\ref{fig: MassFractionandSigmaRhoandMuVAgainstLookbackTime_m11d_m12b_A1}. The dashed vertical lines indicate the transition from an inner CGM dominated by cool gas and turbulence to an inner CGM dominated by the hot phase. 
    Equivalent widths are high ($\approx 1\,$\AA) before the transition and decrease afterward. The lowest-mass halo shown in the bottom panel remains cool and turbulent at all times, with   $0.1<W_\lambda<1$\AA\ for all ions except \ion{O}{vi}.}
    \label{fig: DistributionofTotalWAgainstLookbackTimeforDifferentIons_m11d_m12b_A1}
\end{figure}

Figure~\ref{fig: DistributionofTotalWAgainstLookbackTimeforDifferentIons_m11d_m12b_A1} plots the evolution of $\langle W_\lambda\rangle$ for various strong UV transitions in the three simulations, similar to the analysis of m12i in Fig.~\ref{fig: DistributionofTotalWAgainstLookbackTimeforDifferentIons_m12i_Part2}. 
In m13A1 (top panel) a clear drop in mean $W_\lambda$ occurs already around $t_{\rm lookback}=11.5\,{\rm Gyr}$, consistent with the early transition to a thermal energy-dominated inner CGM phase in this simulation. Equivalent widths of the \ion{Mg}{ii}, \ion{C}{ii}, \ion{Si}{ii}, \ion{Si}{iii}, and \ion{O}{iii} transitions are $1-3$\AA\ prior to the transition, somewhat higher than the $\langle W_\lambda\rangle\approx$ $0.7-1$\AA\ in m12i prior to the transition. The higher $W_\lambda$ are due to the higher turbulent velocity of $200-300\kms$ (bottom-left panel of Fig.~\ref{fig: MassFractionandSigmaRhoandMuVAgainstLookbackTime_m11d_m12b_A1}) and hence higher $b$ in more massive halos (see eqn.~\ref{eq:ExpectedEW}). In contrast, the \ion{O}{iv}, \ion{C}{iv}, and \ion{Si}{iv} have lower $W_\lambda$ of $0.1-0.3$\AA, likely due to the weak UV background (relative to typical CGM density) at $z\gtrsim3$, which implies that cool photoionized gas does not reach high ionization levels. 

The CGM of m12b shown in the middle panel of Fig.~\ref{fig: DistributionofTotalWAgainstLookbackTimeforDifferentIons_m11d_m12b_A1} exhibits similar UV absorption figures as m12i (shown in Fig.~\ref{fig: DistributionofTotalWAgainstLookbackTimeforDifferentIons_m12i_Part1}), albeit the transition happens earlier at $t_{\text{lookback}}\approx7\,$Gyr consistent with the earlier formation of a quasi-static hot inner CGM phase in this simulation. As in m12i, $\langle W_\lambda\rangle\approx0.3-1$\AA\ prior to ICV in most absorption lines while \ion{O}{vi} has a lower $\langle W_\lambda\rangle\approx 0.1-0.2$\AA. The values of $\langle W_\lambda\rangle$ for all ions drop following ICV.

The dwarf galaxy halo shown in the bottom panel of Fig.~\ref{fig: DistributionofTotalWAgainstLookbackTimeforDifferentIons_m11d_m12b_A1} exhibits $\langle W_\lambda\rangle \approx 0.1-0.7$\AA\ down to $z=0$, again with the exception of \ion{O}{vi} which has a lower $\langle W_\lambda\rangle \approx 0.03$\AA. As mentioned above this halo remains in the turbulence-dominated stage at all times. These $W_\lambda$ are somewhat lower than the $W_\lambda\sim 0.2-2$\AA\ in m12i and m12b during their turbulent-dominated stage. The difference is mainly a result of the lower turbulent velocities of $\approx70\kms$ in m11d (bottom-right panel of Fig.~\ref{fig: MassFractionandSigmaRhoandMuVAgainstLookbackTime_m11d_m12b_A1}), which results in narrower absorption features and thus lower $W_\lambda$ (see eqn.~\ref{eq:ExpectedEW}).

\begin{figure*}
    \includegraphics[width=\textwidth]     {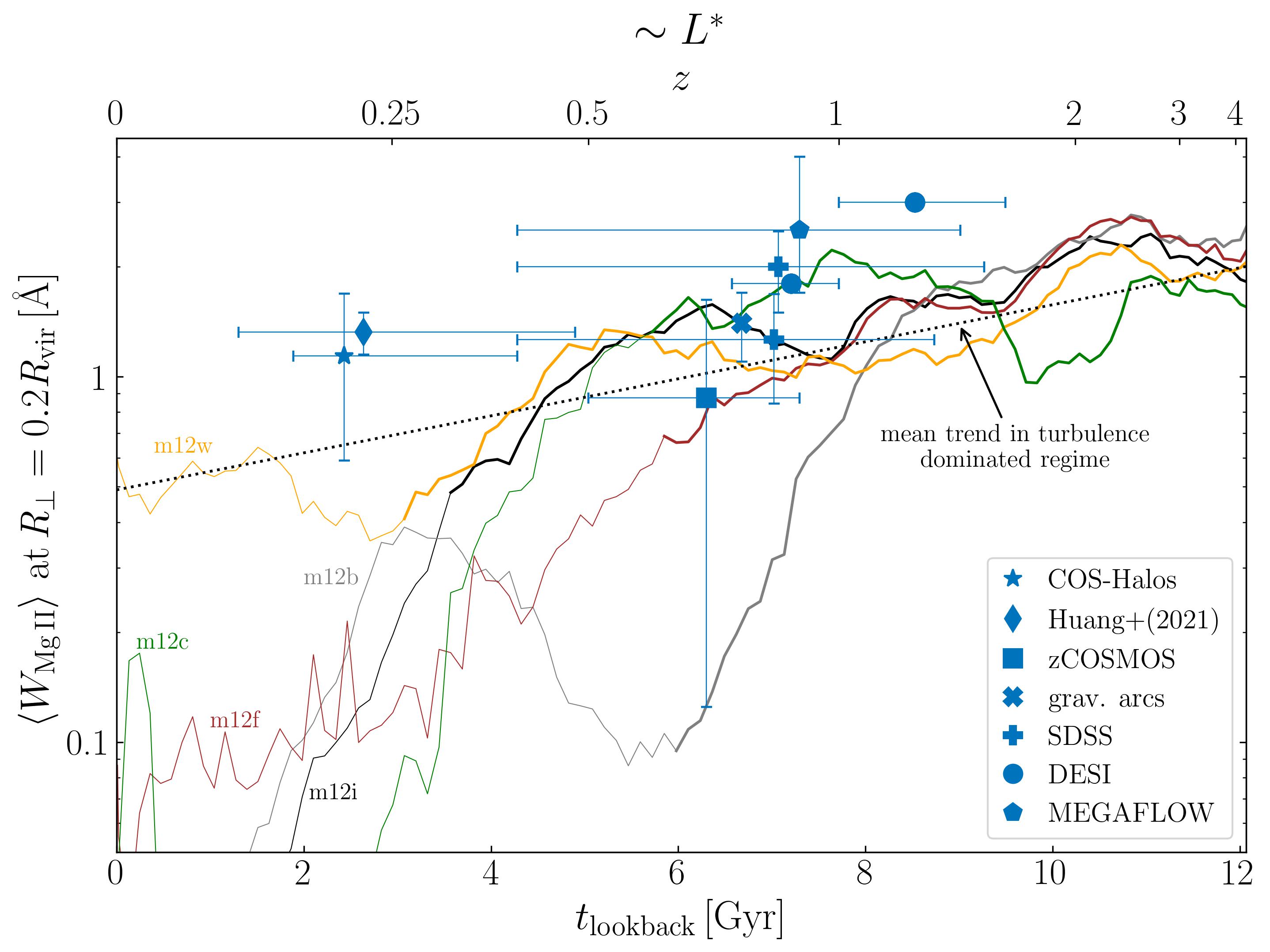}
    \caption{Observed \ion{Mg}{ii} absorption at  $R_\perp = 0.2R_{\rm vir}$ from blue $\sim L^\star$ galaxies versus FIRE predictions.
    Solid colored lines show the evolution of the mean \ion{Mg}{ii} equivalent width $\langle W_{2796} + W_{2803} \rangle$ in five simulated $\sim L^\star$ galaxies. Thicker segments of each curve indicate pre-ICV ($f_{\rm cool}>0.5$, turbulence-dominated) times, and the dotted black line shows the mean trend in this regime. Thin segments show post-ICV ($f_{\rm cool}<0.5$, thermally-supported) times which in most cases predict substantially lower $\langle W_{\ion{Mg}{ii}}\rangle$. Observed values from the blue $\sim L^\star$ surveys listed in Table~\ref{table:ObservationsTable} are plotted as markers and errorbars. Observed $\EWmgii$ at $R_\perp = 0.2\,R_{\rm vir}$ in star-forming $\sim L^\star$ galaxies are consistent with the predictions of turbulence-dominated inner CGM to a factor of $\approx2$.}
    \label{fig: DistributionofTotalWAgainstLookbackTimeForMgIIComparisonWithObservation_L_Stars}
\end{figure*}

\begin{figure*}
    \includegraphics[width=\textwidth]{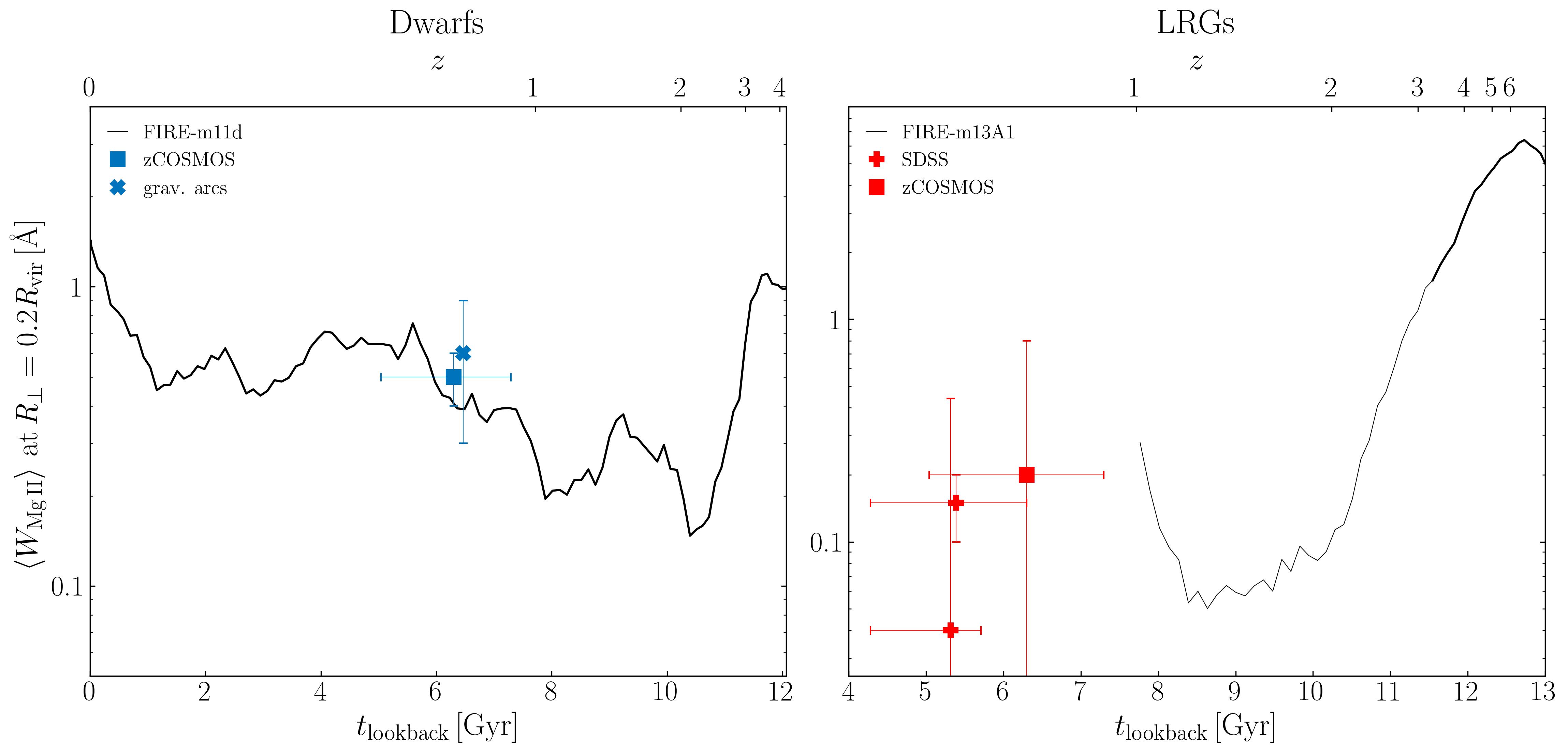}
    \caption{Observed \ion{Mg}{ii} absorption in dwarf galaxies (left) and in LRGs (right) versus FIRE predictions. Solid lines show predicted $\langle W_{2796 + 2803} \rangle$ at $R_\perp = 0.2R_{\rm vir}$, with thicker segments corresponding to pre-ICV ($f_{\rm cool}>0.5$, turbulence-dominated) times. m11d remains turbulence-dominated down to $z=0$. Observed values from the surveys listed in Table~\ref{table:ObservationsTable} are plotted as markers and errorbars, with marker color indicating star-forming (blue) or quiescent (red) galaxies. Blue dwarf galaxy samples are consistent with a turbulence-dominated inner CGM, while LRG observations suggest thermally supported inner CGM.}
    \label{fig: DistributionofTotalWAgainstLookbackTimeForMgIIComparisonWithObservation_Dwarfs+LRGs}
\end{figure*}

\subsection{Dependence on resolution}\label{s:resolution}

To test how our results depend on simulation resolution, we repeat the analysis for two additional simulations of the m12i halo with eight times higher/lower gas mass resolution than in the fiducial simulation (see Table~\ref{table:FIRE-2Simulations}). The results are shown in 
Figure~\ref{fig: m12i_r880_m12i_r7100_m12i_r57000_MassFraction_MuV_With_W_MgII_And_t_cool_t_ff} in the appendix. The top two rows are similar to the two top panels of Fig.~\ref{fig: MassFractionandSigmaRhoandMuVAgainstLookbackTime_m12i}, while the bottom row shows the mean equivalent width of \ion{Mg}{II} and the ratio $\tcool/\tff$ measured at $0.1\Rvir$ (eqns.~\ref{eq:t_cool} -- \ref{eq:t_ff}).
The Figure shows that $v_{\rm c}$ sincrease somewhat more slowly at higher resolution, and thus $\tcool/\tff$ which scales as $\sim v_{\rm c}^4$ (see section~\ref{s:tcool and tff}) exceeds unity later, as also found by \cite{Stern21A}. At all shown resolutions we find that prior to ICV when $\tcool<\tff$, the cool gas mass fraction is $>50\%$, turbulence is on average supersonic ($\sigma_{\rm turb}>\langle c_s\rangle_\rho$), and the equivalent width of \ion{Mg}{ii} is $\gtrsim1$\AA. This  suggests that the conditions for the development of turbulence-dominated CGM and their predicted equivalent widths are independent of resolution at the range probed. This lack of sensitivity to resolution is because turbulence-dominated CGM are predominantly cool even at low resolution. 

At $z=0$ when the inner CGM is dominated by thermal pressure, Fig.~\ref{fig: m12i_r880_m12i_r7100_m12i_r57000_MassFraction_MuV_With_W_MgII_And_t_cool_t_ff} shows that $f_{\rm cool}$ increases with increasing resolution, consistent with the conclusion of previous studies  \citep{vandeVoort19, Hummels19, Peeples19, Ramesh24}. However, it is not clear if the increasing $f_{\rm cool}$ in FIRE is a direct result of the increasing resolution, or rather a result of the lower $\tcool/\tff$ at higher resolution which produces a higher $f_{\rm cool}$. 

\section{COMPARISON WITH UV ABSORPTION OBSERVATIONS}\label{s:obs}

A common paradigm for circumgalactic UV absorbers is that the absorption originates from a cool `cloud',  or from the interface between a cool cloud and the ambient hot medium \citep[e.g.,][]{Tumlinson17,FaucherGiguere23}. 
Our results suggest an alternative origin for $W_\lambda\sim1$\AA\ absorbers, in which they trace the volume-filling medium of the inner CGM, in halos with $\tcool<\tff$ where the inner CGM is predominantly cool and supersonically turbulent.

In this section we compare this alternative paradigm for $W_\lambda\sim1$\AA\ absorbers with several observational constraints. 
This allows us to infer which types of galaxies in the real universe have a turbulence-dominated inner CGM rather than a thermal energy-dominated inner CGM, and thus constrain how the formation of a thermal-energy dominated inner CGM affects galaxy evolution. 

\subsection{\texorpdfstring{\ion{Mg}{ii}}{Mg II} equivalent widths}

The \ion{Mg}{ii} absorption doublet is observable from the ground at $z\gtrsim0.1$ \citep[e.g.,][]{Chen10,Nielsen13, Werk13}, and thus has been the subject of a large number of circumgalactic UV absorption surveys at the redshifts we focus on in this paper ($0\leq z\lesssim2$). 
Figure~\ref{fig: DistributionofTotalWAgainstLookbackTimeForMgIIComparisonWithObservation_L_Stars} 
compares observed mean rest-frame equivalent widths $\EWmgii\equiv \langle W_{2796}+W_{2803}\rangle$ at impact parameters of $R_\perp=0.2\,\Rvir$ to those predicted by the FIRE simulations. We focus on average values since some of the observed $\EWmgii$ are measured on stacked spectra, though we note that the predicted dispersion between sightlines is not large during the turbulence-dominated phase, of order $\approx 0.3\,{\rm dex}$ (see Fig.~\ref{fig: DistributionofTotalWAgainstLookbackTimeforDifferentIons_m12i_Part1}). 
Predictions of the five m12 FIRE simulations are shown as solid colored lines, where $\EWmgii$ are calculated over 200 mock sightlines within $1200\,{\rm Myr}$ running windows (five sightlines per snapshot). We use larger windows than above in order to reduce clutter in the figure. Times when the inner CGM is dominated by cool gas ($f_{\rm cool}>0.5$) are emphasized with thicker lines. 
A drop in the predicted $\EWmgii$ is evident when the inner CGM becomes thermal pressure-dominated at $3<t_{\rm lookback}<6\,{\rm Gyr}$ ($0.25<z<0.6$) in four of the halos (m12i, m12b, m12c, m12f). The fifth halo m12w does not show a drop in $\EWmgii$ despite reaching $f_{\rm cool}=0.5$ at $t_{\rm lookback}=3\,{\rm Gyr}$, likely since it remains with $f_{\rm cool}\approx0.5$ and $\tcool\approx\tff$ down to $z=0$, i.e. its does not fully transition to thermal-energy dominated regime. This is potentially due to its somewhat lower halo mass than the other m12's and the strong dependence of $\tcool/\tff$ on halo mass \citep[see][and Table~\ref{table:FIRE-2Simulations}]{Stern21A}.

During times when turbulence dominates, all five simulations predict similar $\EWmgii$, as expected based on the analytic arguments in section~\ref{sec:analytic}. We fit the predictions in these pre-ICV snapshots with a log-linear relation:
\begin{equation}\label{e:predicted_EWMgII}
 \log\frac{\EWmgii_{\rm pre-ICV}}{\text{\AA}} =  -0.31 + 0.05\,\frac{t_{\rm lookback}}{\rm Gyr} ~.
\end{equation}
This relation is consistent with the order of magnitude analytic estimate of $\EWmgii_{\rm pre-ICV}\sim1$\AA\ in eqn.~(\ref{eq:ExpectedEW}) and is plotted as a dotted line in Fig.~\ref{fig: DistributionofTotalWAgainstLookbackTimeForMgIIComparisonWithObservation_L_Stars}. The main prediction of this paper is that $\sim L^*$ galaxies roughly follow this relation as long as the energetics of their inner CGM remains dominated by turbulence. 

Observed $\EWmgii$ at $R_\perp=0.2\,\Rvir$ from blue $\sim L^\star$ galaxies are plotted as errorbars in Fig.~\ref{fig: DistributionofTotalWAgainstLookbackTimeForMgIIComparisonWithObservation_L_Stars} and listed in Table~\ref{table:ObservationsTable}. Most observational surveys shown are based on matching background sources with foreground galaxies, such that each foreground galaxy has a single sightline through its CGM. The exceptions are the measurements based on gravitational arcs which have multiple sightlines through a single foreground galaxy, since the background source is a lensed galaxy. To derive $\EWmgii(R_\perp=0.2\Rvir)$ we interpolate the $\EWmgii$ versus $R_\perp/\Rvir$ relation deduced in each survey. 
The horizontal error bar spans the redshift range of the sample, with the marker located at the median redshift. Blue galaxy samples with mean stellar $M_\star>10^{10}\msun$ are considered $\sim L^\star$ and plotted in the figure.    
Additional details on this calculation for each survey are given in appendix~\ref{a:obs}. 
 
Fig.~\ref{fig: DistributionofTotalWAgainstLookbackTimeForMgIIComparisonWithObservation_L_Stars}
demonstrates that the observed $\EWmgii$ of star forming $\sim L^\star$ galaxies at $R_\perp=0.2\Rvir$ is $\gtrsim1$\AA\ at $0.1\lesssim z \lesssim1.5$. These observed values are consistent with the predictions of turbulence-dominated inner CGM (dotted line, eqn.~\ref{e:predicted_EWMgII}) to within a factor of $\approx2$. We emphasize that since turbulence-dominated CGM make specific predictions for UV absorption strength (section~\ref{sec:analytic}), this match between predictions and observations is highly non-trivial.
We thus conclude that turbulence-dominated CGM correctly predict $\EWmgii$ observed in inner CGM surrounding blue $\sim L^*$ galaxies.

The blue galaxy observations shown in Fig.~\ref{fig: DistributionofTotalWAgainstLookbackTimeForMgIIComparisonWithObservation_L_Stars} do not show a drop in $\EWmgii$ at low redshift similar to that predicted by the four m12 simulations  in which ICV occurs. Rather, the observations more resemble the m12w simulation which predicts only a mild decrease in $\EWmgii$ with redshift since it remains with $f_{\rm cool}\gtrsim0.5$ down to $z=0$. This difference is potentially due to the somewhat lower masses in the $\sim L^\star$ observations than in the m12 simulations, since when and if ICV occurs strongly depends on mass. A simulation sample better matched in mass to the observations could test this possibility.

The drop in $\EWmgii$ exhibited by four of the m12 simulations predicts that the most massive blue galaxies at low redshift should  exhibit $\EWmgii(0.2\,R_{\rm vir})\ll1$\AA, due to their thermal pressure-dominated inner CGM, in contrast with the blue $\sim L^\star$ galaxies observed by the surveys shown in Fig.~\ref{fig: DistributionofTotalWAgainstLookbackTimeForMgIIComparisonWithObservation_L_Stars}. Tentative evidence for this distinction is seen in figure~4 of \cite{Lan20}, where the most massive galaxy bin at $0.4<z<0.7$ has $\EWmgii(0.2\,\Rvir)\ll1$\AA. Additional support for this comes from the low \ion{Si}{ii} and \ion{C}{iv} columns observed in the CGM of the Milky-Way and M31, as discussed below. 

Figure~\ref{fig: DistributionofTotalWAgainstLookbackTimeForMgIIComparisonWithObservation_Dwarfs+LRGs} compares the FIRE predictions for $\EWmgii$ with observations of blue dwarf galaxies ($M_\star<10^{10}\msun$) and luminous red galaxies (LRGs), also listed in Table~\ref{table:ObservationsTable}.
The inner CGM in the m11d simulation shown in the left panel remains cool and turbulent at all times as discussed above. The predicted $\EWmgii$ at $0.2\Rvir$ largely remains within the range $0.3-1$\AA\ consistent with the observed values. The predicted $\EWmgii$ in m13A1 shown in the right panel drops at early times, as expected given the early ICV in this massive simulation. The mean observed values of the LRG samples are also $\ll1$\AA, inconsistent with the predictions of turbulent-dominated inner CGM. Fig.~\ref{fig: DistributionofTotalWAgainstLookbackTimeForMgIIComparisonWithObservation_Dwarfs+LRGs} thus suggests that the inner CGM of blue dwarf galaxies are turbulent dominated, while the inner CGM of LRGs are thermal-energy dominated, as expected.

The difference in predicted \ion{Mg}{ii} absorption between turbulence-dominated and thermal energy-dominated inner CGM is also evident in Figure~\ref{fig: WMgIIAgainstR_R_virComparisonWithObservations_m12i}, which plots $W_{2796}$ versus impact parameter at $t_{\rm lookback}=2\,{\rm Gyr}\ (z=0.15)$ and at $t_{\rm lookback}=5\,{\rm Gyr}\ (z=0.5)$ in the m12i simulation, corresponding to just after and just before ICV. We plot the mean and scatter of $W_{2796}$ from five consecutive snapshots centered on each of these two times, using ten sightlines per $0.025\,{\rm dex}$-wide bin in $R_\perp$, and then smoothing the curve with a Savitzky–Golay filter to reduce clutter.
The panel shows that at $z=0.5$ when turbulence dominates high equivalent widths of a few tenths of \AA\ extend into the inner CGM, in contrast with being limited to disk radii ($<0.1\Rvir$) at $z=0.15$ when thermal energy dominates. The observed $W_{2796}$ from blue $\sim L^\star$ galaxies in the COS-Halos and \cite{Huang21} samples shown in the left panel exhibit high $W_{2796}$ at inner CGM radii consistent with the turbulence-dominated regime, albeit with a larger dispersion. This difference is potentially due to the range in mass and redshift spanned by the observations in contrast with the single galaxy and redshift in the FIRE predictions shown in Fig.~\ref{fig: WMgIIAgainstR_R_virComparisonWithObservations_m12i}.

The right panel of Fig.~\ref{fig: WMgIIAgainstR_R_virComparisonWithObservations_m12i} is similar to the left panel, but showing observations from red galaxies in the COS-Halos and \cite{Huang21} samples. Red galaxies 
exhibit on average lower $W_{2796}$ than blue galaxies at inner CGM radii, with approximately two-thirds of sightlines at $0.1-0.3\Rvir$ having $W_{2796}<0.2$\AA, in contrast with only $15\%$ of sightlines in the blue galaxy sample. High $W_{2796}$ absorbers thus exist around red galaxies but are not ubiquitous as they are around blue galaxies, ruling out a cool phase-dominated  turbulent inner CGM for red galaxies. The $\approx1/3$ of sightlines with high equivalent widths around red galaxies are not predicted by the thermal-dominated FURE snapshot shown in Fig.~\ref{fig: WMgIIAgainstR_R_virComparisonWithObservations_m12i}, potentially since localized cool clouds are unresolved in the simulation during the post-ICV stage (see section~\ref{s:resolution}).

\subsection{\texorpdfstring{\ion{C}{iv}}{C IV} equivalent widths}\label{s:civ}

The FIRE simulations predict a mean $W_{1548} \sim 1$\AA\ from turbulence-dominated inner CGM of $\sim L^\star$ galaxies (Figs.~\ref{fig: DistributionofTotalWAgainstLookbackTimeforDifferentIons_m12i_Part2} and \ref{fig: DistributionofTotalWAgainstLookbackTimeforDifferentIons_m11d_m12b_A1}). For comparison, \cite{Garza24} recently found a mean $W_{1548}=1\pm0.16$\AA\ in SF COS-Halos galaxies at $R_\perp<0.3\Rvir$, compared with a lower mean $W_{1548}=0.2\pm0.08$\AA\ in quiescent galaxies. \ion{C}{iv} absorbers with $W_{1548}>0.7$\AA\ are also observed with $>50\%$ covering factor out to $R_\perp=23^{+62}_{-16}\kpc$ around [\ion{O}{ii}]- emitting galaxies observed with MUSE at $z\sim1.2$ \citep{Schroetter21}. This corresponds to an impact parameter of $0.17^{+0.52}_{-0.12}\Rvir$ for the characteristic $z=1.2$ and $M_\star=10^{10}\msun$ in the \citeauthor{Schroetter21} sample assuming the stellar to halo mass relation from \cite{Behroozi19}. Observed $\langle W_{1548}\rangle$ of star-forming galaxies are thus also consistent with turbulence-dominated inner CGM, as found above for observations of $\EWmgii$.

In lower mass dwarfs, mean $W_{1548}$ of $0.4$\AA\ are predicted at $z=0$ (bottom panel of  Fig.~\ref{fig: DistributionofTotalWAgainstLookbackTimeforDifferentIons_m11d_m12b_A1}). These values are lower than in more massive turbulence-dominated inner CGM due to the lower turbulent velocities (bottom-right panel of Fig~\ref{fig: MassFractionandSigmaRhoandMuVAgainstLookbackTime_m11d_m12b_A1}). For comparison, \cite{Bordoloi14} measured a mean $W_{1548}=0.45\pm0.1$\AA\ at $0.1<R_\perp< 0.3\,\Rvir$ around $z\sim0.1$ star-forming dwarf galaxies. Similar values were found at small impact parameter in the samples of \cite{Liang14}, \cite{Johnson17}, and \cite{Manuwal21}. 
Observed low-mass blue galaxies are thus consistent with having a cool and turbulent inner CGM, as previously mentioned by \cite{Li21} and similar to more massive blue galaxies.  

The observed prevalence of $W_{1548}\sim1$\AA\ in the inner CGM of blue galaxies at $0<z<1$ is in contrast with the absence of similar absorbers in the inner CGM of red galaxies (\citealt{Garza24}, and see also \citealt{Bordoloi14}). This dichotomy is also evident in \ion{Mg}{ii} absorption as discussed in the previous section. Our results thus suggest that the distinction in inner CGM absorption properties of blue and red galaxies is a manifestation of the difference between a cool,  turbulence-dominated inner CGM around blue galaxies, versus a hot, thermal energy-dominated inner CGM around red galaxies. If this interpretation is correct, this distinction should manifest in the $\tcool/\tff$ ratio inferred for the CGM of these galaxies, as further discussed in section~\ref{sec:tcool-tff} below. 

We note also that \ion{C}{iv} absorption in the inner CGM of the Milky-Way appears to be significantly lower than that in external blue $\lesssim L^*$ galaxies. \cite{Bish21} deduced a covering fraction of only 20\%\ for $W_{1548}>0.2$\AA\ absorbers in the Milky-Way CGM, using sightlines to galactic stars with small angular separations from background quasars in order to account for absorption by the ISM. Such low covering fraction is inconsistent with the turbulence-dominated inner CGM scenario, suggesting that the Milky-Way inner CGM is dominated by a quasi-static hot phase, as suggested also by X-ray observations (see Introduction). The \cite{Bish21} results thus suggest that the Milky-Way is post-ICV similar to the m12 simulations at $z=0$, and in contrast with typical blue $\sim L^\star$ galaxies in the samples shown in Fig.~\ref{fig: DistributionofTotalWAgainstLookbackTimeForMgIIComparisonWithObservation_L_Stars}.

\begin{table*}
\centering
\scriptsize
\resizebox{\textwidth}{!}{
\begin{tabular}{|c|c|c|c|c|c|c|c|c|c|}
\hline
Survey & $\log\ M_{\star}$ [${\rm M}_{\odot}$] & $\log\ M_{\rm halo}$ [${\rm M}_{\odot}$] & Mass group & Type & $z_{\text{range}}$ & $\langle z \rangle$ & $R_{\perp}$ [kpc] & $\EWmgii$ [\AA] & Ref. \\
(1) & (2) & (3) & (4) & (5) & (6) & (7) & (8) & (9) & (10) \\
\hline\hline
\multirow{3}{*}{zCOSMOS} & 9.6 & 11.5 & Dwarf & SF & \multirow{3}{*}{0.5 -- 0.9} & \multirow{3}{*}{0.7} & 25 & 0.5 $\pm$ 0.1 & \multirow{3}{*}{\protect\cite{Bordoloi11}} \\
 & 10.2 & 11.9 & $L^*$ & SF & & & 33 & 0.9 $\pm$ 0.75 & \\
 & 10.9 & 13.0 & LRG & Q & & & 81 & 0.2 $\pm$ 0.6 & \\
\hline
COS-Halos & 10.3 & 11.9 & $L^*$ & SF & 0.15 -- 0.4 & 0.2 & 46 & 1.14 $\pm$ 0.55 & \protect\cite{Werk13} \\
\hline
\multirow{4}{*}{SDSS} & 10.5 & 12.2 & $L^*$ & SF & 0.4 -- 1.5 & 0.85 & 39 & 2.0 $\pm$ 0.5 & \protect\cite{Lan18} \\
 & 10.4 & 12.1 & $L^*$ & SF & 0.4 -- 1.3 & 0.84 & 36 & 1.27 $\pm$ 0.42 & \protect\cite{Anand21} \\
 & 11.2 & 13.5 & LRG & Q & 0.4 -- 0.7 & 0.55 & 124 & 0.15 $\pm$ 0.05 & \protect\cite{Lan18} \\
 & 11.5 & 14.0 & LRG & Q & 0.4 -- 0.6 & 0.54 & 183 & <0.4 & \protect\cite{Anand21} \\
\hline
\multirow{2}{*}{DESI} & 10.0 & 11.8 & \multirow{2}{*}{$L^*$} & \multirow{2}{*}{SF} & 0.75 -- 1.0 & 0.88 & 28 & 1.8 $\pm$ 0.0 & \multirow{2}{*}{\protect\cite{Wu24}} \\
 & 10.1 & 11.8 & & & 1.0 -- 1.6 & 1.23 & 24 & 3.0 $\pm$ 0.0 & \\
\hline
Huang\plus'21 & 10.3 & 11.9 & $L^*$ & SF & 0.1 -- 0.5 & 0.22 & 47 & 1.2 $\pm$ 0.1 & \protect\cite{Huang21} \\
\hline
\multirow{2}{*}{Gravitational Arcs} & 9.7 & 11.7 & Dwarf & SF & --- & 0.73 & 25 & 0.6 $\pm$ 0.3 & \protect\cite{Lopez20} \\
 & 10.1 & 11.9 & $L^*$ & SF & --- & 0.77 & 27 & 1.4 $\pm$ 0.3 & \protect\cite{Mortensen21} \\
 \hline
 MEGAFLOW & 10.1 & 11.8 & $L^*$ & SF & 0.4 -- 1.4 & 0.9 & 26 & 2.5 $\pm$ 1.15 & \protect\cite{Cherrey25} \\
\hline
\end{tabular}
}
\caption{Details of circumgalactic $\EWmgii$ measurements shown in Figs.~\protect\ref{fig: DistributionofTotalWAgainstLookbackTimeForMgIIComparisonWithObservation_L_Stars} and \ref{fig: DistributionofTotalWAgainstLookbackTimeForMgIIComparisonWithObservation_Dwarfs+LRGs}. Columns: (1) survey name; (2) mean stellar mass; (3) mean halo mass, derived from mean $M_\protect\star$, mean $z$, and the $M_{\rm halo}-M_\protect\star$ relation in \protect\cite{Behroozi19}; (4) mass group of foreground galaxies (dwarf, $\sim L^*$ or LRG); (5) foreground galaxy type (SF: star-forming, Q: quiescent); (6) sample redshift range; (7) sample median redshift; (8) impact parameter equal to $0.2\,R_{\rm vir}$; (9) measured mean \ion{Mg}{ii} equivalent width; (10) reference paper. See additional details in appendix~\protect\ref{a:obs}.}
\label{table:ObservationsTable}
\end{table*}

\subsection{Other UV transitions}\label{s:other UV}

In the COS-Halos sample, observations with small impact parameters from blue galaxies and high $\EWmgii$ also typically have strong saturated absorption in other transitions, including \ion{O}{vi}, \ion{Si}{iii}, \ion{C}{ii}, \ion{C}{iii}, and \ion{Si}{ii} \citep{Werk13,Stern16_CGM}. This implies that such absorbers span a large range in ionization, qualitatively consistent with the wide density and temperature distribution expected in turbulence-dominated inner CGM (bottom panel of Fig.~\ref{fig: AllGasMassWeightedlog(rho)Distribution_m12i_z_0_z_0.75}). We leave a more quantitative test of the turbulence-dominated CGM regime against the full range of observed absorbers to future work (see section~\ref{s:cloudy} below). 

\cite{Lehner25} recently found that M31 exhibits $5-6$ times lower \ion{Si}{ii} columns than those in blue COS-Halos galaxies at $R_\perp<0.3\Rvir$, confirming earlier conclusions from \cite{Lehner20} with better statistics. UV absorption from the inner CGM of M31 thus appears inconsistent with the turbulence-dominated scenario in contrast with blue COS-Halos galaxies, and suggests M31 is post-ICV similar to our m12 simulations at $z=0$, and as expected since M31 is a massive disk that most likely has a quasi-static hot CGM. 

\begin{figure*}
    \includegraphics[width=\textwidth]{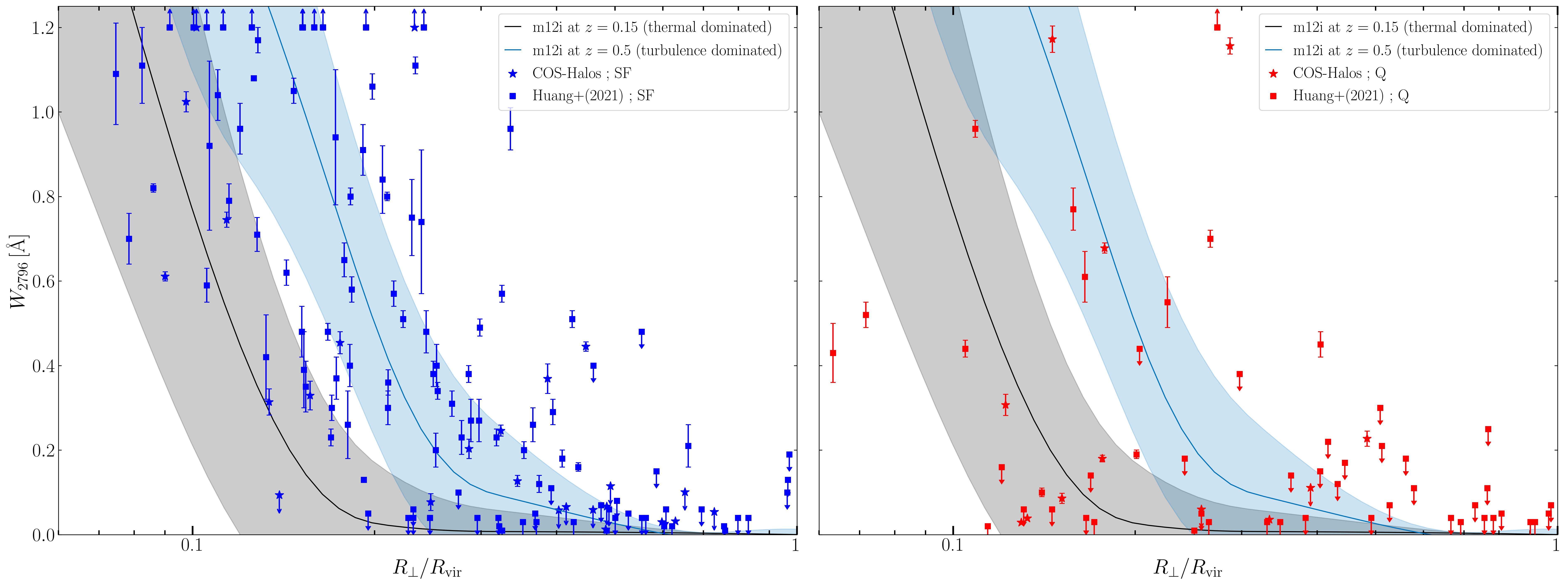}
    \caption{
    The difference in predicted \ion{Mg}{ii} equivalent width between turbulence-dominated and thermal energy-dominated inner CGM, versus impact parameter. 
    Solid lines and shaded regions represent the mean and  scatter of sightlines through the $z=0.5$ snapshot (blue, turbulence-dominated) and the $z=0.15$ snapshot (black, thermally supported) of the m12i simulation. Equivalent widths of a few tenths of \AA\ are predicted from the inner CGM when it is turbulence dominated, while such values are limited to disk radii when the inner CGM is thermally supported. Markers represent observed $W_{2796}$ around star-forming (left panel) and quiescent (right) galaxies. The ubiquity of high $W_{2796}$ in the inner CGM of star forming galaxies supports the turbulence-dominated scenario for these objects.
    }
    \label{fig: WMgIIAgainstR_R_virComparisonWithObservations_m12i}
\end{figure*}

\subsection{\texorpdfstring{\boldmath The $\tcool<\tff$ ratio in systems with $\EWmgii\sim1$\AA}{The tcool<tff ratio in systems with EW(Mg II) ~ 1 Å}}\label{sec:tcool-tff}

Inner CGM filled with a turbulent cool phase are expected when $\tcool<\tff$. We demonstrate in this section that this timescale ratio can be constrained from observations, and thus we can test our association of $\EWmgii\sim1$\AA\ absorbers with turbulence-dominated CGM by checking if these systems indeed have $\tcool<\tff$.  

Using equations~(\ref{eq:t_cool}) -- (\ref{eq:t_ff}), we get  
\begin{equation}
    \frac{\tcool}{\tff} = \frac{(9/10)m_{\rm p}v_{\rm c}^3 }{\sqrt{2}X r \langle n_{\rm H}\rangle \Lambda} ~.
\end{equation}
Further using $T^{\rm (s)}/10^6\,{\rm K}=0.45\,v_{100}^2$ where $v_{100}=v_{\rm c}/100\,{\rm km}\,{\rm s}^{-1}$ (eqn.~\ref{eq:T^s}) and the \cite{Gnat07} approximation for the cooling function\footnote{The typical definition of $\Lambda$ assumes energy losses per unit volume are $\mathcal{L}=n_{\rm e}n_{\rm H}\Lambda$, where $n_{\rm e}$ is the electron density. We converted to our definition ($\mathcal{L}=n_{\rm H}^2\Lambda$) using $n_{\rm e}=1.3\,n_{\rm H}$ appropriate for ionized solar metallicity gas.} for gas with metallicity close to solar:
\begin{equation}\label{eq:Lambda approx}   
\Lambda(T,Z)=1.7\cdot10^{-22}(T/10^6\,{\rm K})^{-0.54}(Z/{\rm Z}_\odot)\,{\rm erg}\,{\rm cm}^3\,{\rm s}^{-1}~,
\end{equation}
we get
\begin{equation}\label{e:tratio obs}
    \frac{\tcool}{\tff} \approx 0.8\, v_{100}^{4.1} \left(\frac{N_{\rm H}}{10^{19}\,{\rm cm}^{-2}}\cdot\frac{Z}{{\rm Z}_\odot}\right)^{-1}~,
\end{equation}
where we used $T=T^{({\rm s})}$ in eqn.~(\ref{eq:Lambda approx}) and replaced $\langle n_{\rm H}\rangle r$ with $\approx(2/\pi)N_{\rm H}$, a relation which is exact for $\langle n_{\rm H}\rangle \propto r^{-2}$. Thus, eqn.~(\ref{e:tratio obs}) demonstrates that one can estimate $\tcool/\tff$ based on a measurement of $v_{\rm c}$ and the metal column density $N_{\rm H}Z$. 

To estimate $\NH Z$ we use the mean Si$^+$ column $N_{\rm Si^+}=6\pm1\cdot 10^{14}\,{\rm cm}^{-2}$ measured by \cite{Lan17}, based on mean \ion{Si}{ii}~$\lambda 1808$ absorption associated with $\EWmgii>1$\AA\ absorbers at $z\approx1$. The \ion{Si}{ii}~$\lambda 1808$ transition is typically optically thin due to its low oscillator strength ($\tau=0.37$ at line center for the sightline shown in Figs.~\ref{fig: TemperatureProjectionsWithSi_IIISpectrum_m12i_z_0_z_0.75} and \ref{fig: EUVIons(Mg_II,O_III,C_IV,O_VI)_m12i_z_0.75}) and thus the deduced ion column is more robust than when using saturated absorption lines. 
Given the solar Si/H ratio of $10^{-4.5}$, 
we thus infer $N_{\rm H}(Z/{\rm Z}_\odot)=2\pm0.3 \cdot 10^{19}N_{\rm Si}/N_{\rm Si^{+}}$. Additionally, the same sample has $v_{\rm c}\approx 140\pm20\kms$, derived from the 1D velocity dispersion of the dark matter halo $\sigma_{\rm dm}$ inferred by \cite{Lan18} based on abundance matching, and using $v_{\rm c}= \sqrt{2}\sigma_{\rm dm}$ appropriate for an isothermal potential. We thus get
\begin{equation}\label{e:tratio obs2}
\frac{\tcool}{\tff}=1.7 \frac{N_{\rm Si^{+}}}{N_{\rm Si}}\left(\frac{v_{\rm c}}{140\kms}\right)^{4.1}\left(\frac{N_{\rm Si^+}}{6\cdot 10^{14}\,{\rm cm}^{-2}}\right)^{-1} ~,
\end{equation}
where the uncertainties on the derived $v_{\rm c}$ and metal column mentioned above imply a factor of $\approx2$ uncertainty on the derived $\tcool/\tff$. For a typical $N_{\rm Si^{+}}/N_{\rm Si}$ fraction of 0.2 in turbulence-dominated inner CGM in m12i at $z=1$, the observationally-inferred $\tcool/\tff$ is thus $\approx0.2 - 0.7$. This value is smaller than unity, which provides further supporting evidence for our conclusion that $\EWmgii\sim1$\AA\ absorbers trace turbulence-dominated inner CGM. The result that the inferred ratio is close to unity suggests that these are the most massive galaxies with a turbulence dominated inner CGM at this redshift. 

\cite{Lan18} also fit a line of sight gas dispersion of $\sigma_{\rm gas}=105\pm10\kms$ to the mean \ion{Mg}{ii} absorption profile of their sample. The ratio $\sigma_{\rm gas}/\sigma_{\rm dm}\approx1$ they deduce can be compared to the $\sigma_{\rm turb}/v_{\rm c}\approx0.7 - 0.9 $ we find in m11 and m12 FIRE galaxies at $z\approx1$ (Figs.~\ref{fig: MassFractionandSigmaRhoandMuVAgainstLookbackTime_m12i} and \ref{fig: MassFractionandSigmaRhoandMuVAgainstLookbackTime_m11d_m12b_A1}). Since $\sigma_{\rm turb}\approx\sqrt{3}\sigma_{\rm gas}$ and $v_{\rm c}\approx\sqrt{2}\sigma_{\rm dm}$, we get $\sigma_{\rm gas}/\sigma_{\rm dm}\approx 0.6-0.7$ in FIRE, somewhat lower than deduced by \cite{Lan18} from the observations. This difference may be a result of stacking saturated absorption features around galaxies which span a range in mass, which could bias the measured $\sigma_{\rm gas}$ somewhat high relative to the median value in the sample. Alternatively, the turbulent velocity in FIRE could be somewhat underpredicted relative to real galaxies. We leave a more accurate comparison of observed \ion{Mg}{ii} equivalent widths with those in FIRE to future work. 

\section{DISCUSSION}\label{s:discussion}

\subsection{How does hot CGM formation affect galaxy evolution?}\label{How does hot CGM formation affect galaxy evolution?}

It has long been predicted that the formation of a quasi-static hot phase in the CGM is required for galaxy quenching \citep{keres05, dekel06,Bower06,Croton06,Somerville08}. 
Our results provide direct evidence that these processes are at least correlated, by showing that sightlines through inner CGM prior to hot phase formation are expected to exhibit  $\EWmgii\sim\EWciv\sim1$\AA, in contrast with after hot phase formation where lower average equivalent widths of $\ll1$\AA\ are expected (Figs.~\ref{fig: DistributionofTotalWAgainstLookbackTimeforDifferentIons_m12i_Part2} and \ref{fig: DistributionofTotalWAgainstLookbackTimeforDifferentIons_m11d_m12b_A1}). 
Thus the observed $\EWmgii\sim1$\AA\ around
star forming galaxies in contrast with $\EWmgii\ll1$\AA\ around red galaxies \citep[see Figs.~\ref{fig: DistributionofTotalWAgainstLookbackTimeForMgIIComparisonWithObservation_L_Stars} and ~\ref{fig: DistributionofTotalWAgainstLookbackTimeForMgIIComparisonWithObservation_Dwarfs+LRGs}]{Bordoloi11, Bowen11, Lan14, Huang15, Lan18, Anand21} and evidence for a similar dichotomy in $\EWciv$ \citep{Bordoloi14, Garza24} indicates that in typical $\sim L^*$ blue galaxies a quasi-static hot phase has not yet formed in the inner CGM, while in red galaxies it has. 

More recent studies based on the FIRE simulations have argued that hot phase formation in the inner CGM also facilitates the formation of thin galactic disks, while prior to hot phase formation thick disk and irregular morphologies are expected \citep{Stern21A,Yu21,Yu23,Hafen22,Gurvich23}. Observations suggest that thin disks have formed mainly at late times ($z<1$) and in relatively massive star forming galaxies \citep{kassin12,Tiley21}. If this connection between `disk settling' and ICV is true, then we expect that massive thin disk galaxies should exhibit $\EWmgii\ll1$\AA\ and $\EWciv\ll1$\AA\ in their inner CGM, in contrast with the $W_\lambda\sim1$\AA\ seen in the $M_\star\sim 10^{10}\msun$ galaxies which dominate the samples shown in Fig.~\ref{fig: DistributionofTotalWAgainstLookbackTimeForMgIIComparisonWithObservation_L_Stars}. Such low equivalent widths are indeed inferred in the Milky-Way CGM by \citep{Bish21}, in M31 by \cite{Lehner25}, and in the most massive blue galaxies at $0.4<z<0.7$ by \cite{Lan20} as discussed above.
There is thus tentative evidence to support this suggested scenario, though further comparison of SF galaxy morphologies and $\EWmgii$ at small impact parameters are required. The large sample of quasar-absorber pairs upcoming in DESI \citep{Wu24} could be useful for this test.

\subsection{\texorpdfstring{\boldmath Implications for \cloudy\ modelling of UV absorbers}{Implications for CLOUDY modelling of UV absorbers}}\label{s:cloudy}

Observational inference studies based on circumgalactic UV absorbers usually use \cloudy\ \citep{Ferland17,Chatzikos23} to derive properties of absorbing gas structures, including volume and column densities, cloud physical sizes and gas metallicity \citep[e.g.,][]{Prochaska99}. These studies almost always assume that along a given line of sight and a given velocity relative to the galaxy, the absorbing gas originates in one or more independent clouds, where each cloud has uniform properties, i.e.\ a single density and metallicity. 
While this approach is reasonably motivated when UV absorbers are cool clouds embedded in a hot medium (e.g.~left panels of Fig.~\ref{fig: TemperatureProjectionsWithSi_IIISpectrum_m12i_z_0_z_0.75}) it is inadequate if the hot phase is subdominant and a cool turbulent medium fills the volume. As shown in the right panel of Fig.~\ref{fig: AllGasMassWeightedlog(rho)Distribution_m12i_z_0_z_0.75},  when the cool gas dominates it has a continuous lognormal density distribution, and thus a different modeling approach is required.

An alternative would be to account for the inherently lognormal density distributio of UV-absorbing gas within the \cloudy\ model \footnote{\cloudy\ can model arbitrary density profiles in the absorbers, such as lognormals or power-laws as done in \cite{Stern16_CGM}.}. By comparing observations to a grid of such models with varying mean density $\langle n_{\rm H}\rangle$, density width $\sigma_{\log\rho}$, and total gas mass, one can find the parameters which best describe a given observation set. These best-fit parameters can then be compared to those in the simulations, thus providing an observational test of the cool and turbulent CGM scenario. 

A similar modelling approach was recently suggested by \cite{Dutta24}, though they advocated applying this technique regardless of whether the absorbers originate in thermal pressure-dominated or turbulence-dominated CGM. We emphasize that the density distributions of cool gas in these two regimes are inherently different (see Fig.~\ref{fig: AllGasMassWeightedlog(rho)Distribution_m12i_z_0_z_0.75}), since in turbulence-dominated CGM the density distribution is a result of the supersonic turbulence, while in thermal-pressure dominated CGM the cool gas density distribution is set by the interaction between cool-clouds and the hot background \citep[e.g., the turbulent radiative mixing layers described in][]{Fielding20}. The free parameter space required to explore absorption from turbulence-dominated CGM as suggested here is thus smaller than the general case proposed by \cite{Dutta24}.  

\subsection{Implications for small-scale physics of the CGM}\label{s:small scale}

Recent theoretical studies of small-scale physics in the CGM have focused on the thermal pressure-dominated scenario, in which cool clouds are embedded in a hot volume-filling background (e.g., \citealt{McCourt18}, \citealt{Gronke18}, \citealt{Fielding20}, \citealt{Afruni21}, and many others, see review in \citealt{FaucherGiguere23}). 
The physics of small-scale structure when a quasi-static hot phase is subdominant as discussed in this work are likely qualitatively different than when it is dominant. For example, when turbulence dominates, the interaction between different temperature phases is driven primarily by ram pressure -- rather than by thermal pressure as commonly assumed \citep[e.g.,][]{McCourt18,Gronke22,Abruzzo22}. Consequently, observable properties such as cloud coherence scales \citep[e.g.,][]{Afruni23} and velocity structure functions \citep[e.g.,][]{Chen23} are expected to differ. Given our results above that turbulence-dominated CGM may be prevalent at inner halo radii of star-forming galaxies (Fig.~\ref{fig: DistributionofTotalWAgainstLookbackTimeForMgIIComparisonWithObservation_L_Stars}), further theoretical exploration of small-scale physics in turbulence-dominated CGM would be beneficial.

\subsection{Caveats: Resolution, Cosmic Rays, and AGN feedback}

It has been argued that the cool gas mass fraction $f_{\rm cool}$ in  thermal-pressure dominated CGM depends on simulation resolution \citep{vandeVoort19,Hummels19,Peeples19,Ramesh24}. We thus cannot strictly rule out a thermal-pressure dominated inner CGM origin for $W_\lambda\sim1$\AA\ absorbers.  However, in the turbulence-dominated regime we find $f_{\rm cool}\approx1$ independent of simulation resolution (Fig.~\ref{fig: m12i_r880_m12i_r7100_m12i_r57000_MassFraction_MuV_With_W_MgII_And_t_cool_t_ff} and section~\ref{s:resolution}), and thus the predicted absorption equivalent widths induced by cool gas are robust in this regime and consistent with analytic estimates (section~\ref{sec:analytic}). The turbulence-dominated inner CGM scenario is thus possible to rule out with cool gas observations. The success of this scenario to reproduce observations thus supports its applicability. 

We note also that the version of the FIRE simulations used in this study assume ideal hydrodynamics, and in particular do not include cosmic ray (CR) physics. Other FIRE zoom simulations that include CRs with constant diffusion coefficients have shown that CRs can potentially prevent the formation of a volume-filling hot phase at the Milky-Way mass scale at $z<1$, replacing thermal pressure support with CR pressure support \citep[e.g.,][]{hopkins20_whatabout,Hopkins21_CRtransport,Ji20,Ji21}. However, in lower mass galaxies and at $z\gtrsim1$, CR pressure remains subdominant in such FIRE simulations with CRs. At these masses and redshifts, we find above that inner CGM are dominated by turbulence. The previous analyses of FIRE simulations including CRs suggest that this regime is not qualitatively affected by CR feedback.

A dominance of CR pressure over thermal pressure at the Milky-Way mass scale implies a high mass fraction of the cool CGM phase, and thus has been suggested as an explanation for the substantial cool gas reservoirs seen in CGM observations of low-redshift $\lesssim L^*$ galaxies \citep{Salem16,Butsky18, Buck20, Ji20, Butsky20, Butsky22,DeFelippis24}. Our results suggest an alternative explanation, where the high observed cool gas columns are a result of short cooling times in the hot phase, rather than by CR pressure.

The simulations analyzed in this work also do not include feedback from active galactic nuclei (AGN), and thus our results on cool, pre-ICV CGM are strictly applicable only if AGN feedback does not qualitatively change the CGM  in this regime. We note that it has long been argued that cool CGM are not strongly affected by AGN feedback, either because the CGM is less susceptible to energy deposition or due to limited black hole growth during this early phase (see Introduction above and discussion in \citealt{Byrne23}). The ability of pre-ICV CGM to correctly reproduce observed MgII equivalent width observations around blue galaxies at $0 \lesssim z \lesssim 1.5$ (Fig.~\ref{fig: DistributionofTotalWAgainstLookbackTimeForMgIIComparisonWithObservation_L_Stars}) may suggest that the effect of AGN feedback on the CGM pre-ICV is indeed small. 

\section{SUMMARY}\label{s:summary}

This paper continues the investigation of a qualitative transition in the inner CGM ($\sim0.2\Rvir$) of galaxies simulated in FIRE, at which a quasi-static and volume-filling hot phase forms at inner halo radii. This transition has been dubbed `inner CGM virialization' or ICV, and occurs when the cooling time of hot gas in the inner CGM exceeds the free-fall time, corresponding to the halo mass exceeding a threshold of  $\approx10^{12}\msun$. Prior to this transition hot gas which formed via accretion or feedback shocks rapidly cools in the inner CGM, so the inner CGM is dominated by cool inflows and outflows \citep{vandeVoort16,Stern21A,Stern21B,Yu21,Yu23,Hafen22,Gurvich23,Byrne23}.
The current study focuses on implications of this transition for circumgalactic UV absorption at $0\leq z \lesssim2$, and specifically we characterize the distinct CGM absorption signatures in pre-ICV galaxies. 
Our main results can be summarized as follows:

\begin{enumerate}[leftmargin=*, labelwidth=1.5em, labelsep=0.5em]
    \item ICV in FIRE corresponds to a transition from supersonic CGM turbulence ($\sigma_{\rm turb}\gg\langle c_{\rm s}\rangle_{\rho}$) to subsonic CGM turbulence ($\sigma_{\rm turb}\lesssim\langle c_{\rm s}\rangle_\rho$), where $\sigma_{\rm turb}$ and $\langle c_{\rm s}\rangle_\rho$ are respectively the 3D turbulent velocity and mass-weighted sound speed in a given radial shell (Figs.~\ref{fig: MassFractionandSigmaRhoandMuVAgainstLookbackTime_m12i}, \ref{fig: MassFractionandSigmaRhoandMuVAgainstLookbackTime_m11d_m12b_A1}). Equivalently, ICV is a transition between turbulence pressure-dominated and thermal pressure-dominated inner CGM (Fig.~\ref{fig: (P_turb):(P_turb+P_thermal)AgainstLookbackTime _m12i.png}). This conclusion is consistent with previous idealized studies on hot phase formation which accounted for stellar feedback \citep{Fielding17,Pandya23}. 
    \item The transition to subsonic turbulence is driven by an increase in $\langle c_{\rm s}\rangle_\rho$ from $20-30\kms$ to $\gtrsim100\kms$ when the hot phase becomes dominant, in contrast with $\sigma_{\rm turb}$ which is roughly constant or even decreases. During the supersonic phase we find $\sigma_{\rm turb}(0.2\,\Rvir)\sim v_{\rm c}$, where $v_{\rm c}$ is the circular velocity (Figs.~\ref{fig: MassFractionandSigmaRhoandMuVAgainstLookbackTime_m12i}, \ref{fig: MassFractionandSigmaRhoandMuVAgainstLookbackTime_m11d_m12b_A1}).
    \item When turbulence is subsonic, the gas density distribution at a given radius consists of a narrow peak of hot gas with a high density tail due to cool clouds, as in the common CGM paradigm.  In contrast, when turbulence is supersonic, the density distribution forms a single wide, roughly lognormal with FWHM approaching $2\,{\rm dex}$ (Fig.~\ref{fig: AllGasMassWeightedlog(rho)Distribution_m12i_z_0_z_0.75}). The widths of the density distributions are consistent with expectations from idealized simulations of isothermal turbulence with compressive driving (bottom panel of Fig.~\ref{fig: MassFractionandSigmaRhoandMuVAgainstLookbackTime_m12i}).
    \item The nature of UV-absorbing gas changes upon ICV. After ICV, absorbers trace localized cool clouds embedded in a hot CGM as is commonly assumed \citep[e.g.,][]{Tumlinson17}. Prior to ICV, absorption features trace the turbulent and cool volume-filling phase of the inner CGM (Figs.~\ref{fig: TemperatureProjectionsWithSi_IIISpectrum_m12i_z_0_z_0.75}, \ref{fig: EUVIons(Mg_II,O_III,C_IV,O_VI)_m12i_z_0.75}). 
    \item For pre-ICV $\lesssim L^\star$ galaxies at $0\leq z \lesssim 2$ with turbulence-dominated inner CGM, we predict mean equivalent widths of $\langle W_\lambda\rangle \sim 1\text{\AA}$ at impact parameters $\approx0.2\Rvir$, across a broad range of strong UV transitions (\ion{Mg}{ii}, \ion{C}{ii-iv}, \ion{Si}{ii-iv}, \ion{O}{iii-v}),  (Figs.~\ref{fig: DistributionofTotalWAgainstLookbackTimeforDifferentIons_m12i_Part1}, \ref{fig: DistributionofTotalWAgainstLookbackTimeforDifferentIons_m11d_m12b_A1}). These high $\langle W_\lambda\rangle$ are due to the dominance of the cool phase, the large turbulent velocity, and the wide density distribution which entails a wide range of ionization. 
    \item Simulation resolution, at the range probed of  $880-57\,000\,\msun$, does not significantly affect the predicted $\langle W_\lambda\rangle$ prior to ICV. This follows since most of the inner CGM is cool even at low resolution, so increasing resolution does not significantly further increase the mass of the cool phase (Fig.~\ref{fig: m12i_r880_m12i_r7100_m12i_r57000_MassFraction_MuV_With_W_MgII_And_t_cool_t_ff}). 
    \item Available UV absorption surveys, including COS-Halos \citep{Werk13}, zCOSMOS \citep{Bordoloi11}, SDSS \citep{Lan18, Anand21}, and DESI \citep{Wu24} indicate that star-forming $\lesssim L^*$ galaxies have order-unity covering fraction of $W_{\ion{Mg}{ii}} \sim 1\text{\AA}$ absorbers at $R_\perp\approx0.2\Rvir$, consistent with the prediction of turbulence-dominated inner CGM (Fig.~\ref{fig: DistributionofTotalWAgainstLookbackTimeForMgIIComparisonWithObservation_L_Stars}).  
    Red galaxies exhibit significantly lower mean $W_{\ion{Mg}{ii}}$, inconsistent with turbulence-dominated CGM predictions. This provides direct evidence for a connection between the formation of a quasi-static hot CGM phase and quenching of star formation, as postulated by many previous studies \citep{keres05,dekel06,Croton06,Somerville08,Bower17,Byrne23}. 
    \item The Milky-Way exhibits $\langle W_{1548}\rangle\ll1$\AA\ \citep{Bish21}, inconsistent with a turbulence dominated inner CGM, and tentative evidence suggests that low $\langle W_{\lambda}\rangle$  are common also in M31 \citep{Lehner20,Lehner25} and in the most massive star forming galaxies at $z\approx0.5$ \citep{Lan20}. UV absorption signatures thus suggest that massive, low $z$ disks have thermal-energy dominated inner CGM as suggested also by X-ray emission and absorption observations. The inner CGM of these massive low $z$ disks is thus qualitatively distinct from those of blue $\sim L^\star$ galaxies in the survey shown in Fig.~\ref{fig: DistributionofTotalWAgainstLookbackTimeForMgIIComparisonWithObservation_L_Stars}.
\end{enumerate}
 
Our results provide a means to identify which galaxies are not surrounded by a quasi-static hot phase, but rather by a predominantly cool, turbulence-dominated  inner CGM: such galaxies should exhibit $\sim1$\AA\ equivalent width UV absorption in their inner CGM accross a wide range of ions. Based on these results and available observations, we conclude that a quasi-static hot phase in the inner CGM exists only around the most massive star forming galaxies (including the Milky-Way and M31) and around quenched galaxies. This result constrains the relation between CGM thermodynamics and the evolution of the central galaxy.

We conclude by noting that most contemporary observational inference studies of the CGM, and studies of small-scale CGM physics, assume the existence of a quasi-static volume-filling hot phase  (sections~\ref{s:cloudy} -- \ref{s:small scale}). This prevalent assumption is despite that CGM which lack such a hot phase have been predicted in low-mass halos for decades now, in multiple theoretical studies based on analytic methods, idealized simulations, and cosmological simulations. Given our results above that cool phase-dominated inner CGM are common around star forming galaxies, we encourage further study of the theoretical predictions and observational implications of this important, cool and turbulent CGM regime. 

\section*{ACKNOWLEDGEMENTS}

We thank A.~Fox, S.~D.~Johnson, and    J.~X.~Prochaska for insightful conversations which  motivated this work, and A.~Sternberg for providing comments on a draft version of the manuscript. AK, JS, and RG were supported by the Israel Science Foundation (grant No.~2584/21). 
CAFG was supported by NSF through grants AST-2108230 and AST-2307327; by NASA through grants 21-ATP21-0036 and 23-ATP23-0008; and by STScI through grant JWST-AR-03252.001-A. Part of this work was performed at the Aspen Center for Physics, which is supported by National Science Foundation grant PHY-2210452. 
Numerical calculations were run on the Northwestern computer cluster Quest, the Caltech computer cluster Wheeler, Frontera allocation FTA-Hopkins/AST20016 supported by the NSF and TACC, XSEDE/ACCESS allocations ACI-1548562, TGAST140023, and TG-AST140064 also supported by the NSF and TACC, and NASA HEC allocations SMD-16-7561, SMD-17-1204, and SMD-16-7592.
In the analysis, we utilized the \textsc{yt v4.4} toolkit (\citealt{Turk2011, yt_project}), the \textsc{Trident v1.3} code (\citealt{Hummels2017, trident_docs}), and the \textsc{Amiga Halo Finder} (\textsc{AHF}; \citealt{Knollmann2009}). All analysis scripts were developed in \textsc{Python 3.10}.

\section*{DATA AVAILABILITY}

A public version of the GIZMO code is available at \url{http://www.tapir.caltech.edu/~phopkins/Site/GIZMO.html}. FIRE data products, including FIRE-2 simulation snapshots, initial conditions, and derived data products are available at \url{http://fire.northwestern.edu/data/}.
\bibliographystyle{mnras}
\bibliography{References}

\begin{thebibliography}{}
\makeatletter
\relax
\def\mn@urlcharsother{\let\do\@makeother \do\$\do\&\do\#\do\^\do\_\do\%\do\~}
\def\mn@doi{\begingroup\mn@urlcharsother \@ifnextchar [ {\mn@doi@} {\mn@doi@[]}}
\def\mn@doi@[#1]#2{\def\@tempa{#1}\ifx\@tempa\@empty \href {http://dx.doi.org/#2} {doi:#2}\else \href {http://dx.doi.org/#2} {#1}\fi \endgroup}
\def\mn@eprint#1#2{\mn@eprint@#1:#2::\@nil}
\def\mn@eprint@arXiv#1{\href {http://arxiv.org/abs/#1} {{\tt arXiv:#1}}}
\def\mn@eprint@dblp#1{\href {http://dblp.uni-trier.de/rec/bibtex/#1.xml} {dblp:#1}}
\def\mn@eprint@#1:#2:#3:#4\@nil{\def\@tempa {#1}\def\@tempb {#2}\def\@tempc {#3}\ifx \@tempc \@empty \let \@tempc \@tempb \let \@tempb \@tempa \fi \ifx \@tempb \@empty \def\@tempb {arXiv}\fi \@ifundefined {mn@eprint@\@tempb}{\@tempb:\@tempc}{\expandafter \expandafter \csname mn@eprint@\@tempb\endcsname \expandafter{\@tempc}}}

\bibitem[\protect\citeauthoryear{{Abruzzo}, {Bryan}  \& {Fielding}}{{Abruzzo} et~al.}{2022}]{Abruzzo22}
{Abruzzo} M.~W.,  {Bryan} G.~L.,   {Fielding} D.~B.,  2022, \mn@doi [\apj] {10.3847/1538-4357/ac3c48}, \href {https://ui.adsabs.harvard.edu/abs/2022ApJ...925..199A} {925, 199}

\bibitem[\protect\citeauthoryear{{Afruni}, {Fraternali}  \& {Pezzulli}}{{Afruni} et~al.}{2021}]{Afruni21}
{Afruni} A.,  {Fraternali} F.,   {Pezzulli} G.,  2021, \mn@doi [\mnras] {10.1093/mnras/staa3759}, \href {https://ui.adsabs.harvard.edu/abs/2021MNRAS.501.5575A} {501, 5575}

\bibitem[\protect\citeauthoryear{{Afruni} et~al.,}{{Afruni} et~al.}{2023}]{Afruni23}
{Afruni} A.,  et~al., 2023, \mn@doi [\aap] {10.1051/0004-6361/202347867}, \href {https://ui.adsabs.harvard.edu/abs/2023A&A...680A.112A} {680, A112}

\bibitem[\protect\citeauthoryear{Aghanim, Akrami, Ashdown  et~al.}{Aghanim et~al.}{2018}]{Plank2018}
Aghanim P. C.~N.,  Akrami Y.,  Ashdown M.,   et~al., 2018, \mn@doi [A&A, A6, 641] {https://doi.org/10.1051/0004-6361/201833910e}

\bibitem[\protect\citeauthoryear{{Anand}, {Nelson}  \& {Kauffmann}}{{Anand} et~al.}{2021}]{Anand21}
{Anand} A.,  {Nelson} D.,   {Kauffmann} G.,  2021, \mn@doi [\mnras] {10.1093/mnras/stab871}, \href {https://ui.adsabs.harvard.edu/abs/2021MNRAS.504...65A} {504, 65}

\bibitem[\protect\citeauthoryear{{Anderson}, {Churazov}  \& {Bregman}}{{Anderson} et~al.}{2016}]{Anderson16}
{Anderson} M.~E.,  {Churazov} E.,   {Bregman} J.~N.,  2016, \mn@doi [\mnras] {10.1093/mnras/stv2314}, \href {https://ui.adsabs.harvard.edu/abs/2016MNRAS.455..227A} {455, 227}

\bibitem[\protect\citeauthoryear{Angl{\'e}s-Alc{\'a}zar, Faucher-Gigu{\`e}re  \& Quataert}{Angl{\'e}s-Alc{\'a}zar et~al.}{2017}]{Angles17b}
Angl{\'e}s-Alc{\'a}zar D.,  Faucher-Gigu{\`e}re C.-A.,   Quataert E.,  2017, \mn@doi [MNRAS, 472, L109] {https://doi.org/10.1093/mnrasl/slx161}

\bibitem[\protect\citeauthoryear{{Behroozi}, {Wechsler}, {Hearin}  \& {Conroy}}{{Behroozi} et~al.}{2019}]{Behroozi19}
{Behroozi} P.,  {Wechsler} R.~H.,  {Hearin} A.~P.,   {Conroy} C.,  2019, \mn@doi [\mnras] {10.1093/mnras/stz1182}, \href {https://ui.adsabs.harvard.edu/abs/2019MNRAS.488.3143B} {488, 3143}

\bibitem[\protect\citeauthoryear{Bhattarai, Loebman, Ness, Cunningham, Wetzel  \& Benincasa}{Bhattarai et~al.}{2022}]{Bhattarai22}
Bhattarai B.,  Loebman S.,  Ness M.,  Cunningham E.,  Wetzel A.,   Benincasa S.,  2022, Bulletin of the American Astronomical Society, 54

\bibitem[\protect\citeauthoryear{{Birnboim} \& {Dekel}}{{Birnboim} \& {Dekel}}{2003}]{Birnboim03}
{Birnboim} Y.,  {Dekel} A.,  2003, \mn@doi [\mnras] {10.1046/j.1365-8711.2003.06955.x}, \href {https://ui.adsabs.harvard.edu/abs/2003MNRAS.345..349B} {345, 349}

\bibitem[\protect\citeauthoryear{{Bish}, {Werk}, {Peek}, {Zheng}  \& {Putman}}{{Bish} et~al.}{2021}]{Bish21}
{Bish} H.~V.,  {Werk} J.~K.,  {Peek} J.,  {Zheng} Y.,   {Putman} M.,  2021, \mn@doi [\apj] {10.3847/1538-4357/abeb6b}, \href {https://ui.adsabs.harvard.edu/abs/2021ApJ...912....8B} {912, 8}

\bibitem[\protect\citeauthoryear{Bordoloi et~al.,}{Bordoloi et~al.}{2011}]{Bordoloi11}
Bordoloi R.,  et~al., 2011, \mn@doi [The Astrophysical Journal, 743, 10] {https://doi.org/10.1088/0004-637X/743/1/10}

\bibitem[\protect\citeauthoryear{{Bordoloi} et~al.,}{{Bordoloi} et~al.}{2014}]{Bordoloi14}
{Bordoloi} R.,  et~al., 2014, \mn@doi [\apj] {10.1088/0004-637X/796/2/136}, \href {https://ui.adsabs.harvard.edu/abs/2014ApJ...796..136B} {796, 136}

\bibitem[\protect\citeauthoryear{Bowen \& Chelouche}{Bowen \& Chelouche}{2010}]{Bowen11}
Bowen D.~V.,  Chelouche D.,  2010, \mn@doi [The Astrophysical Journal] {10.1088/0004-637X/727/1/47}, 727, 47

\bibitem[\protect\citeauthoryear{{Bower}, {Benson}, {Malbon}, {Helly}, {Frenk}, {Baugh}, {Cole}  \& {Lacey}}{{Bower} et~al.}{2006}]{Bower06}
{Bower} R.~G.,  {Benson} A.~J.,  {Malbon} R.,  {Helly} J.~C.,  {Frenk} C.~S.,  {Baugh} C.~M.,  {Cole} S.,   {Lacey} C.~G.,  2006, \mn@doi [\mnras] {10.1111/j.1365-2966.2006.10519.x}, \href {https://ui.adsabs.harvard.edu/abs/2006MNRAS.370..645B} {370, 645}

\bibitem[\protect\citeauthoryear{{Bower}, {Schaye}, {Frenk}, {Theuns}, {Schaller}, {Crain}  \& {McAlpine}}{{Bower} et~al.}{2017}]{Bower17}
{Bower} R.~G.,  {Schaye} J.,  {Frenk} C.~S.,  {Theuns} T.,  {Schaller} M.,  {Crain} R.~A.,   {McAlpine} S.,  2017, \mn@doi [\mnras] {10.1093/mnras/stw2735}, \href {https://ui.adsabs.harvard.edu/abs/2017MNRAS.465...32B} {465, 32}

\bibitem[\protect\citeauthoryear{{Bregman}, {Anderson}, {Miller}, {Hodges-Kluck}, {Dai}, {Li}, {Li}  \& {Qu}}{{Bregman} et~al.}{2018}]{Bregman18}
{Bregman} J.~N.,  {Anderson} M.~E.,  {Miller} M.~J.,  {Hodges-Kluck} E.,  {Dai} X.,  {Li} J.-T.,  {Li} Y.,   {Qu} Z.,  2018, \mn@doi [\apj] {10.3847/1538-4357/aacafe}, \href {https://ui.adsabs.harvard.edu/abs/2018ApJ...862....3B} {862, 3}

\bibitem[\protect\citeauthoryear{{Bregman}, {Hodges-Kluck}, {Qu}, {Pratt}, {Li}  \& {Yun}}{{Bregman} et~al.}{2022}]{Bregman22}
{Bregman} J.~N.,  {Hodges-Kluck} E.,  {Qu} Z.,  {Pratt} C.,  {Li} J.-T.,   {Yun} Y.,  2022, \mn@doi [\apj] {10.3847/1538-4357/ac51de}, \href {https://ui.adsabs.harvard.edu/abs/2022ApJ...928...14B} {928, 14}

\bibitem[\protect\citeauthoryear{{Bryan} \& {Norman}}{{Bryan} \& {Norman}}{1998}]{Bryan98}
{Bryan} G.~L.,  {Norman} M.~L.,  1998, \mn@doi [\apj] {10.1086/305262}, \href {https://ui.adsabs.harvard.edu/abs/1998ApJ...495...80B} {495, 80}

\bibitem[\protect\citeauthoryear{{Buck}, {Pfrommer}, {Pakmor}, {Grand}  \& {Springel}}{{Buck} et~al.}{2020}]{Buck20}
{Buck} T.,  {Pfrommer} C.,  {Pakmor} R.,  {Grand} R. J.~J.,   {Springel} V.,  2020, \mn@doi [\mnras] {10.1093/mnras/staa1960}, \href {https://ui.adsabs.harvard.edu/abs/2020MNRAS.497.1712B} {497, 1712}

\bibitem[\protect\citeauthoryear{{Butsky} \& {Quinn}}{{Butsky} \& {Quinn}}{2018}]{Butsky18}
{Butsky} I.~S.,  {Quinn} T.~R.,  2018, \mn@doi [\apj] {10.3847/1538-4357/aaeac2}, \href {https://ui.adsabs.harvard.edu/abs/2018ApJ...868..108B} {868, 108}

\bibitem[\protect\citeauthoryear{{Butsky}, {Fielding}, {Hayward}, {Hummels}, {Quinn}  \& {Werk}}{{Butsky} et~al.}{2020}]{Butsky20}
{Butsky} I.~S.,  {Fielding} D.~B.,  {Hayward} C.~C.,  {Hummels} C.~B.,  {Quinn} T.~R.,   {Werk} J.~K.,  2020, \mn@doi [\apj] {10.3847/1538-4357/abbad2}, \href {https://ui.adsabs.harvard.edu/abs/2020ApJ...903...77B} {903, 77}

\bibitem[\protect\citeauthoryear{{Butsky} et~al.,}{{Butsky} et~al.}{2022}]{Butsky22}
{Butsky} I.~S.,  et~al., 2022, \mn@doi [\apj] {10.3847/1538-4357/ac7ebd}, \href {https://ui.adsabs.harvard.edu/abs/2022ApJ...935...69B} {935, 69}

\bibitem[\protect\citeauthoryear{{Byrne}, {Faucher-Gigu{\`e}re}, {Stern}, {Angl{\'e}s-Alc{\'a}zar}, {Wellons}, {Gurvich}  \& {Hopkins}}{{Byrne} et~al.}{2023}]{Byrne23}
{Byrne} L.,  {Faucher-Gigu{\`e}re} C.-A.,  {Stern} J.,  {Angl{\'e}s-Alc{\'a}zar} D.,  {Wellons} S.,  {Gurvich} A.~B.,   {Hopkins} P.~F.,  2023, \mn@doi [\mnras] {10.1093/mnras/stad171}, \href {https://ui.adsabs.harvard.edu/abs/2023MNRAS.tmp..182B} {}

\bibitem[\protect\citeauthoryear{{Chatzikos} et~al.,}{{Chatzikos} et~al.}{2023}]{Chatzikos23}
{Chatzikos} M.,  et~al., 2023, \mn@doi [\rmxaa] {10.22201/ia.01851101p.2023.59.02.12}, \href {https://ui.adsabs.harvard.edu/abs/2023RMxAA..59..327C} {59, 327}

\bibitem[\protect\citeauthoryear{{Chen}, {Helsby}, {Gauthier}, {Shectman}, {Thompson}  \& {Tinker}}{{Chen} et~al.}{2010}]{Chen10}
{Chen} H.-W.,  {Helsby} J.~E.,  {Gauthier} J.-R.,  {Shectman} S.~A.,  {Thompson} I.~B.,   {Tinker} J.~L.,  2010, \mn@doi [\apj] {10.1088/0004-637X/714/2/1521}, \href {https://ui.adsabs.harvard.edu/abs/2010ApJ...714.1521C} {714, 1521}

\bibitem[\protect\citeauthoryear{{Chen} et~al.,}{{Chen} et~al.}{2023}]{Chen23}
{Chen} H.-W.,  et~al., 2023, \mn@doi [\apjl] {10.3847/2041-8213/acf85b}, \href {https://ui.adsabs.harvard.edu/abs/2023ApJ...955L..25C} {955, L25}

\bibitem[\protect\citeauthoryear{{Cherrey} et~al.,}{{Cherrey} et~al.}{2025}]{Cherrey25}
{Cherrey} M.,  et~al., 2025, \mn@doi [\aap] {10.1051/0004-6361/202451165}, \href {https://ui.adsabs.harvard.edu/abs/2025A&A...694A.117C} {694, A117}

\bibitem[\protect\citeauthoryear{{Croton} et~al.,}{{Croton} et~al.}{2006}]{Croton06}
{Croton} D.~J.,  et~al., 2006, \mn@doi [\mnras] {10.1111/j.1365-2966.2005.09675.x}, \href {https://ui.adsabs.harvard.edu/abs/2006MNRAS.365...11C} {365, 11}

\bibitem[\protect\citeauthoryear{{DeFelippis}, {Bournaud}, {Bouch{\'e}}, {Tollet}, {Farcy}, {Rey}, {Rosdahl}  \& {Blaizot}}{{DeFelippis} et~al.}{2024}]{DeFelippis24}
{DeFelippis} D.,  {Bournaud} F.,  {Bouch{\'e}} N.,  {Tollet} E.,  {Farcy} M.,  {Rey} M.,  {Rosdahl} J.,   {Blaizot} J.,  2024, \mn@doi [\mnras] {10.1093/mnras/stae837}, \href {https://ui.adsabs.harvard.edu/abs/2024MNRAS.530...52D} {530, 52}

\bibitem[\protect\citeauthoryear{{Dekel} \& {Birnboim}}{{Dekel} \& {Birnboim}}{2006}]{dekel06}
{Dekel} A.,  {Birnboim} Y.,  2006, \mn@doi [\mnras] {10.1111/j.1365-2966.2006.10145.x}, \href {https://ui.adsabs.harvard.edu/abs/2006MNRAS.368....2D} {368, 2}

\bibitem[\protect\citeauthoryear{Draine}{Draine}{2011}]{Draine2011}
Draine B.~T.,  2011, Physics of the Interstellar and Intergalactic Medium.
Princeton University Press, Princeton, NJ, \url {https://press.princeton.edu/books/paperback/9780691122144/physics-of-the-interstellar-and-intergalactic-medium}

\bibitem[\protect\citeauthoryear{Dutta, Bisht, Sharma, Ghosh, Roy  \& Nath}{Dutta et~al.}{2024}]{Dutta24}
Dutta A.,  Bisht M.~S.,  Sharma P.,  Ghosh R.,  Roy M.,   Nath B.~B.,  2024, \mn@doi [Monthly Notices of the Royal Astronomical Society] {10.1093/mnras/stae977}, 531, 5117

\bibitem[\protect\citeauthoryear{Dutton \& Macci{\`o}}{Dutton \& Macci{\`o}}{2014}]{Dutton14}
Dutton A.~A.,  Macci{\`o} A.~V.,  2014, \mn@doi [Monthly Notices of the Royal Astronomical Society] {10.1093/mnras/stu742}, 441, 3359

\bibitem[\protect\citeauthoryear{El-Badry, Quataert  \& Wetzel}{El-Badry et~al.}{2018}]{El-Badry2018A}
El-Badry K.,  Quataert E.,   Wetzel A.,  2018, \mn@doi [MNRAS, 473, 1930] {https://doi.org/10.1093/mnras/stx2482}

\bibitem[\protect\citeauthoryear{{Escala} et~al.,}{{Escala} et~al.}{2018}]{Escala18}
{Escala} I.,  et~al., 2018, \mn@doi [\mnras] {10.1093/mnras/stx2858}, \href {https://ui.adsabs.harvard.edu/abs/2018MNRAS.474.2194E} {474, 2194}

\bibitem[\protect\citeauthoryear{{Faerman}, {Sternberg}  \& {McKee}}{{Faerman} et~al.}{2020}]{Faerman20}
{Faerman} Y.,  {Sternberg} A.,   {McKee} C.~F.,  2020, \mn@doi [\apj] {10.3847/1538-4357/ab7ffc}, \href {https://ui.adsabs.harvard.edu/abs/2020ApJ...893...82F} {893, 82}

\bibitem[\protect\citeauthoryear{Faucher-Gigu{\`e}re \& Oh}{Faucher-Gigu{\`e}re \& Oh}{2023}]{FaucherGiguere23}
Faucher-Gigu{\`e}re C.-A.,  Oh S.~P.,  2023, \mn@doi [Annual Review of Astronomy and Astrophysics] {10.1146/annurev-astro-052920-125203}, 61, 131

\bibitem[\protect\citeauthoryear{Faucher-Gigu{\`e}re, Lidz, Zaldarriaga  \& Hernquist}{Faucher-Gigu{\`e}re et~al.}{2009}]{FaucherGiguere2009}
Faucher-Gigu{\`e}re C.-A.,  Lidz A.,  Zaldarriaga M.,   Hernquist L.,  2009, \mn@doi [PASP, 703, 1416] {https://doi.org/10.1088/0004-637X/703/2/1416}

\bibitem[\protect\citeauthoryear{Ferland, Korista  \& Verner}{Ferland et~al.}{1998}]{Ferland1998}
Ferland G.~J.,  Korista K.~T.,   Verner D.~A.,  1998, \mn@doi [PASP, 110, 761] {https://doi.org/10.1086/316190}

\bibitem[\protect\citeauthoryear{{Ferland} et~al.,}{{Ferland} et~al.}{2017}]{Ferland17}
{Ferland} G.~J.,  et~al., 2017, \mn@doi [\rmxaa] {10.48550/arXiv.1705.10877}, \href {https://ui.adsabs.harvard.edu/abs/2017RMxAA..53..385F} {53, 385}

\bibitem[\protect\citeauthoryear{{Fielding}, {Quataert}, {McCourt}  \& {Thompson}}{{Fielding} et~al.}{2017}]{Fielding17}
{Fielding} D.,  {Quataert} E.,  {McCourt} M.,   {Thompson} T.~A.,  2017, \mn@doi [\mnras] {10.1093/mnras/stw3326}, \href {https://ui.adsabs.harvard.edu/abs/2017MNRAS.466.3810F} {466, 3810}

\bibitem[\protect\citeauthoryear{{Fielding}, {Ostriker}, {Bryan}  \& {Jermyn}}{{Fielding} et~al.}{2020}]{Fielding20}
{Fielding} D.~B.,  {Ostriker} E.~C.,  {Bryan} G.~L.,   {Jermyn} A.~S.,  2020, \mn@doi [\apjl] {10.3847/2041-8213/ab8d2c}, \href {https://ui.adsabs.harvard.edu/abs/2020ApJ...894L..24F} {894, L24}

\bibitem[\protect\citeauthoryear{Garrison-Kimmel et~al.,}{Garrison-Kimmel et~al.}{2017}]{Garrison-Kimmel2017}
Garrison-Kimmel S.,  et~al., 2017, \mn@doi [Monthly Notices of the Royal Astronomical Society] {10.1093/mnras/stx1710}, 471, 1709

\bibitem[\protect\citeauthoryear{Garrison-Kimmel et~al.,}{Garrison-Kimmel et~al.}{2019}]{Garrison-Kimmel2019}
Garrison-Kimmel S.,  et~al., 2019, \mn@doi [Monthly Notices of the Royal Astronomical Society] {10.1093/mnras/stz1317}, 487, 1380

\bibitem[\protect\citeauthoryear{{Garza}, {Werk}, {Berg}, {Faerman}, {Oppenheimer}, {Bordoloi}  \& {Ellison}}{{Garza} et~al.}{2024}]{Garza24}
{Garza} S.~L.,  {Werk} J.~K.,  {Berg} T. A.~M.,  {Faerman} Y.,  {Oppenheimer} B.~D.,  {Bordoloi} R.,   {Ellison} S.~L.,  2024, \mn@doi [arXiv e-prints] {10.48550/arXiv.2412.12302}, \href {https://ui.adsabs.harvard.edu/abs/2024arXiv241212302G} {p. arXiv:2412.12302}

\bibitem[\protect\citeauthoryear{{Gnat} \& {Sternberg}}{{Gnat} \& {Sternberg}}{2007}]{Gnat07}
{Gnat} O.,  {Sternberg} A.,  2007, \mn@doi [\apjs] {10.1086/509786}, \href {https://ui.adsabs.harvard.edu/abs/2007ApJS..168..213G} {168, 213}

\bibitem[\protect\citeauthoryear{{Gronke} \& {Oh}}{{Gronke} \& {Oh}}{2018}]{Gronke18}
{Gronke} M.,  {Oh} S.~P.,  2018, \mn@doi [\mnras] {10.1093/mnrasl/sly131}, \href {https://ui.adsabs.harvard.edu/abs/2018MNRAS.480L.111G} {480, L111}

\bibitem[\protect\citeauthoryear{{Gronke}, {Oh}, {Ji}  \& {Norman}}{{Gronke} et~al.}{2022}]{Gronke22}
{Gronke} M.,  {Oh} S.~P.,  {Ji} S.,   {Norman} C.,  2022, \mn@doi [\mnras] {10.1093/mnras/stab3351}, \href {https://ui.adsabs.harvard.edu/abs/2022MNRAS.511..859G} {511, 859}

\bibitem[\protect\citeauthoryear{Gurvich et~al.,}{Gurvich et~al.}{2023}]{Gurvich23}
Gurvich A.~B.,  et~al., 2023, \mn@doi [MNRAS, 519, 2] {https://doi.org/10.1093/mnras/stac3712}

\bibitem[\protect\citeauthoryear{Haardt \& Madau}{Haardt \& Madau}{2012}]{Madau2012}
Haardt F.,  Madau P.,  2012, \mn@doi [ApJ, 746, 125] {10.1088/0004-637X/746/2/125}

\bibitem[\protect\citeauthoryear{{Hafen} et~al.,}{{Hafen} et~al.}{2022}]{Hafen22}
{Hafen} Z.,  et~al., 2022, \mn@doi [\mnras] {10.1093/mnras/stac1603}, \href {https://ui.adsabs.harvard.edu/abs/2022MNRAS.514.5056H} {514, 5056}

\bibitem[\protect\citeauthoryear{{Holguin}, {Hayward}, {Ma}, {Angl{\'e}s-Alc{\'a}zar}  \& {Cochrane}}{{Holguin} et~al.}{2024}]{Holguin24}
{Holguin} F.,  {Hayward} C.~C.,  {Ma} X.,  {Angl{\'e}s-Alc{\'a}zar} D.,   {Cochrane} R.~K.,  2024, \mn@doi [arXiv e-prints] {10.48550/arXiv.2405.13110}, \href {https://ui.adsabs.harvard.edu/abs/2024arXiv240513110H} {p. arXiv:2405.13110}

\bibitem[\protect\citeauthoryear{{Hopkins}}{{Hopkins}}{2015}]{hopkins15}
{Hopkins} P.~F.,  2015, \mn@doi [\mnras] {10.1093/mnras/stv195}, \href {https://ui.adsabs.harvard.edu/abs/2015MNRAS.450...53H} {450, 53}

\bibitem[\protect\citeauthoryear{{Hopkins}}{{Hopkins}}{2017}]{Hopkins17}
{Hopkins} P.~F.,  2017, \mn@doi [\mnras] {10.1093/mnras/stw3306}, \href {https://ui.adsabs.harvard.edu/abs/2017MNRAS.466.3387H} {466, 3387}

\bibitem[\protect\citeauthoryear{{Hopkins}, {Narayanan}  \& {Murray}}{{Hopkins} et~al.}{2013}]{hopkins13}
{Hopkins} P.~F.,  {Narayanan} D.,   {Murray} N.,  2013, \mn@doi [\mnras] {10.1093/mnras/stt723}, \href {https://ui.adsabs.harvard.edu/abs/2013MNRAS.432.2647H} {432, 2647}

\bibitem[\protect\citeauthoryear{{Hopkins}, {Kere{\v{s}}}, {O{\~n}orbe}, {Faucher-Gigu{\`e}re}, {Quataert}, {Murray}  \& {Bullock}}{{Hopkins} et~al.}{2014}]{hopkins14}
{Hopkins} P.~F.,  {Kere{\v{s}}} D.,  {O{\~n}orbe} J.,  {Faucher-Gigu{\`e}re} C.-A.,  {Quataert} E.,  {Murray} N.,   {Bullock} J.~S.,  2014, \mn@doi [\mnras] {10.1093/mnras/stu1738}, \href {https://ui.adsabs.harvard.edu/abs/2014MNRAS.445..581H} {445, 581}

\bibitem[\protect\citeauthoryear{{Hopkins} et~al.,}{{Hopkins} et~al.}{2018}]{Hopkins18}
{Hopkins} P.~F.,  et~al., 2018, \mn@doi [\mnras] {10.1093/mnras/sty1690}, \href {https://ui.adsabs.harvard.edu/abs/2018MNRAS.480..800H} {480, 800}

\bibitem[\protect\citeauthoryear{{Hopkins} et~al.,}{{Hopkins} et~al.}{2020}]{hopkins20_whatabout}
{Hopkins} P.~F.,  et~al., 2020, \mn@doi [\mnras] {10.1093/mnras/stz3321}, \href {https://ui.adsabs.harvard.edu/abs/2020MNRAS.492.3465H} {492, 3465}

\bibitem[\protect\citeauthoryear{{Hopkins}, {Squire}, {Chan}, {Quataert}, {Ji}, {Kere{\v{s}}}  \& {Faucher-Gigu{\`e}re}}{{Hopkins} et~al.}{2021}]{Hopkins21_CRtransport}
{Hopkins} P.~F.,  {Squire} J.,  {Chan} T.~K.,  {Quataert} E.,  {Ji} S.,  {Kere{\v{s}}} D.,   {Faucher-Gigu{\`e}re} C.-A.,  2021, \mn@doi [\mnras] {10.1093/mnras/staa3691}, \href {https://ui.adsabs.harvard.edu/abs/2021MNRAS.501.4184H} {501, 4184}

\bibitem[\protect\citeauthoryear{{Hopkins} et~al.,}{{Hopkins} et~al.}{2023}]{hopkins23}
{Hopkins} P.~F.,  et~al., 2023, \mn@doi [\mnras] {10.1093/mnras/stac3489}, \href {https://ui.adsabs.harvard.edu/abs/2023MNRAS.519.3154H} {519, 3154}

\bibitem[\protect\citeauthoryear{Huang, Chen, Johnson  \& Weiner}{Huang et~al.}{2015}]{Huang15}
Huang Y.-H.,  Chen H.-W.,  Johnson S.~D.,   Weiner B.~J.,  2015, \mn@doi [Monthly Notices of the Royal Astronomical Society] {10.1093/mnras/stv2327}, 455, 1713

\bibitem[\protect\citeauthoryear{Huang, Chen, Shectman, Johnson, Zahedy, Helsby, Gauthier  \& Thompson}{Huang et~al.}{2021}]{Huang21}
Huang Y.-H.,  Chen H.-W.,  Shectman S.~A.,  Johnson S.~D.,  Zahedy F.~S.,  Helsby J.~E.,  Gauthier J.-R.,   Thompson I.~B.,  2021, \mn@doi [Monthly Notices of the Royal Astronomical Society, 502, 4743-4761] {https://doi.org/10.1093/mnras/stab360}

\bibitem[\protect\citeauthoryear{Hummels, Smith  \& Silvia}{Hummels et~al.}{2017}]{Hummels2017}
Hummels C.~B.,  Smith B.~D.,   Silvia D.~W.,  2017, \mn@doi [ApJ, 847, 59] {10.3847/1538-4357/aa7e2d}

\bibitem[\protect\citeauthoryear{{Hummels} et~al.,}{{Hummels} et~al.}{2019}]{Hummels19}
{Hummels} C.~B.,  et~al., 2019, \mn@doi [\apj] {10.3847/1538-4357/ab378f10.48550/arXiv.1811.12410}, \href {https://ui.adsabs.harvard.edu/abs/2019ApJ...882..156H} {882, 156}

\bibitem[\protect\citeauthoryear{{Ji} et~al.,}{{Ji} et~al.}{2020}]{Ji20}
{Ji} S.,  et~al., 2020, \mn@doi [\mnras] {10.1093/mnras/staa1849}, \href {https://ui.adsabs.harvard.edu/abs/2020MNRAS.496.4221J} {496, 4221}

\bibitem[\protect\citeauthoryear{{Ji}, {Kere{\v{s}}}, {Chan}, {Stern}, {Hummels}, {Hopkins}, {Quataert}  \& {Faucher-Gigu{\`e}re}}{{Ji} et~al.}{2021}]{Ji21}
{Ji} S.,  {Kere{\v{s}}} D.,  {Chan} T.~K.,  {Stern} J.,  {Hummels} C.~B.,  {Hopkins} P.~F.,  {Quataert} E.,   {Faucher-Gigu{\`e}re} C.-A.,  2021, \mn@doi [\mnras] {10.1093/mnras/stab1264}, \href {https://ui.adsabs.harvard.edu/abs/2021MNRAS.505..259J} {505, 259}

\bibitem[\protect\citeauthoryear{{Johnson}, {Chen}, {Mulchaey}, {Schaye}  \& {Straka}}{{Johnson} et~al.}{2017}]{Johnson17}
{Johnson} S.~D.,  {Chen} H.-W.,  {Mulchaey} J.~S.,  {Schaye} J.,   {Straka} L.~A.,  2017, \mn@doi [\apjl] {10.3847/2041-8213/aa9370}, \href {https://ui.adsabs.harvard.edu/abs/2017ApJ...850L..10J} {850, L10}

\bibitem[\protect\citeauthoryear{{Kassin} et~al.,}{{Kassin} et~al.}{2012}]{kassin12}
{Kassin} S.~A.,  et~al., 2012, \mn@doi [\apj] {10.1088/0004-637X/758/2/106}, \href {https://ui.adsabs.harvard.edu/abs/2012ApJ...758..106K} {758, 106}

\bibitem[\protect\citeauthoryear{{Kere{\v{s}}}, {Katz}, {Weinberg}  \& {Dav{\'e}}}{{Kere{\v{s}}} et~al.}{2005}]{keres05}
{Kere{\v{s}}} D.,  {Katz} N.,  {Weinberg} D.~H.,   {Dav{\'e}} R.,  2005, \mn@doi [\mnras] {10.1111/j.1365-2966.2005.09451.x}, \href {https://ui.adsabs.harvard.edu/abs/2005MNRAS.363....2K} {363, 2}

\bibitem[\protect\citeauthoryear{Knollmann \& Knebe}{Knollmann \& Knebe}{2009}]{Knollmann2009}
Knollmann S.~R.,  Knebe A.,  2009, \mn@doi [ApJS, 182, 608] {https://doi.org/10.1088/0067-0049/182/2/608}

\bibitem[\protect\citeauthoryear{Kroupa}{Kroupa}{2001}]{Kroupa2001}
Kroupa P.,  2001, \mn@doi [MNRAS, 322, 231] {https://doi.org/10.1046/j.1365-8711.2001.04022.x}

\bibitem[\protect\citeauthoryear{Krumholz}{Krumholz}{2014}]{Krumholz2014}
Krumholz M.~R.,  2014, \mn@doi [Physics Reports, 539, 49] {10.1016/j.physrep.2014.02.001}

\bibitem[\protect\citeauthoryear{{Lan}}{{Lan}}{2020}]{Lan20}
{Lan} T.-W.,  2020, \mn@doi [\apj] {10.3847/1538-4357/ab989a}, \href {https://ui.adsabs.harvard.edu/abs/2020ApJ...897...97L} {897, 97}

\bibitem[\protect\citeauthoryear{{Lan} \& {Fukugita}}{{Lan} \& {Fukugita}}{2017}]{Lan17}
{Lan} T.-W.,  {Fukugita} M.,  2017, \mn@doi [\apj] {10.3847/1538-4357/aa93eb}, \href {https://ui.adsabs.harvard.edu/abs/2017ApJ...850..156L} {850, 156}

\bibitem[\protect\citeauthoryear{Lan \& Mo}{Lan \& Mo}{2018}]{Lan18}
Lan T.-W.,  Mo H.,  2018, \mn@doi [The Astrophysical Journal, 866, 36] {https://doi.org/10.3847/1538-4357/aadc08}

\bibitem[\protect\citeauthoryear{{Lan}, {M{\'e}nard}  \& {Zhu}}{{Lan} et~al.}{2014}]{Lan14}
{Lan} T.-W.,  {M{\'e}nard} B.,   {Zhu} G.,  2014, \mn@doi [\apj] {10.1088/0004-637X/795/1/31}, \href {https://ui.adsabs.harvard.edu/abs/2014ApJ...795...31L} {795, 31}

\bibitem[\protect\citeauthoryear{Lehner et~al.,}{Lehner et~al.}{2020}]{Lehner20}
Lehner N.,  et~al., 2020, \mn@doi [The Astrophysical Journal, 900, 9] {https://doi.org/10.3847/1538-4357/aba49c}

\bibitem[\protect\citeauthoryear{{Lehner} et~al.,}{{Lehner} et~al.}{2025}]{Lehner25}
{Lehner} N.,  et~al., 2025, \mn@doi [arXiv e-prints] {10.48550/arXiv.2506.16573}, \href {https://ui.adsabs.harvard.edu/abs/2025arXiv250616573L} {p. arXiv:2506.16573}

\bibitem[\protect\citeauthoryear{Leitherer, Schaerer  \& Goldader}{Leitherer et~al.}{1999}]{Leitherer1999}
Leitherer C.,  Schaerer D.,   Goldader J.~D.,  1999, \mn@doi [ApJS, 123, 3] {https://doi.org/10.1086/313233}

\bibitem[\protect\citeauthoryear{{Li} et~al.,}{{Li} et~al.}{2021}]{Li21}
{Li} F.,  et~al., 2021, \mn@doi [\mnras] {10.1093/mnras/staa3322}, \href {https://ui.adsabs.harvard.edu/abs/2021MNRAS.500.1038L} {500, 1038}

\bibitem[\protect\citeauthoryear{{Liang} \& {Chen}}{{Liang} \& {Chen}}{2014}]{Liang14}
{Liang} C.~J.,  {Chen} H.-W.,  2014, \mn@doi [\mnras] {10.1093/mnras/stu1901}, \href {https://ui.adsabs.harvard.edu/abs/2014MNRAS.445.2061L} {445, 2061}

\bibitem[\protect\citeauthoryear{{Lochhaas}, {Bryan}, {Li}, {Li}  \& {Fielding}}{{Lochhaas} et~al.}{2020}]{Lochhaas20}
{Lochhaas} C.,  {Bryan} G.~L.,  {Li} Y.,  {Li} M.,   {Fielding} D.,  2020, \mn@doi [\mnras] {10.1093/mnras/staa358}, \href {https://ui.adsabs.harvard.edu/abs/2020MNRAS.493.1461L} {493, 1461}

\bibitem[\protect\citeauthoryear{Lopez et~al.,}{Lopez et~al.}{2020}]{Lopez20}
Lopez S.,  et~al., 2020, \mn@doi [Monthly Notices of the Royal Astronomical Society, 491, 4442-4461] {https://doi.org/10.1093/mnras/stz3183}

\bibitem[\protect\citeauthoryear{{Manuwal}, {Narayanan}, {Udhwani}, {Srianand}, {Savage}, {Charlton}  \& {Misawa}}{{Manuwal} et~al.}{2021}]{Manuwal21}
{Manuwal} A.,  {Narayanan} A.,  {Udhwani} P.,  {Srianand} R.,  {Savage} B.~D.,  {Charlton} J.~C.,   {Misawa} T.,  2021, \mn@doi [\mnras] {10.1093/mnras/stab1556}, \href {https://ui.adsabs.harvard.edu/abs/2021MNRAS.505.3635M} {505, 3635}

\bibitem[\protect\citeauthoryear{{McCourt}, {Oh}, {O'Leary}  \& {Madigan}}{{McCourt} et~al.}{2018}]{McCourt18}
{McCourt} M.,  {Oh} S.~P.,  {O'Leary} R.,   {Madigan} A.-M.,  2018, \mn@doi [\mnras] {10.1093/mnras/stx2687}, \href {https://ui.adsabs.harvard.edu/abs/2018MNRAS.473.5407M} {473, 5407}

\bibitem[\protect\citeauthoryear{{Mo}, {Chen}  \& {Wang}}{{Mo} et~al.}{2024}]{Mo24}
{Mo} H.,  {Chen} Y.,   {Wang} H.,  2024, \mn@doi [\mnras] {10.1093/mnras/stae1727}, \href {https://ui.adsabs.harvard.edu/abs/2024MNRAS.532.3808M} {532, 3808}

\bibitem[\protect\citeauthoryear{Mortensen, C., Jones, Faucher-Gigu{\`e}re, Sanders, Ellis, Leethochawalit  \& Stark}{Mortensen et~al.}{2021}]{Mortensen21}
Mortensen K.,  C. K. V.~G.,  Jones T.,  Faucher-Gigu{\`e}re C.-A.,  Sanders R.~L.,  Ellis R.~S.,  Leethochawalit N.,   Stark D.~P.,  2021, \mn@doi [The Astrophysical Journal, 914, 92] {https://doi.org/10.3847/1538-4357/abfa11}

\bibitem[\protect\citeauthoryear{{Nielsen}, {Churchill}  \& {Kacprzak}}{{Nielsen} et~al.}{2013}]{Nielsen13}
{Nielsen} N.~M.,  {Churchill} C.~W.,   {Kacprzak} G.~G.,  2013, \mn@doi [\apj] {10.1088/0004-637X/776/2/115}, \href {https://ui.adsabs.harvard.edu/abs/2013ApJ...776..115N} {776, 115}

\bibitem[\protect\citeauthoryear{Oren, Sternberg, McKee, Faerman  \& Genel}{Oren et~al.}{2024}]{Oren2024}
Oren Y.,  Sternberg A.,  McKee C.~F.,  Faerman Y.,   Genel S.,  2024, \mn@doi [The Astrophysical Journal] {https://doi.org/10.48550/arXiv.2403.09476}

\bibitem[\protect\citeauthoryear{{Pandya} et~al.,}{{Pandya} et~al.}{2023}]{Pandya23}
{Pandya} V.,  et~al., 2023, \mn@doi [\apj] {10.3847/1538-4357/acf3ea}, \href {https://ui.adsabs.harvard.edu/abs/2023ApJ...956..118P} {956, 118}

\bibitem[\protect\citeauthoryear{{Peeples} et~al.,}{{Peeples} et~al.}{2019}]{Peeples19}
{Peeples} M.~S.,  et~al., 2019, \mn@doi [\apj] {10.3847/1538-4357/ab065410.48550/arXiv.1810.06566}, \href {https://ui.adsabs.harvard.edu/abs/2019ApJ...873..129P} {873, 129}

\bibitem[\protect\citeauthoryear{{Pezzulli}, {Fraternali}  \& {Binney}}{{Pezzulli} et~al.}{2017}]{Pezzulli17}
{Pezzulli} G.,  {Fraternali} F.,   {Binney} J.,  2017, \mn@doi [\mnras] {10.1093/mnras/stx029}, \href {https://ui.adsabs.harvard.edu/abs/2017MNRAS.467..311P} {467, 311}

\bibitem[\protect\citeauthoryear{{Prochaska}}{{Prochaska}}{1999}]{Prochaska99}
{Prochaska} J.~X.,  1999, \mn@doi [\apjl] {10.1086/311849}, \href {https://ui.adsabs.harvard.edu/abs/1999ApJ...511L..71P} {511, L71}

\bibitem[\protect\citeauthoryear{{Ramesh} \& {Nelson}}{{Ramesh} \& {Nelson}}{2024}]{Ramesh24}
{Ramesh} R.,  {Nelson} D.,  2024, \mn@doi [\mnras] {10.1093/mnras/stae237}, \href {https://ui.adsabs.harvard.edu/abs/2024MNRAS.528.3320R} {528, 3320}

\bibitem[\protect\citeauthoryear{{Rees} \& {Ostriker}}{{Rees} \& {Ostriker}}{1977}]{rees77}
{Rees} M.~J.,  {Ostriker} J.~P.,  1977, \mn@doi [\mnras] {10.1093/mnras/179.4.541}, \href {https://ui.adsabs.harvard.edu/abs/1977MNRAS.179..541R} {179, 541}

\bibitem[\protect\citeauthoryear{{Salem}, {Bryan}  \& {Corlies}}{{Salem} et~al.}{2016}]{Salem16}
{Salem} M.,  {Bryan} G.~L.,   {Corlies} L.,  2016, \mn@doi [\mnras] {10.1093/mnras/stv2641}, \href {https://ui.adsabs.harvard.edu/abs/2016MNRAS.456..582S} {456, 582}

\bibitem[\protect\citeauthoryear{{Sales}, {Navarro}, {Theuns}, {Schaye}, {White}, {Frenk}, {Crain}  \& {Dalla Vecchia}}{{Sales} et~al.}{2012}]{Sales12}
{Sales} L.~V.,  {Navarro} J.~F.,  {Theuns} T.,  {Schaye} J.,  {White} S. D.~M.,  {Frenk} C.~S.,  {Crain} R.~A.,   {Dalla Vecchia} C.,  2012, \mn@doi [\mnras] {10.1111/j.1365-2966.2012.20975.x}, \href {https://ui.adsabs.harvard.edu/abs/2012MNRAS.423.1544S} {423, 1544}

\bibitem[\protect\citeauthoryear{Samuel et~al.,}{Samuel et~al.}{2020}]{Samuel2020}
Samuel J.,  et~al., 2020, \mn@doi [Monthly Notices of the Royal Astronomical Society] {10.1093/mnras/stz3054}, 491, 1471

\bibitem[\protect\citeauthoryear{{Schroetter} et~al.,}{{Schroetter} et~al.}{2021}]{Schroetter21}
{Schroetter} I.,  et~al., 2021, \mn@doi [\mnras] {10.1093/mnras/stab1447}, \href {https://ui.adsabs.harvard.edu/abs/2021MNRAS.506.1355S} {506, 1355}

\bibitem[\protect\citeauthoryear{{Silk}}{{Silk}}{1977}]{silk77}
{Silk} J.,  1977, \mn@doi [\apj] {10.1086/154972}, \href {https://ui.adsabs.harvard.edu/abs/1977ApJ...211..638S} {211, 638}

\bibitem[\protect\citeauthoryear{{Singh}, {Lau}, {Faerman}, {Stern}  \& {Nagai}}{{Singh} et~al.}{2024}]{Singh24}
{Singh} P.,  {Lau} E.~T.,  {Faerman} Y.,  {Stern} J.,   {Nagai} D.,  2024, \mn@doi [\mnras] {10.1093/mnras/stae1695}, \href {https://ui.adsabs.harvard.edu/abs/2024MNRAS.532.3222S} {532, 3222}

\bibitem[\protect\citeauthoryear{{Somerville}, {Hopkins}, {Cox}, {Robertson}  \& {Hernquist}}{{Somerville} et~al.}{2008}]{Somerville08}
{Somerville} R.~S.,  {Hopkins} P.~F.,  {Cox} T.~J.,  {Robertson} B.~E.,   {Hernquist} L.,  2008, \mn@doi [\mnras] {10.1111/j.1365-2966.2008.13805.x}, \href {https://ui.adsabs.harvard.edu/abs/2008MNRAS.391..481S} {391, 481}

\bibitem[\protect\citeauthoryear{{Sormani}, {Sobacchi}, {Pezzulli}, {Binney}  \& {Klessen}}{{Sormani} et~al.}{2018}]{Sormani18}
{Sormani} M.~C.,  {Sobacchi} E.,  {Pezzulli} G.,  {Binney} J.,   {Klessen} R.~S.,  2018, \mn@doi [\mnras] {10.1093/mnras/sty2500}, \href {https://ui.adsabs.harvard.edu/abs/2018MNRAS.481.3370S} {481, 3370}

\bibitem[\protect\citeauthoryear{{Spitzer}}{{Spitzer}}{1956}]{Spitzer56}
{Spitzer} Lyman J.,  1956, \mn@doi [\apj] {10.1086/146200}, \href {https://ui.adsabs.harvard.edu/abs/1956ApJ...124...20S} {124, 20}

\bibitem[\protect\citeauthoryear{Springel}{Springel}{2005}]{Springel2005}
Springel V.,  2005, \mn@doi [MNRAS, 364, 1105] {https://doi.org/10.1111/j.1365-2966.2005.09655.x}

\bibitem[\protect\citeauthoryear{{Stern}, {Hennawi}, {Prochaska}  \& {Werk}}{{Stern} et~al.}{2016}]{Stern16_CGM}
{Stern} J.,  {Hennawi} J.~F.,  {Prochaska} J.~X.,   {Werk} J.~K.,  2016, \mn@doi [\apj] {10.3847/0004-637X/830/2/87}, \href {https://ui.adsabs.harvard.edu/abs/2016ApJ...830...87S} {830, 87}

\bibitem[\protect\citeauthoryear{{Stern}, {Fielding}, {Faucher-Gigu{\`e}re}  \& {Quataert}}{{Stern} et~al.}{2019}]{Stern19}
{Stern} J.,  {Fielding} D.,  {Faucher-Gigu{\`e}re} C.-A.,   {Quataert} E.,  2019, \mn@doi [\mnras] {10.1093/mnras/stz1859}, \href {https://ui.adsabs.harvard.edu/abs/2019MNRAS.488.2549S} {488, 2549}

\bibitem[\protect\citeauthoryear{{Stern}, {Fielding}, {Faucher-Gigu{\`e}re}  \& {Quataert}}{{Stern} et~al.}{2020}]{Stern20}
{Stern} J.,  {Fielding} D.,  {Faucher-Gigu{\`e}re} C.-A.,   {Quataert} E.,  2020, \mn@doi [\mnras] {10.1093/mnras/staa198}, \href {https://ui.adsabs.harvard.edu/abs/2020MNRAS.492.6042S} {492, 6042}

\bibitem[\protect\citeauthoryear{Stern et~al.,}{Stern et~al.}{2021b}]{Stern21B}
Stern J.,  et~al., 2021b, \mn@doi [MNRAS, 507, 2869] {https://doi.org/10.1093/mnras/stab2240}

\bibitem[\protect\citeauthoryear{Stern et~al.,}{Stern et~al.}{2021a}]{Stern21A}
Stern J.,  et~al., 2021a, \mn@doi [ApJ, 911, 88] {10.3847/1538-4357/abd776}

\bibitem[\protect\citeauthoryear{{Stern}, {Fielding}, {Hafen}, {Su}, {Naor}, {Faucher-Gigu{\`e}re}, {Quataert}  \& {Bullock}}{{Stern} et~al.}{2024}]{Stern24}
{Stern} J.,  {Fielding} D.,  {Hafen} Z.,  {Su} K.-Y.,  {Naor} N.,  {Faucher-Gigu{\`e}re} C.-A.,  {Quataert} E.,   {Bullock} J.,  2024, \mn@doi [\mnras] {10.1093/mnras/stae824}, \href {https://ui.adsabs.harvard.edu/abs/2024MNRAS.530.1711S} {530, 1711}

\bibitem[\protect\citeauthoryear{{Sultan}, {Faucher-Gigu{\`e}re}, {Stern}, {Rotshtein}, {Byrne}  \& {Wijers}}{{Sultan} et~al.}{2024}]{Sultan24}
{Sultan} I.,  {Faucher-Gigu{\`e}re} C.-A.,  {Stern} J.,  {Rotshtein} S.,  {Byrne} L.,   {Wijers} N.,  2024, \mn@doi [arXiv e-prints] {10.48550/arXiv.2410.16359}, \href {https://ui.adsabs.harvard.edu/abs/2024arXiv241016359S} {p. arXiv:2410.16359}

\bibitem[\protect\citeauthoryear{{The yt Project}}{{The yt Project}}{2025}]{yt_project}
{The yt Project} 2025, yt: Pythonic data analysis and visualization for astrophysical simulation data, \url {https://yt-project.org/doc/index.html}

\bibitem[\protect\citeauthoryear{{Theuns}}{{Theuns}}{2021}]{Theuns21}
{Theuns} T.,  2021, \mn@doi [\mnras] {10.1093/mnras/staa3412}, \href {https://ui.adsabs.harvard.edu/abs/2021MNRAS.500.2741T} {500, 2741}

\bibitem[\protect\citeauthoryear{{Tiley} et~al.,}{{Tiley} et~al.}{2021}]{Tiley21}
{Tiley} A.~L.,  et~al., 2021, \mn@doi [\mnras] {10.1093/mnras/stab1692}, \href {https://ui.adsabs.harvard.edu/abs/2021MNRAS.506..323T} {506, 323}

\bibitem[\protect\citeauthoryear{{Trident}}{{Trident}}{2025}]{trident_docs}
{Trident} 2025, Trident Documentation, \url {https://trident.readthedocs.io/en/latest/}

\bibitem[\protect\citeauthoryear{{Tumlinson}, {Peeples}  \& {Werk}}{{Tumlinson} et~al.}{2017}]{Tumlinson17}
{Tumlinson} J.,  {Peeples} M.~S.,   {Werk} J.~K.,  2017, \mn@doi [\araa] {10.1146/annurev-astro-091916-055240}, \href {https://ui.adsabs.harvard.edu/abs/2017ARA&A..55..389T} {55, 389}

\bibitem[\protect\citeauthoryear{Turk, Smith, Oishi, Skory, Skillman, Abel  \& Norman}{Turk et~al.}{2011}]{Turk2011}
Turk M.~J.,  Smith B.~D.,  Oishi J.~S.,  Skory S.,  Skillman S.,  Abel T.,   Norman M.~L.,  2011, \mn@doi [The Astrophysical Journal Supplement Series] {10.1088/0067-0049/192/1/9}, 192, 9

\bibitem[\protect\citeauthoryear{Werk, Prochaska, Thom, Tumlinson, Tripp, O'Meara  \& Peeples}{Werk et~al.}{2013}]{Werk13}
Werk J.~K.,  Prochaska J.~X.,  Thom C.,  Tumlinson J.,  Tripp T.~M.,  O'Meara J.~M.,   Peeples M.~S.,  2013, \mn@doi [The Astrophysical Journal Supplement, 204, 17] {https://doi.org/10.1088/0067-0049/204/2/17}

\bibitem[\protect\citeauthoryear{Wetzel, Hopkins  \& Kim}{Wetzel et~al.}{2016}]{Wetzel2016}
Wetzel A.~R.,  Hopkins P.~F.,   Kim J.-h.,  2016, \mn@doi [ApJL, 827, L23] {https://doi.org/10.3847/2041-8205/827/2/L23}

\bibitem[\protect\citeauthoryear{{White} \& {Frenk}}{{White} \& {Frenk}}{1991}]{white91}
{White} S. D.~M.,  {Frenk} C.~S.,  1991, \mn@doi [\apj] {10.1086/170483}, \href {https://ui.adsabs.harvard.edu/abs/1991ApJ...379...52W} {379, 52}

\bibitem[\protect\citeauthoryear{{White} \& {Rees}}{{White} \& {Rees}}{1978}]{white78}
{White} S.~D.~M.,  {Rees} M.~J.,  1978, \mn@doi [\mnras] {10.1093/mnras/183.3.341}, \href {https://ui.adsabs.harvard.edu/abs/1978MNRAS.183..341W} {183, 341}

\bibitem[\protect\citeauthoryear{{Wiersma}, {Schaye}  \& {Smith}}{{Wiersma} et~al.}{2009}]{Wiersma09}
{Wiersma} R. P.~C.,  {Schaye} J.,   {Smith} B.~D.,  2009, \mn@doi [\mnras] {10.1111/j.1365-2966.2008.14191.x}, \href {https://ui.adsabs.harvard.edu/abs/2009MNRAS.393...99W} {393, 99}

\bibitem[\protect\citeauthoryear{Wu et~al.,}{Wu et~al.}{2024}]{Wu24}
Wu X.,  et~al., 2024, \mn@doi [eprint arXiv:2407.17809] {https://doi.org/10.48550/arXiv.2407.17809}

\bibitem[\protect\citeauthoryear{{Yu} et~al.,}{{Yu} et~al.}{2021}]{Yu21}
{Yu} S.,  et~al., 2021, arXiv e-prints, \href {https://ui.adsabs.harvard.edu/abs/2021arXiv210303888Y} {p. arXiv:2103.03888}

\bibitem[\protect\citeauthoryear{{Yu} et~al.,}{{Yu} et~al.}{2023}]{Yu23}
{Yu} S.,  et~al., 2023, \mn@doi [\mnras] {10.1093/mnras/stad1806}, \href {https://ui.adsabs.harvard.edu/abs/2023MNRAS.523.6220Y} {523, 6220}

\bibitem[\protect\citeauthoryear{{van de Voort}, {Quataert}, {Hopkins}, {Faucher-Gigu{\`e}re}, {Feldmann}, {Kere{\v{s}}}, {Chan}  \& {Hafen}}{{van de Voort} et~al.}{2016}]{vandeVoort16}
{van de Voort} F.,  {Quataert} E.,  {Hopkins} P.~F.,  {Faucher-Gigu{\`e}re} C.-A.,  {Feldmann} R.,  {Kere{\v{s}}} D.,  {Chan} T.~K.,   {Hafen} Z.,  2016, \mn@doi [\mnras] {10.1093/mnras/stw2322}, \href {https://ui.adsabs.harvard.edu/abs/2016MNRAS.463.4533V} {463, 4533}

\bibitem[\protect\citeauthoryear{{van de Voort}, {Springel}, {Mandelker}, {van den Bosch}  \& {Pakmor}}{{van de Voort} et~al.}{2019}]{vandeVoort19}
{van de Voort} F.,  {Springel} V.,  {Mandelker} N.,  {van den Bosch} F.~C.,   {Pakmor} R.,  2019, \mn@doi [\mnras] {10.1093/mnrasl/sly19010.48550/arXiv.1808.04369}, \href {https://ui.adsabs.harvard.edu/abs/2019MNRAS.482L..85V} {482, L85}

\makeatother
\end{thebibliography}
\bsp

\appendix

\section{\texorpdfstring{\boldmath Gas Density Distributions at \(0.5R_{\rm vir}\)}{Gas Density Distributions at 0.5 Rvir}}\label{sec:halfRvir}

\begin{figure}
    \includegraphics[width=\columnwidth]{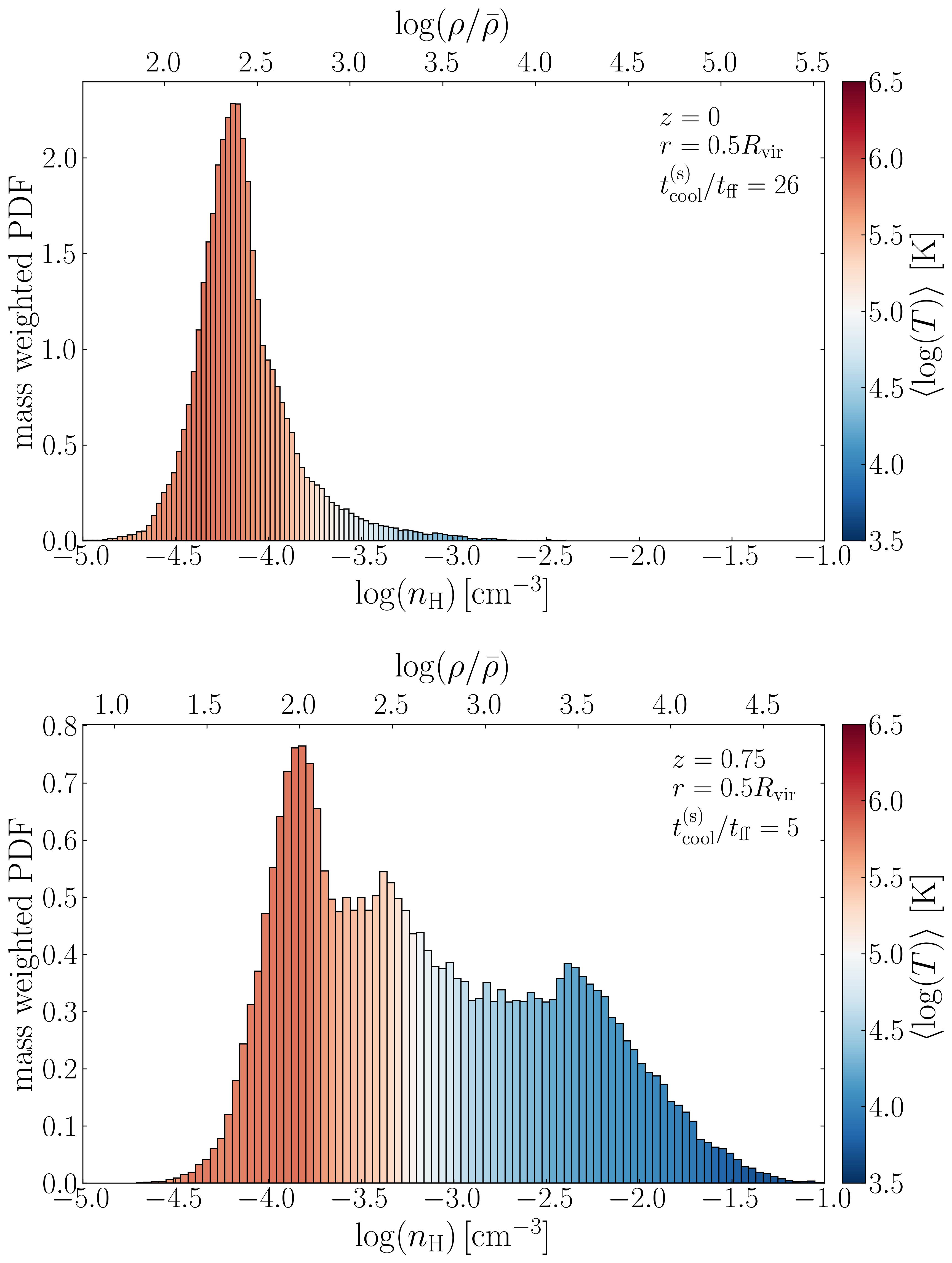}
    \caption{Similar to Fig.~\ref{fig: AllGasMassWeightedlog(rho)Distribution_m12i_z_0_z_0.75}, for $r=0.5\,\Rvir$. Panels show mass-weighted gas density distributions in a thin shell in the m12i simulation at $z=0$ (\textit{top}) and at $z=0.75$ (\textit{bottom}). Color indicates average temperature at each density bin.  In both snapshots hot gas dominates the mass consistent with $\tcool>\tff$ (noted), though with a significantly more prominent cool gas phase at $z=0.75$.}
    \label{fig: AllGasMassWeightedlog_rho_Distribution_m12i_z_0_z_0.75_For_0.5Rvir}
\end{figure}

Figure~\ref{fig: AllGasMassWeightedlog_rho_Distribution_m12i_z_0_z_0.75_For_0.5Rvir} plots the mass-weighted gas density distributions at $r = 0.5R_{\rm vir}$ for the two snapshots shown in Fig.~\ref{fig: AllGasMassWeightedlog(rho)Distribution_m12i_z_0_z_0.75}. As above, we include all gas resolution elements in a thin spherical shell of width $\delta r = 0.01\,R_{\rm vir}$ without any selection on temperature. Color denotes the mass-weighted average $\log T$ of gas at each density bin, and the top axis notes the overdensity $\rho/\bar{\rho}$.
In the post-ICV $z=0$ snapshot (top panel), the hot low density gas dominates with a relatively narrow  dispersion of $\sigma_{\log \rho} \approx 0.2$ dex, as expected due to the high $\tcool=26\,\tff$ at this radius and redshift. 
In the pre-ICV snapshot at $z=0.75$ (bottom panel), the hot gas is still dominant since the outer CGM has already virialized \citep{Stern21A}. There is  though a significantly more prominent cool gas phase than seen at $z=0$, perhaps due to the lower $\tcool/\tff=5$, or due to the effect of bursty feedback on the outer CGM prior to ICV. 

\section{\texorpdfstring{\boldmath Resolution Dependence}{Resolution Dependence}}

\begin{figure*}
    \includegraphics[width=\textwidth]{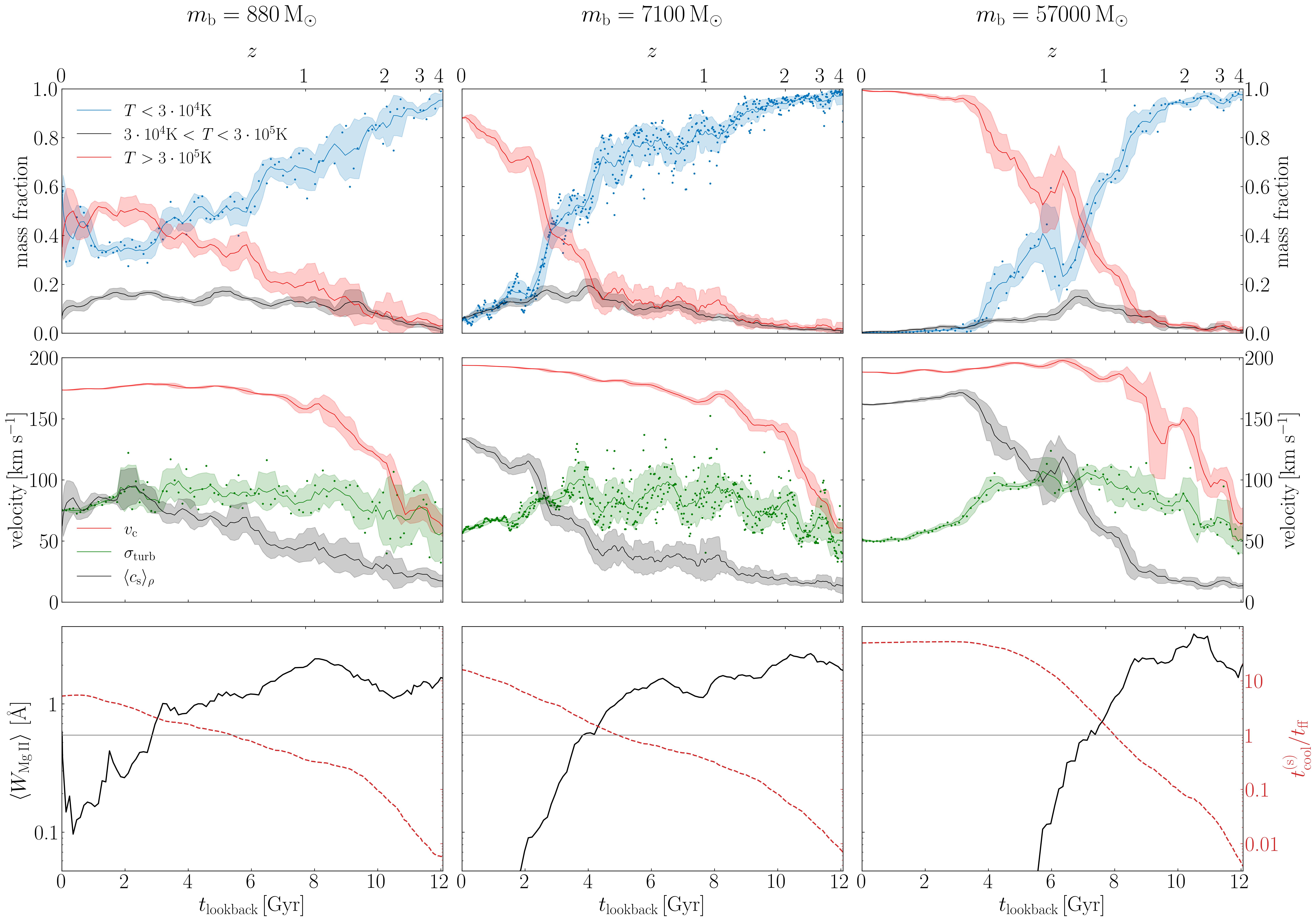}
    \caption{The evolution of gas properties at $r=0.2\,R_{\rm vir}$ versus simulation resolution (noted on top)
    in the FIRE-m12i halo. 
    \textit{Top row:} Gas mass fractions in three temperature bins. Lines and shaded regions indicate running means and dispersions, while dots indicate individual snapshots (shown only for the cool phase). 
    \textit{Middle row:} Circular velocity (red), turbulent velocity (green), and mass-weighted sound speed (black). 
    \textit{Bottom row:} Mean equivalent width of \ion{Mg}{II} (solid black, left axis) and the ratio $t^{\rm (s)}_{\rm cool}/t_{\rm ff}$ (dashed red, right axis). Grey horizontal line marks $\tcool=\tff$.
    The panels show that $v_{\rm c}$ and thus $\tcool/\tff\propto v_{\rm c}^4$ increase somewhat more slowly at higher resolution, and hence ICV occurs somewhat later. 
    At all resolutions we find that prior to ICV when $\tcool<\tff$, cool gas dominates the mass, turbulence is supersonic, and $\langle W_{\ion{Mg}{ii}}\rangle\gtrsim1$\AA. Our main conclusions are thus independent of resolution at the range probed.}
    \label{fig: m12i_r880_m12i_r7100_m12i_r57000_MassFraction_MuV_With_W_MgII_And_t_cool_t_ff}
\end{figure*}

Figure~\ref{fig: m12i_r880_m12i_r7100_m12i_r57000_MassFraction_MuV_With_W_MgII_And_t_cool_t_ff} explores how our results depend on simulation resolution as discussed in section~\ref{s:resolution}, utilizing the three FIRE-m12i simulations with baryon mass resolution of $m_{\rm b}=880\msun$ (left panels), $7100\msun$ (fiducial value, middle panels) and $57\,000\msun$ (right panels). Top two rows are similar to the two top panels of Fig.~\ref{fig: MassFractionandSigmaRhoandMuVAgainstLookbackTime_m12i}, while the bottom row shows the mean equivalent width of \ion{Mg}{II} and the ratio $t^{\rm (s)}_{\rm cool}/t_{\rm ff}$ derived using eqns.~(\ref{eq:t_cool}) -- (\ref{eq:t_ff}).

\section{\texorpdfstring{\boldmath Observed $\EWmgii$}{Observed EW (Mg II)}}\label{a:obs}

In this appendix we provide details on observed $\EWmgii$ gathered from the literature and plotted in Figures~\ref{fig: DistributionofTotalWAgainstLookbackTimeForMgIIComparisonWithObservation_L_Stars} and \ref{fig: DistributionofTotalWAgainstLookbackTimeForMgIIComparisonWithObservation_Dwarfs+LRGs}. The relevant values are listed in Table~\ref{table:ObservationsTable}. Calculations of $\Rvir$ are based on \cite{Behroozi19} to estimate $M_{\rm halo}$ from $M_\star$ and $z$, and assume an NFW profile with concentration parameter from \cite{Dutton14} to calculate $R_{\rm vir}$ from $M_{\rm halo}$.
\begin{itemize}[leftmargin=*]
    \item zCOSMOS \citep{Bordoloi11}: $\EWmgii$ is measured on co-added background \textit{galaxy} spectra obtained with the LR blue grism on VLT (resolution $R\sim200$). Co-added spectra are derived for different impact parameters from the foreground galaxy, and for different foreground galaxy masses and types. For both mass bins of the SF galaxies and the low mass bin of the red galaxies the median $0.2\,\Rvir$ is in the range $26-38\,$kpc, so we use their $\EWmgii$ measurement in the smallest impact parameter bin of $R_\perp=0-50\,{\rm kpc}$. The high mass red galaxy bin has a median $0.2\,\Rvir=81\,$kpc, so we use the $R_\perp=65-80\,{\rm kpc}$ bin.
    \item COS-Halos \citep{Werk13}: $W_{\ion{Mg}{ii}}$ is measured with Keck/HIRES spectra of the background quasars (resolution $6\kms$, FWHM). This sample includes 26 star forming galaxies, of which only three are at $0.15<R_\perp/R_{\rm vir}<0.25$ (see Fig.~\ref{fig: WMgIIAgainstR_R_virComparisonWithObservations_m12i}). We use $\EWmgii=1.14 \pm 0.55$\AA\ based on these three objects. The seven objects with lower $R_\perp/R_{\rm vir}$ have a similar $\EWmgii=1.2 \pm 0.3$\AA, while the five objects with $0.25<R_\perp/R_{\rm vir}<0.35$ have a lower $\EWmgii=0.23 \pm 0.07$\AA. 
    \item SDSS \citep{Lan18}: $\EWmgii$ are measured on composite SDSS spectra (resolution $150\kms$) from the DR14 quasar catalog. Foreground galaxies include emission line galaxies and LRGs from the BOSS and eBOSS surveys.  We use the median $M_\star$ of each survey to calculate $R_\perp=0.2\,\Rvir$ and then derive $\EWmgii(R_\perp=0.2\Rvir)$ by interpolating the $\EWmgii$ versus $R_\perp$ relation deduced in \citeauthor{Lan18}. 
    \item SDSS \citep{Anand21}: \ion{Mg}{ii} is detected in individual spectra. We calculate $\EWmgii$ by taking the reported average $W_{\ion{Mg}{ii}}$ for detected objects ($>0.4$\AA) and multiplying by the reported covering factor. \item DESI \citep{Wu24}: Background quasars and foreground SF galaxies are selected from the DESI internal release `Iron', with the foreground galaxies divided into two redshift bins.  $\EWmgii$ are measured both on stacked and individual spectra (resolution $\approx100\kms$). We use the median $M_\star$ and $z$ of each redshift bin to calculate $R_\perp=0.2\,\Rvir$, and then derive $\EWmgii(R_\perp=0.2\Rvir)$ by interpolating the $\EWmgii$ versus $R_\perp$ relation deduced in \citeauthor{Wu24}
    \item \cite{Huang21}: $\EWmgii$ measured in MagE spectra of background quasars (resolution $\approx70\kms$, FWHM) on the Magellan Clay Telescope.  Foreground galaxies are selected from SDSS DR6 (see \citealt{Chen10} for survey design). We use the value and uncertainty from their derived  $W_{\lambda2796}$ vs.\ $R_\perp/R_{\rm vir}$ relation to derive  $W_{\lambda2796}(0.2R_{\rm vir})$ and multiply by two to get $\EWmgii$, as expected if the lines are fully saturated. Their mean relation for isolated galaxies is consistent with that found by the MAGIICAT survey \citep{Nielsen13}.
    \item MEGAFLOW \citep{Cherrey25}: A sample based on the combination of Multi Unit Spectrograph Explorer (MUSE) galaxy observations and Ultraviolet and Visible Echelle Spectrograph (UVES) observations of 22 quasar fields. We use the median $M_\star$ of objects at $R_\perp<50\kpc$ to calculate $R_\perp=0.2\,\Rvir$ and then derive $W_{2796}(R_\perp=0.2\Rvir)$ by interpolating their $W_{2796}$ versus $R_\perp$ relation and multiplying by two to get $\EWmgii$.  
    \item Gravitational arcs \citep{Lopez20,Mortensen21}: two galaxies which inner CGM are probed by multiple sightlines towards background gravitationally-lensed arcs. We use the $W_{2796}(R_\perp)$ relation deduced from the SW arc in \cite{Lopez20} and both arcs in \cite{Mortensen21}, while for $0.2\Rvir$ we use the estimates of $25\kpc$ and $27\kpc$ in the papers. We calculate $\EWmgii$ using $\EWmgii=2W_{2796}(0.2\Rvir)$, as expected if the lines are fully saturated.  
\end{itemize}

\label{lastpage}
\end{document}